\documentclass[epsfig,12pt]{article}
\usepackage{makeidx}
\usepackage{amsmath}
\usepackage{amsfonts}
\usepackage{amssymb}
\usepackage{graphicx}

\input epsf.sty
\newtheorem{theorem}{Theorem}
\newtheorem{acknowledgement}[theorem]{Acknowledgement}

\makeatletter \@addtoreset{equation}{section}
\renewcommand{\theequation}{\thesection.\arabic{equation}}
\input{tcilatex}
\begin{document}

\title{\vspace{-2cm}\rightline{\mbox{\small {LPHE-MS rabat}}
\vspace {1cm}} \textbf{Twisted }$\emph{3D}$ $\mathcal{N}=\mathbf{4}$ \textbf{%
Supersymmetric YM on deformed }$\mathbb{A}_{\mathbf{3}}^{\mathbf{\ast }}$%
\textbf{\ Lattice}}
\author{El Hassan Saidi$\thanks{%
h-saidi@fsr.ac ma}$ \\
{\small 1. LPHE-Modeling and Simulations, Faculty Of Sciences, Rabat, Morocco%
}\\
{\small 2. Centre of Physics and Mathematics, CPM- Morocco}}
\maketitle

\begin{abstract}
We study a class of twisted \emph{3D} $\mathcal{N}=4$ supersymmetric
Yang-Mills (SYM) theory on particular 3- dimensional lattice denoted as $%
\mathcal{L}_{3D}^{su_{3}\times u_{1}}$ and given by non trivial fibration $%
\mathcal{L}_{1D}^{u_{1}}\times \mathcal{L}_{2D}^{su_{3}}$ with base $%
\mathcal{L}_{2D}^{su_{3}}=\mathbb{A}_{2}^{\ast }$, the weight lattice of $%
SU\left( 3\right) $. We first, develop the twisted \emph{3D} $\mathcal{N}=4$
SYM in continuum by using superspace method where the scalar supercharge $Q$
is manifestly exhibited. Then, we show how to engineer the 3D lattice $%
\mathcal{L}_{3D}^{su_{3}\times u_{1}}$ that host this theory. After that we
build the lattice action $\mathcal{S}_{latt}$ invariant under the 3
following: $\left( i\right) $ $U\left( N\right) $ gauge invariance, $\left(
ii\right) $ BRST symmetry, $\left( iii\right) $ the hidden $SU\left(
3\right) \times U\left( 1\right) $ symmetry of $\mathcal{L}%
_{3D}^{su_{3}\times u_{1}}$. Other features such as reduction to twisted 
\emph{2D} supersymmetry with \emph{8} supercharges living on $\mathcal{L}%
_{2D}^{su_{2}\times u_{1}}$, the extension to twisted maximal \emph{5D} SYM
with \emph{16} supercharges on lattice $\mathcal{L}_{5D}^{su_{4}\times
u_{1}} $ as well as the relation with known results are also given.

\  \  \  \  \  \  \  \newline
\textbf{Keywords}: Reduction of chiral \emph{6D} $\mathcal{N}=\left(
1,0\right) $ SYM, BRST symmetry and {Scalar supersymmetry,} Twisted {SYM} on
lattice{, Root and weight lattices of }${SU}\left( {k}\right) $.
\end{abstract}

\tableofcontents

\section{Introduction}

Following \textrm{\cite{1A}-\cite{6A} }and refs therein, the lattice version
of maximal euclidian\textrm{\footnote{%
Euclidian QFTs are generally thought of in terms of a Wick rotation of
corresponding Lorentzian QFTs. However this analytic continuation is not a
soft operation especially for spinors. This issue is not directly addressed
in this paper; but results of the Osterwalder-Schrader (OS) method are used.
For more details on this issue, including the OS method and other approaches
to overcome difficulties induced by analytic continuation, see \textrm{\cite%
{V}} and refs therein; see also eq(\ref{tab}) to fix the ideas. }} four
dimensional $\mathcal{N}=4$ supersymmetric Yang Mills theory (SYM) with $%
U\left( N\right) $ gauge invariance may be approached by twisting
supersymmetry and requiring invariance under the scalar supercharge $Q$ of
the resulting twisted gauge theory. In this method, the $16=2^{4}$
supersymmetric charges $\left( Q_{\alpha}^{i},Q_{\dot{\alpha}i}\right) $,\
transforming in the spinorial representation of $SO_{E}\left( 4\right)
\times SO_{R}\left( 6\right) $, are thought of in terms of $2^{2}\times2^{2}$
matrix $\mathbb{Q}_{4\times4}$ that can be expanded on products of $4\times4$
gamma $\gamma^{\mu}$ matrices;\textrm{\ for }a general review see \textrm{%
\cite{1A,1B} and \cite{2B}-\cite{5B} }for related works. The expansion of $%
\mathbb{Q}_{4\times4}$ leads, on one hand, to the integral spin
decomposition 
\begin{equation}
\mathbb{Q}_{4\times4}=IQ+\gamma^{\mu}Q_{\mu}+\gamma^{\left[ \mu \nu \right]
}Q_{\mu \nu}+\gamma^{\mu}\gamma_{5}\tilde{Q}_{\mu}+\gamma_{5}\tilde{Q}
\label{44}
\end{equation}
where the \emph{16} supercharges are split as $16=1+4+6+4+1$; and, on the
other hand, to a remarkable packaging of the field spectrum of the twisted
4D $\mathcal{N}=4$ SYM theory into $SU\left( 5\right) \times U\left(
1\right) $ representations like%
\begin{equation}
\begin{tabular}{lllll}
bosons & : & $10$ & $\rightarrow$ & $5\oplus \bar{5}$ \\ 
fermions & : & $16$ & $\rightarrow$ & $1\oplus \bar{5}\oplus10$%
\end{tabular}
\   \label{U1}
\end{equation}
Because of the algebraic property $Q^{2}=0$, the scalar supercharge behaves
as a topological object \textrm{\cite{1C,2C}}; a feature that allows to: $%
\left( i\right) $ put the fields of the twisted $\mathcal{N}=4$
supersymmetric Yang Mills on a 4D lattice $\mathbb{A}_{4}^{\ast}$ with a
hidden $SU\left( 5\right) $ symmetry \textrm{\cite{5A,6A}}; and $\left(
ii\right) $ write down a $U\left( N\right) $ gauge invariant lattice field
action $\mathcal{S}_{latt}$ having, in addition to the $SU\left( 5\right) $
symmetry of $\mathbb{A}_{4}^{\ast}$, a BRST symmetry generated by $Q$
governing its quantum properties \textrm{\cite{6A}}.\ 

\  \  \  \  \  \newline
In this paper, we borrow this idea to study the lattice version of the class
of twisted \emph{3D} supersymmetric Yang-Mills theories with $\emph{8}$
supercharges having an $SU\left( 3\right) \times U\left( 1\right) $ symmetry.%
\emph{\ }This twisted \emph{3D} supersymmetric YM theory follows from the
reduction of chiral $\mathcal{N}=\left( 1,0\right) $ SYM in \emph{6D }and
living on a particular \emph{3D} lattice to be built in the present work
(see section 6). Our interest into this class of twisted YM theories has
been motivated by the two following:

\begin{description}
\item[1)] extend the approach of \textrm{\cite{1A,2A}} to the class of
lattice supersymmetric YM models based on twisting SYM theories with \emph{8}
supercharges. It turns out that the twisted \emph{3D} lattice gauge theory
is very suggestive; it lives on a particular crystal denoted here as 
\begin{equation*}
\mathcal{L}_{3D}^{su_{3}\times u_{1}}
\end{equation*}
having a hidden $SU\left( 3\right) \times U\left( 1\right) $ symmetry; and
given by a non trivial fibration $\mathcal{L}_{2D}^{su_{3}}\times \mathcal{L}%
_{1D}^{u_{1}}$ with 2-dimensional base sublattice $\mathcal{L}_{2D}^{su_{3}}=%
\mathbb{A}_{2}^{\ast}$ and fiber $\mathcal{L}_{1D}^{u_{1}}$ isomorphic to $%
\mathbb{Z},$ the set of integers. This kind of fibration, encoded by eq(\ref%
{fbb}), allows also to get more insight into literature results; especially
in the case of twisted maximal supersymmetry living on the lattice $\mathbb{A%
}_{4}^{\ast}$ with$\ SU\left( 5\right) $ symmetry.\newline
\  \  \  \  \  \  \  \newline
To approach the case of twisted SYM with \emph{8} supercharges, and in a
subsequent step the class with \emph{16} supercharges as done in section 8,
we develop a method of engineering $\left( k+1\right) $-dimensional crystals
with $SU\left( k\right) \times U\left( 1\right) $ symmetry; and use results
on the breaking mode of the real $SO\left( 2k\right) $ euclidian symmetries
down to the complex 
\begin{equation*}
SU\left( k\right) \times U\left( 1\right) ,\qquad k\text{ odd integer}
\end{equation*}
to get the packaging of the twisted fields into representations of $SU\left(
k\right) \times U\left( 1\right) $; and also to determine their
interpretation on lattice $\mathcal{L}_{kD}^{su_{k}\times u_{1}}$ in terms
of links and plaquettes.\newline
\  \  \  \  \  \newline
Recall that \emph{4D} $\mathcal{N}=4$ supersymmetric Yang Mills with $%
SO_{E}\left( 4\right) \times SO_{R}\left( 6\right) $ symmetry is a maximal
supersymmetric YM theory that has the same number of conserved supercharges
as $\mathcal{N}=\left( 1,0\right) $ SYM in euclidian 10-dimensions with
isotropy symmetry 
\begin{equation*}
SO_{E}\left( 10\right) ,\qquad k=5
\end{equation*}
Similarly, the twisted \emph{3D} $\mathcal{N}=4$ YM theory we are interested
in here can be obtained in a quite analogous manner; but by dimensional
reduction of the chiral $\mathcal{N}=\left( 1,0\right) $ SYM in 6-dimensions
with euclidian symmetry%
\begin{equation*}
SO_{E}\left( 6\right) ,\qquad k=3
\end{equation*}

\item[2)] explore the role of the extra $U\left( 1\right) $ symmetry that
appears in twisted supersymmetric YM theories; in particular in the case of 
\emph{8} supercharges with $SU\left( 3\right) \times U\left( 1\right) $
symmetry; and also in twisted maximal supersymmetry with an $SU\left(
5\right) \times U\left( 1\right) $. A way to exhibit this global abelian
invariance is through the breaking of $SO_{E}\left( 2k\right) \ $down to $%
SU\left( k\right) \times U\left( 1\right) ,$ which for $k=5$ and $k=3,$ read
respectively as follows 
\begin{equation*}
\begin{tabular}{lll}
$SO_{E}\left( 10\right) $ \  \  \  & $\rightarrow$ & $\  \ SU\left( 5\right)
\times U\left( 1\right) $ \\ 
$SO_{E}\left( 6\right) $ & $\rightarrow$ & $\  \ SU\left( 3\right) \times
U\left( 1\right) $%
\end{tabular}%
\end{equation*}
Under these symmetry breaking modes, real vector \underline{$\mathbf{2k}$}
and spinorial \underline{$\mathbf{2}$}$^{k-1}$ representations of $%
SO_{E}\left( 10\right) $ and $SO_{E}\left( 6\right) $ decompose as sums of
representations with respect to the complex symmetries. For $SO_{E}\left(
10\right) $, the decomposition of the $10_{v}$ and the $16_{s}$ are given by%
\begin{equation*}
\begin{tabular}{lll}
$SO_{E}\left( 10\right) $ \  \  & $\rightarrow$ & $SU\left( 5\right) \times
U\left( 1\right) $ \\ 
$\  \  \  \ 10_{v}$ & \ : & $\  \ 5_{+2q}+\bar{5}_{-2q}$ \\ 
$\  \  \  \ 16_{s}$ & \ : & $\  \ 1_{-5q}+\bar{5}_{+3q}+10_{-q}$ \\ 
&  & 
\end{tabular}%
\end{equation*}
with $q$ a unit $U\left( 1\right) $ charge; and for the case of $%
SO_{E}\left( 6\right) $, their analogue read like%
\begin{equation*}
\begin{tabular}{lll}
$SO_{E}\left( 6\right) $ \  \  & $\rightarrow$ & $SU\left( 3\right) \times
U\left( 1\right) $ \\ 
$\  \  \  \ 6_{v}$ & \ : & $\  \ 3_{+2q}+\bar{3}_{-2q}$ \\ 
$\  \  \  \ 4_{s}$ & \ : & $\  \ 1_{-3q}+3_{+q}$ \\ 
&  & 
\end{tabular}%
\end{equation*}
These breakings show that in twisted supersymmetric YM theories, the twisted
fields and the twisted supersymmetric operators, in particular the BRST
charge $Q^{\left( -kq\right) }$, are in general sections of a $U\left(
1\right) $ bundle. This property teaches us in turns that on lattice side $%
SU\left( 3\right) $ scalars carrying non trivial U$\left( 1\right) $ charges
have also a non trivial interpretation; they are associated with links along
the 1- dimensional fiber and, in some sense, constitute a refining of
results of \textrm{\cite{1A,2A} since a similar }conclusion is also valid
for the case of twisted maximal supersymmetry on the lattice $\mathcal{L}%
_{5D}^{su_{5}\times u_{1}}$.
\end{description}

\  \  \  \newline
In what follows, we focus on the study of the lattice version of twisted 
\emph{3D} $\mathcal{N}=4$ SYM; and, to exhibit the role played by $U\left(
1\right) $ subsymmetry, we distinguish the two cases: the generic $q\neq0$
and the singular $q=0$. A similar analysis can be performed for the case of
twisted maximal supersymmetric YM in \emph{5D} as reported in the section
conclusion and comments.

\  \  \newline
The organization is as follows: In sections 2 and 3, we first review some
useful tools on $SO\left( t,s\right) $ spinors in diverse dimensions $D=t+s$%
. Then, we study the twisted \emph{3D} $\mathcal{N}=4$ supersymmetric YM
theory in continuum. In section 4, we build the action in superspace and
derive its component field expression. In section 5, we study the twisted 
\emph{3D} $\mathcal{N}=4$ supersymmetric YM on the base sublattice $\mathbb{A%
}_{2}^{\ast}$ having $SU\left( 3\right) $ symmetry and corresponding to the
singular limit $q=0$. In section 6, we study twisted \emph{3D} $\mathcal{N}%
=4 $ SYM on the \emph{3D} crystal $\mathcal{L}_{3D}^{su_{3}\times u_{1}}$
with $SU\left( 3\right) \times U\left( 1\right) $ symmetry and $q\neq0$. In
section 7, we build the action of the twisted field on $\mathcal{L}%
_{3D}^{su_{3}\times u_{1}}$. In section 8, we give a conclusion and make two
comments; one on the reduction down to twisted \emph{2D} $\mathcal{N}=4$ and
the second concerns the extension to \emph{5D} $\mathcal{N}=4$ on the
lattice $\mathcal{L}_{5D}^{su_{5}\times u_{1}}$ containing $\mathbb{A}%
_{4}^{\ast}$ as a base sublattice. In section 9, we give an appendix where
we give explicit computations and technical details on the construction of
gauge covariant superfields.

\section{Twisted SYM with 8 supercharges}

After recalling useful tools on $SO\left( t,s\right) $ spinors in diverse
dimensions and briefly describing the reduction of chiral $\mathcal{N}%
=\left( 1,0\right) $ SYM in Lorentzian $\emph{6D}$ down to \emph{3}$D$, we
study twisted \emph{3D} $\mathcal{N}=4$ SYM in continuum. More on continuum
and the lattice version of this SYM theory will be developed in next
sections.

\subsection{Generalities on spinors in D-dimensions}

Here we collect some results on $SO\left( t,s\right) $ spinors living on the
flat space $\mathbb{R}^{\left( t,s\right) }$ with space time dimension $%
D=s+t $ and signature $s-t$ where $s$ and $t$ stand respectively for the
numbers of space like and time like directions. A particular interest will
be given to the Lorentzian ($t=1$) and euclidian ($t=0$) signatures that we
are interested in this work.

\subsubsection{\emph{Classification of} $so\left( t,s\right) $ \emph{spinors}%
}

Generally speaking, spinors $\Psi_{A}$ living on $\mathbb{R}^{\left(
t,s\right) }$ with metric $\eta_{MN}=$ $diag\left( -..-,+..+\right) $ have
complex $2^{\left[ \frac{D}{2}\right] }$ components transforming under the
space isotropy symmetry $SO\left( t,s\right) $ as%
\begin{equation}
\begin{tabular}{llll}
$\Psi_{A}$ & $\rightarrow$ & $\Psi_{A}^{\prime}$ & $=S_{A}^{B}\Psi_{B}$%
\end{tabular}
\label{dir}
\end{equation}
with matrix transformation given by%
\begin{equation}
\begin{tabular}{lll}
$S$ & $=$ & $e^{\frac{i}{4}\omega_{MN}\Sigma^{\left[ MN\right] }}$ \\ 
$\Sigma^{\left[ MN\right] }$ & $=$ & $\Gamma^{M}\Gamma^{N}-\Gamma^{N}%
\Gamma^{M}$%
\end{tabular}
\label{ir}
\end{equation}
In these relations, the $2^{\left[ \frac{D}{2}\right] }\times2^{\left[ \frac{%
D}{2}\right] }$ matrices $\Gamma_{M}$ are the usual gamma matrices
generating the Clifford algebra $Cl\left( t,s\right) $ defined by 
\begin{equation}
\Gamma_{M}\Gamma_{N}+\Gamma_{N}\Gamma_{M}=2\eta_{MN}
\end{equation}
The matrices $\Gamma_{M}$ are generally realized in terms of particular
monomials of tensor products of the usual $2\times2$ hermitian Pauli
matrices $\sigma_{1},$ $\sigma_{2},$ $\sigma_{3}$ that we take as%
\begin{equation}
\begin{tabular}{lll}
$\sigma_{1}=\left( 
\begin{array}{cc}
0 & 1 \\ 
1 & 0%
\end{array}
\right) ,$ & $\sigma_{2}=\left( 
\begin{array}{cc}
0 & -i \\ 
i & 0%
\end{array}
\right) =i\varepsilon,$ & $\sigma_{3}=\left( 
\begin{array}{cc}
1 & 0 \\ 
0 & -1%
\end{array}
\right) $%
\end{tabular}%
\end{equation}
obeying amongst others the property $\sigma_{1}^{T}=\sigma_{1},$ $\sigma
_{3}^{T}=\sigma_{3}$ but $\sigma_{2}^{T}=-\sigma_{2}$. For the example of
the euclidian $\left( 0,6\right) $ signature, where the metric $\eta_{MN}$
coincides with the Kronecker symbol $\delta_{MN}$, the $\Gamma_{M}$'s are $%
8\times8$ matrices realized as%
\begin{equation}
\begin{tabular}{llll}
$\Gamma_{1}$ & $=$ & $\sigma_{1}\otimes \mathbb{I}\otimes \mathbb{I}$ &  \\ 
$\Gamma_{2}$ & $=$ & $\sigma_{2}\otimes \mathbb{I}\otimes \mathbb{I}$ &  \\ 
$\Gamma_{3}$ & $=$ & $\sigma_{3}\otimes \sigma_{1}\otimes \mathbb{I}$ &  \\ 
$\Gamma_{4}$ & $=$ & $\sigma_{3}\otimes \sigma_{2}\otimes \mathbb{I}$ &  \\ 
$\Gamma_{5}$ & $=$ & $\sigma_{3}\otimes \sigma_{3}\otimes \sigma_{1}$ &  \\ 
$\Gamma_{6}$ & $=$ & $\sigma_{3}\otimes \sigma_{3}\otimes \sigma_{2}$ &  \\ 
$\Gamma_{7}$ & $=$ & $\sigma_{3}\otimes \sigma_{3}\otimes \sigma_{3}$ & 
\end{tabular}
\label{rea}
\end{equation}
with the remarkable properties%
\begin{equation}
\begin{tabular}{lll}
$\left( \Gamma_{i}\right) ^{\dagger}$ & $=$ & $+\Gamma_{i}$ \\ 
$\left( \Gamma_{2k+1}\right) ^{T}$ & $=$ & $-\Gamma_{2k+1}$ \\ 
$\left( \Gamma_{2k}\right) ^{T}$ & $=$ & $+\Gamma_{2k}$ \\ 
&  & 
\end{tabular}%
\end{equation}
For the cases of the Lorentzian $\left( 1,5\right) $ and $\left( 2,4\right) $
signatures, the realization of the corresponding $\Gamma_{M}$'s is obtained
from the euclidian representation by using Wick like rotations as follows%
\begin{equation}
\begin{tabular}{ll|llll}
\  \  \  \  \ {\small signature (}${\small 1,5}${\small )} &  &  & 
\multicolumn{3}{l}{\  \  \  \  \  \  \ {\small signature (}${\small 2,4}${\small )}
} \\ \hline
&  &  &  &  &  \\ 
$\Upsilon_{0}=i\Gamma_{1}$ &  &  & $\Upsilon_{0}=i\Gamma_{1}$ & , & $%
\Upsilon_{1}=i\Gamma_{2}$ \\ 
$\Upsilon_{m}=\Gamma_{m+1},$ \ $m>0$ &  &  & \multicolumn{3}{l}{$\Upsilon
_{m}=\Gamma_{m+1},$ \ $m>1$} \\ 
&  &  &  &  &  \\ \hline
\end{tabular}%
\end{equation}%
\begin{equation*}
\end{equation*}
The complex spinorial field $\Psi_{A}$, to which we refer to as $so\left(
t,s\right) $ Dirac spinors, exhibits several features whose useful ones are
summarized below:

\ 

$\left( i\right) $ \emph{adjoint spinors}\newline
Along with the complex $\Psi$, one has three cousin spinors namely $\Psi^{T}$%
, $\Psi^{\ast}$ and $\Psi^{\dagger}=\Psi^{\ast T}$ respectively associated
with the Clifford algebras generated by 
\begin{equation}
\begin{tabular}{lllll}
$\Gamma_{M}^{T}$ & , & $\Gamma_{M}^{\ast}$ & , & $\Gamma_{M}^{\dagger}$%
\end{tabular}%
\end{equation}
and which are related to $\Gamma_{M}$ by similarity transformations as given
below%
\begin{equation}
\begin{tabular}{lll}
$\Gamma_{M}^{\dagger}$ & $=$ & $\left( -\right) ^{t}\mathcal{A}\Gamma _{M}%
\mathcal{A}^{-1}$ \\ 
$\Gamma_{M}^{T}$ & $=$ & $-\eta \mathcal{C}\Gamma_{M}\mathcal{C}^{-1}$ \\ 
$\Gamma_{M}^{\ast}$ & $=$ & $-\eta \left( -\right) ^{t}\mathcal{B}\Gamma _{M}%
\mathcal{B}^{-1}$ \\ 
&  & 
\end{tabular}%
\end{equation}
with%
\begin{equation}
\begin{tabular}{lll}
$\mathcal{A}$ & $=$ & $\Gamma_{1}..\Gamma_{t}$ \\ 
$\mathcal{C}^{T}$ & $=$ & $-\varepsilon \mathcal{C}$ \\ 
$\mathcal{B}^{T}$ & $=$ & $\mathcal{CA}^{-1}$%
\end{tabular}%
\end{equation}
and%
\begin{equation}
\begin{tabular}{lll}
$\mathcal{A}$ & $=$ & $\eta^{t}\left( -\right) ^{\frac{t\left( t+1\right) }{2%
}}\mathcal{C\mathcal{A}C}^{-1}$ \\ 
&  &  \\ 
$\mathcal{B}^{\ast}\mathcal{B}$ & $=$ & $-\varepsilon \eta^{t}\left(
-\right) ^{\frac{t\left( t+1\right) }{2}}$ \\ 
&  & 
\end{tabular}%
\end{equation}
and where $\varepsilon=\pm1$ and $\eta=\pm1$. We also have%
\begin{equation}
\begin{tabular}{lll}
&  &  \\ 
$\left( \mathcal{C}\Gamma_{M}\right) ^{T}$ & $=$ & $\varepsilon \eta 
\mathcal{C}\Gamma_{M}$ \\ 
&  &  \\ 
$\left( \mathcal{C}\Gamma_{1}..\Gamma_{m}\right) ^{T}$ & $=$ & $-\varepsilon
\eta^{m}\left( -\right) ^{\frac{m\left( m+1\right) }{2}}\left( \mathcal{C}%
\Gamma_{1}..\Gamma_{m}\right) $%
\end{tabular}%
\end{equation}%
\begin{equation*}
\end{equation*}
Notice that for odd dimensions, there is one solution for the matrix $%
\mathcal{C}$; but for even dimensions, we distinguish two kinds of possible 
\emph{C} matrices as illutrated below on the example of $D=6:$%
\begin{equation}
\begin{tabular}{llllllll}
$\mathcal{C}_{+}$ & $=$ & $\sigma_{2}\otimes \sigma_{1}\otimes \sigma_{2}$ & 
, & for & $\eta=+1$ & , & $\varepsilon=-1$ \\ 
$\mathcal{C}_{-}$ & $=$ & $\sigma_{1}\otimes \sigma_{2}\otimes \sigma_{1}$ & 
, & for & $\eta=-1$ & , & $\varepsilon=+1$%
\end{tabular}%
\end{equation}%
\begin{equation*}
\end{equation*}
satisfying $\left( \mathcal{C}_{\pm}\right) ^{T}=\pm \mathcal{C}_{\pm}$ and $%
\left( \mathcal{C}_{\pm}\right) ^{2}=I_{id}$. Notice moreover that for 6D,
both of the matrices $\mathcal{C}_{+}$ and $\mathcal{C}_{-}$ have as product 
$\varepsilon \eta=-1$; and so%
\begin{equation}
\left( \mathcal{C}\Gamma_{M}\right) ^{T}=-\mathcal{C}\Gamma_{M}
\end{equation}
which is an undesirable property that requires the use of $SU\left( 2\right) 
$ symplectic spinors $Q_{A}^{i}=\left( Q_{A}^{1},Q_{A}^{2}\right) $ in order
to recover the symmetric feature of the anticommutation relation between the
supercharges of supersymmetric YM theory in 6D namely%
\begin{equation}
Q_{A}^{i}Q_{B}^{j}+Q_{B}^{j}Q_{A}^{i}=\varepsilon^{ij}\left( \mathcal{C}%
\Gamma^{M}\right) _{AB}P_{M}
\end{equation}
where $\varepsilon_{ij}$ is the usual $2\times N$ antisymmetric matrix
obeying $\left( \varepsilon_{ij}\right)
^{\ast}\varepsilon_{jl}=-\delta_{l}^{i}$.

\  \  \  \ 

$\left( ii\right) $ \emph{Weyl spinors}\newline
In odd dimensions, the Dirac fermion $\Psi$ satisfying (\ref{dir}-\ref{ir})
is an irreducible spinor; but in even dimensions, say $D=2k$, it can be
decomposed into two irreducible Weyl spinors $\Psi_{L}$ and $\Psi_{R}$
having each $2^{k-1}$ complex components.%
\begin{equation}
2^{k}=2^{k-1}\oplus2^{k-1}
\end{equation}
The two chiral spinors $\Psi_{L}$ and $\Psi_{R}$ are related to the Dirac $%
\Psi$ through the following projections 
\begin{equation}
\begin{tabular}{lll}
$\Psi_{L}$ & $=$ & $\frac{1}{2}\left( I+\Gamma_{{\small D+1}}\right) \Psi$
\\ 
$\Psi_{R}$ & $=$ & $\frac{1}{2}\left( I-\Gamma_{{\small D+1}}\right) \Psi$%
\end{tabular}%
\end{equation}
with chirality operator $\Gamma_{{\small D+1}}=\left( -i\right)
^{k+t}\Gamma_{1}..\Gamma_{2k}$ which, by using the realization (\ref{rea}),
reads as%
\begin{equation}
\begin{tabular}{llll}
$\Gamma_{{\small D+1}}$ & $=$ & $\left( -i\right) ^{\frac{D}{2}}\Gamma
_{1}..\Gamma_{D}$ &  \\ 
& $=$ & $\underset{k}{\underbrace{\sigma_{3}\otimes \sigma_{3}\otimes
..\otimes \sigma_{3}}}$ & 
\end{tabular}
\label{dd}
\end{equation}
This matrix obeys $\Gamma_{{\small D+1}}\Gamma_{{\small D+1}}=I$ and is
independent of the space time signature. Notice that in even dimensions, the
anticommutation relations between two Weyl supercharges say the left one%
\begin{equation}
\begin{tabular}{lll}
$Q_{L}$ & $=$ & $\frac{1}{2}\left( I+\Gamma_{{\small D+1}}\right) Q$%
\end{tabular}%
\end{equation}
read as follows%
\begin{equation}
\left \{ Q_{L},Q_{L}\right \} =\frac{1-\left( -\right) ^{\frac{D}{2}}}{4}%
\left( I+\Gamma_{{\small D+1}}\right) \mathcal{C}\Gamma^{M}P_{M}  \label{od}
\end{equation}%
\begin{equation*}
\end{equation*}
Therefore, non vanishing anticommutators limits the possible dimensions
where $\left \{ Q_{L},Q_{L}\right \} \neq0$ since the non vanishing
condition requires that $\frac{D}{2}$ has to be odd; i.e: 
\begin{equation}
D=4l+2,\qquad l=1,2,...
\end{equation}
The leading dimensions where this is possible are \emph{D=2}, \emph{D=6} and 
\emph{D=10}.

\  \  \  \  \ 

$\left( iii\right) $ \emph{reality conditions}\newline
Complex Dirac spinors might be also subject to reality conditions such as
Majorana or Majorana Weyl conditions. These conditions are not usually
possible since reality condition depends both on the space time dimension 
\emph{D} and the signature $s-t$. The general result on the possibility of
putting a reality condition on spinors in diverse Lorentzian and euclidian
dimensions is collected in the following table \textrm{\cite{7R}-\cite{10R}},%
\begin{equation*}
\end{equation*}%
\begin{equation}
\begin{tabular}{lllll}
{\small dimension} $D$ &  & {\small Lorentzian} $\mathbb{R}^{1,D-1}$ &  & 
{\small Euclidian} $\mathbb{R}^{D}$ \\ \hline
&  &  &  &  \\ 
${\small \  \  \  \  \  \  \  \  \ }${\small 1} &  & ${\small \  \  \  \  \  \  \  \  \ M}$
&  & ${\small \  \  \  \  \  \  \  \  \ M}$ \\ 
${\small \  \  \  \  \  \  \  \  \ }${\small 2} &  & ${\small \  \  \  \  \  \  \  \  \ MW}$
&  & ${\small \  \  \  \  \  \  \  \  \ M}^{-}$ \\ 
${\small \  \  \  \  \  \  \  \  \ }${\small 3} &  & ${\small \  \  \  \  \  \  \  \  \ M}$
&  & ${\small \  \  \  \  \  \  \  \  \ SM}$ \\ 
${\small \  \  \  \  \  \  \  \  \ }${\small 4} &  & ${\small \  \  \  \  \  \  \  \  \ M}%
^{+}$ &  & ${\small \  \  \  \  \  \  \  \  \ SMW}$ \\ 
${\small \  \  \  \  \  \  \  \  \ }${\small 5} &  & ${\small \  \  \  \  \  \  \  \  \ SM}$
&  & ${\small \  \  \  \  \  \  \  \  \ SM}$ \\ 
${\small \  \  \  \  \  \  \  \  \ }${\small 6} &  & ${\small \  \  \  \  \  \  \  \  \ SMW}$
&  & ${\small \  \  \  \  \  \  \  \  \ M}^{+}$ \\ 
${\small \  \  \  \  \  \  \  \  \ }${\small 7} &  & ${\small \  \  \  \  \  \  \  \  \ SM}$
&  & ${\small \  \  \  \  \  \  \  \  \ M}$ \\ 
${\small \  \  \  \  \  \  \  \  \ }${\small 8} &  & ${\small \  \  \  \  \  \  \  \  \ M}%
^{-}$ &  & ${\small \  \  \  \  \  \  \  \  \ MW}$ \\ 
${\small \  \  \  \  \  \  \  \  \ }${\small 9} &  & ${\small \  \  \  \  \  \  \  \  \ M}$
&  & ${\small \  \  \  \  \  \  \  \  \ M}$ \\ 
${\small \  \  \  \  \  \  \  \  \ }${\small 10} &  & ${\small \  \  \  \  \  \  \  \  \ MW}$
&  & ${\small \  \  \  \  \  \  \  \  \ M}^{-}$ \\ 
${\small \  \  \  \  \  \  \  \  \ }${\small 11} &  & ${\small \  \  \  \  \  \  \  \  \ M}$
&  & ${\small \  \  \  \  \  \  \  \  \ SM}$ \\ 
&  &  &  &  \\ \hline
\end{tabular}
\label{tab}
\end{equation}%
\begin{equation*}
\end{equation*}
with $M$ standing for Majorana and $MW$ for Majorana Weyl spinors. We also
have $M^{\pm}$ referring to Majorana spinors with charge conjugation $%
C_{\pm} $. Notice that in the case where there is no Majorana spinor, one
can have symplectic Majorana spinors or symplectic Majorana Weyl spinors
referred in the table respectively by the symbols $SM$ and \emph{SMW}. By
symplectic Majorana spinor, we mean a set of 2N Dirac spinors $%
\Psi_{A}^{1},..,\Psi _{A}^{2N}$ constrained as follows 
\begin{equation}
\begin{tabular}{llll}
$\left( \Psi_{A}^{i}\right) ^{\ast}$ & $=$ & $\Omega_{ij}\mathcal{B}%
_{A}^{B}\Psi_{B}^{j}$ &  \\ 
&  &  & 
\end{tabular}%
\end{equation}
where $\Omega_{ij}$ is the usual $2N\times2N$ antisymmetric symplectic
matrix obeying $\left( \Omega_{ij}\right) ^{\ast}\Omega_{jl}=-\delta_{l}^{i}$
and where $\left( \mathcal{B}_{A}^{B}\right) ^{\ast}\mathcal{B}%
_{C}^{B}=-\delta_{C}^{A}$. We also have \emph{SMW }whenever $\mathcal{B}$
and $\Gamma_{{\small D+1}}$ commute knowing that 
\begin{equation}
\left( \Gamma_{{\small D+1}}\right) ^{\ast}=\left( -\right) ^{t+\frac {D}{2}}%
\mathcal{B}\Gamma_{{\small D+1}}\mathcal{B}^{-1}
\end{equation}%
\begin{equation*}
\end{equation*}
Since $\left( \Gamma_{{\small D+1}}\right) ^{\ast}=\Gamma_{{\small D+1}}$
due to (\ref{dd}), it follows that $\mathcal{B}$ and $\Gamma_{{\small D+1}}$
commute for $t+\frac{D}{2}=0$ $\func{mod}2$. From the classification table (%
\ref{tab}), we learn a set of interesting features in particular:

\begin{itemize}
\item there is no Majorana spinor in euclidian \emph{3D, 4D, 5D}; and nor in
the Lorentzian \emph{5D, 6D }and\emph{\ 7D}. Therefore, when studying the
euclidian \emph{3D} $\mathcal{N}=4$ SYM and euclidian \emph{5D} $\mathcal{N}%
=4$ SYM, one is constrained to use symplectic Majorana spinors.

\item the Majorana and Majorana-Weyl conditions are not preserved by
analytic continuation from Lorentzian to euclidian signature. This is a well
known problem that has been considered from various view points \textrm{\cite%
{V} and refs therein}; in particular from the approach of
Osterwalder-Schrader where the hermiticity in euclidian space is abandoned 
\textrm{\cite{W}}.
\end{itemize}

\subsubsection{\emph{Chiral supersymmetric YM in} \emph{6D} \emph{and} \emph{%
10D}}

Like for the well known case of Lorentzian maximal SYM theories with \emph{16%
} real supercharges, supersymmetric QFTs with \emph{8} real supercharges can
be formulated in diverse dimensions. These are the \emph{1D} $\mathcal{N}=8,$
\emph{2D} $\mathcal{N}=\left( 4,4\right) ,$ \emph{3D} $\mathcal{N}=4$, \emph{%
4D} $\mathcal{N}=2$ and \emph{5D} $\mathcal{N}=2$ theories; they may be
obtained by reduction of the chiral 
\begin{equation*}
\emph{6D}\text{, }\mathcal{N}=\left( 1,0\right)
\end{equation*}
supersymmetric YM theory\textrm{\footnote{%
Notice that for those dimensions \emph{D} where there is no Mjaorana spinor
like in $D=1+4$ or euclidian \emph{3D }and\emph{\ 5D}, we shall also use the
standard conventional notations \emph{5D} $\mathcal{N}=4$ (\emph{3D} $%
\mathcal{N}=4$) to refer to the 16 real ( 8 real) supercharges although
strictly speaking this convention is not rigorous.}}. This Lorentzian \emph{%
6D} SYM theory can be then viewed as the mother of supersymmetric theories
with \emph{8} supercharges. From this point of view, chiral $\mathcal{N}%
=\left( 1,0\right) $ SYM in \emph{6D} shares a kind of maternity property
with $\mathcal{N}=\left( 1,0\right) $ SYM in \emph{10D} which is the mother
theory of supersymmetric QFTs' \textrm{having 16 supercharges.}

\  \  \  \ 

\emph{Twisted} \emph{3D} $\mathcal{N}=4$ \emph{SYM}\newline
In twisted \emph{3D} $\mathcal{N}=4$ supersymmetric YM theory, the 8 real
supersymmetric charges are represented by a complex $2\times2$ matrix $%
\mathbb{Q}_{2\times2}$ that can be expanded in terms of the \emph{3} Pauli
matrices $\sigma^{\mu}$ as follows 
\begin{equation}
\mathbb{Q}_{2\times2}=QI+Q_{\mu}\sigma^{\mu}  \label{1U}
\end{equation}
Similarly as for the case of eq(\ref{U1}) of twisted maximal supersymmetric
YM theory, the field content of the spectrum of the twisted \emph{3D} $%
\mathcal{N}=4$ SYM theory can be packaged into $SU\left( 3\right) \times
U\left( 1\right) $ representations like%
\begin{equation}
\begin{tabular}{lllll}
bosons & : & $6$ & $\rightarrow$ & $3\oplus \bar{3}$ \\ 
fermions & : & $4$ & $\rightarrow$ & $1\oplus3$%
\end{tabular}
\label{2U}
\end{equation}

\  \  \ 

\emph{Correspondence 3D} $\mathcal{N}=4$ and \emph{5D} $\mathcal{N}=4$%
\newline
Pushing forward the similarity between twisted SYM with \emph{8}
supercharges and twisted maximal supersymmetric YM (\emph{see footnote 2}),
one finds the following correspondence to be established throughout this
study:

\begin{itemize}
\item \emph{twisted SYM with 16 supercharges}%
\begin{equation}
\begin{tabular}{l|l|l}
& \  \ lattice \  \  \  \  \  \  \  \  & hidden symmetry \\ \hline
& \  \  \  \  \  \  \  \  \  \  \  \  \  &  \\ 
\emph{5D} $\mathcal{N}=4$ & $\  \  \  \  \  \mathcal{L}_{5D}^{su_{5}\times u_{1}}$
& $\ SU\left( 5\right) \times U\left( 1\right) $ \\ 
&  &  \\ 
\emph{4D} $\mathcal{N}=4$ & $\  \  \mathcal{L}_{4D}^{su_{5}}=\mathbb{A}%
_{4}^{\ast}$ & $\  \ SU\left( 5\right) $ \\ 
&  &  \\ \hline
\end{tabular}%
\end{equation}%
\begin{equation*}
\end{equation*}
with 5- dimensional lattice $\mathcal{L}_{5D}^{su_{5}\times u_{1}}$ given by
the fibration%
\begin{equation}
\begin{tabular}{lll}
&  &  \\ 
$\mathcal{L}_{1D}^{u_{1}}$ & $\rightarrow$ & $\  \  \mathcal{L}%
_{5D}^{su_{5}\times u_{1}}$ \\ 
&  & $\  \  \  \  \downarrow$ \\ 
&  & $\  \  \mathcal{L}_{4D}^{su_{5}}$ \\ 
&  & 
\end{tabular}%
\end{equation}
The base sublattice $\mathcal{L}_{4D}^{su_{5}}$ is given by the 4-
dimensional lattice \textrm{\cite{1A,2A}}%
\begin{equation}
\mathcal{L}_{4D}^{su_{5}}=\mathbb{A}_{4}^{\ast}
\end{equation}
generated by the 4 fundamental weight vectors%
\begin{equation}
\begin{tabular}{lllll}
$\vec{\Omega}_{1},$ & $\vec{\Omega}_{2},$ & $\vec{\Omega}_{3},$ & $\vec {%
\Omega}_{4}$ & 
\end{tabular}%
\end{equation}
of the $SU\left( 5\right) $ symmetry group. These weight vectors are the
dual of the 4 simple roots of the Lie algebra of $SU\left( 5\right) $;%
\begin{equation}
\begin{tabular}{lllll}
$\vec{a}_{1},$ & $\vec{a}_{2},$ & $\vec{a}_{3},$ & $\vec{a}_{4}$ & 
\end{tabular}%
\end{equation}
generating the 4-dimensional root lattice $\mathbb{A}_{4}$ of \emph{SU}$%
\left( 5\right) $. So the crystal $\mathbb{A}_{4}^{\ast}$ is the dual of $%
\mathbb{A}_{4}$; see later on for further details and \textrm{\cite{1S}-\cite%
{6S}} for related constructions.

\item \emph{twisted SYM with 8 supercharges}%
\begin{equation}
\begin{tabular}{l|l|l}
& \  \ lattice \  \  \  \  \  \  \  \  & hidden symmetry \\ \hline
& \  \  \  \  \  \  \  \  \  \  \  \  \  &  \\ 
\emph{3D} $\mathcal{N}=4$ & $\  \  \  \  \  \mathcal{L}_{3D}^{su_{3}\times u_{1}}$
& $\ SU\left( 3\right) \times U\left( 1\right) $ \\ 
&  &  \\ 
\emph{2D} $\mathcal{N}=4$ & $\  \  \  \  \  \  \mathcal{L}_{2D}^{su_{3}}$ & $\  \
SU\left( 3\right) $ \\ 
&  &  \\ \hline
\end{tabular}%
\end{equation}
with%
\begin{equation}
\begin{tabular}{lll}
$\mathcal{L}_{1D}^{u_{1}}$ & $\rightarrow$ & $\  \  \mathcal{L}%
_{3D}^{su_{3}\times u_{1}}$ \\ 
&  & $\  \  \  \  \downarrow$ \\ 
&  & $\  \  \mathcal{L}_{2D}^{su_{3}}$ \\ 
&  & 
\end{tabular}
\label{bff}
\end{equation}
and base sublattice given by the 2- dimensional lattice \textrm{\cite{7S}-%
\cite{8S}}%
\begin{equation}
\mathcal{L}_{2D}=\mathbb{A}_{2}^{\ast}
\end{equation}
generated by the 2 fundamental weight vectors 
\begin{equation}
\begin{tabular}{lll}
$\vec{\omega}_{1}$ & $,$ & $\vec{\omega}_{2}$%
\end{tabular}%
\end{equation}
of the $SU\left( 3\right) $ symmetry. These weight vectors are the dual of
the \emph{2 }simple roots $\vec{\alpha}_{1},$ $\vec{\alpha}_{2}$ of $%
SU\left( 3\right) $; then $\mathbb{A}_{2}^{\ast}$ is the dual of the
2-dimensional root lattice $\mathbb{A}_{2}$ of SU$\left( 3\right) $ which
may be thought of as the \emph{2D} honeycomb, see fig. \ref{TW} for
illustration. 
\begin{figure}[ptbh]
\begin{center}
\hspace{0cm} \includegraphics[width=10cm]{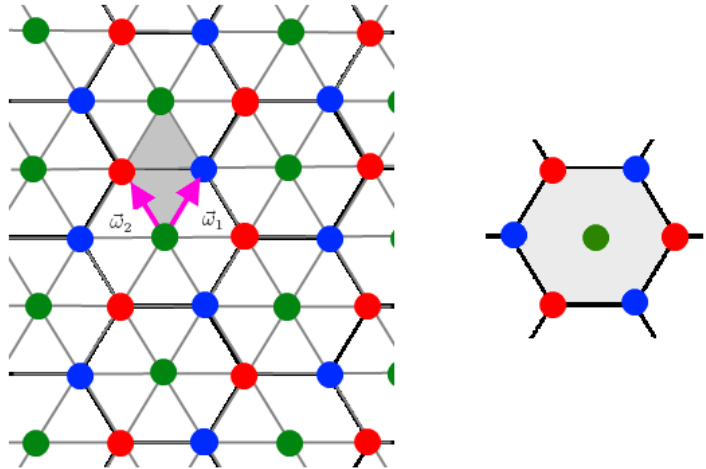}
\end{center}
\par
\vspace{0 cm}
\caption{the 2D lattice $\mathbb{A}_{2}^{\ast}$ generated by $\vec{\protect%
\omega}_{1},$ $\vec{\protect \omega}_{2}$; the 2 fundamental weight vectors
of SU$\left( 3\right) $. Each (green) node in $\mathbb{A}_{2}^{\ast}$ has $%
3+3$ first nearest neighbors forming respectively a triplet (red sites) and
an anti-triplet (blue sites) of $SU\left( 3\right) .$}
\label{TW}
\end{figure}
\end{itemize}

\  \  \  \newline
In what follows, we study the twisted \emph{3D} $\mathcal{N}=4$ SYM in
continuum.\emph{\ }First, we describe some special features on SYM in \emph{%
6D}; then we derive the $SU\left( 3\right) \times U\left( 1\right) $
covariant spectrum of twisted $\mathcal{N}=4$ SYM in \emph{3D.} Next, we
give the twisted $\mathcal{N}=4$ superalgebra having an $SU\left( 3\right)
\times U\left( 1\right) $ isotropy symmetry with supersymmetric generators as%
\begin{equation}
\begin{tabular}{lll}
$Q^{{\small (+3q)}}$ & $,$ & $Q_{a}^{{\small (-q)}}$%
\end{tabular}%
\end{equation}
transforming respectively as a complex scalar and a complex triplet of $%
SU\left( 3\right) $. After, we use superspace method to realize the scalar
supersymmetric charge $Q^{{\small (+3q)}}$ which may be also thought of as a
BRST charge operator.

\subsection{Twisted $\mathcal{N}=4$ SYM in \emph{3D }from $\mathcal{N}=1$ SYM%
$_{6}$}

In Lorentzian \emph{6D} one distinguishes 3 kinds of supersymmetric YM
theories: two of them have \emph{16} real conserved supercharges and the
third one has \emph{8} real supercharges \textrm{\cite{1R}- \cite{3R}}. The
field theories having \emph{16} supercharges are given by the well known non
chiral \emph{6D}$\  \mathcal{N}=\left( 1,1\right) $ and the chiral \emph{6D}$%
\  \mathcal{N}=\left( 2,0\right) $. The field theory with \emph{8}
supercharges is given by the chiral \emph{6D}$\  \mathcal{N}=\left(
1,0\right) $ SYM or equivalently $\mathcal{N}=\left( 0,1\right) $; it is the
gauge theory we consider below.

\subsubsection{\emph{the} \emph{6D}$\  \mathcal{N}=1$ \emph{vector multiplet }%
}

First, recall that in Lorentzian \emph{6D} there are only Weyl $%
\Psi_{6D}^{W} $ and Dirac fermions $\Psi_{6D}^{Dirac}=$ $\left(
\Psi_{6D}^{WL},\Psi_{6D}^{WR}\right) $; so that the smallest supermultiplet
contains a left $\Psi_{6D}^{WL}$ fermion, transforming in the $SO\left(
1,5\right) $ spinor representation $4_{+}$, or right $\Psi_{6D}^{WR}$ Weyl
fermion transforming in $4_{-}$. Recall also that in the language of \emph{4D%
} fermions, a Weyl spinor in \emph{6D}, say the left one, 
\begin{equation*}
\Psi_{6D}^{WL}\sim4_{+}
\end{equation*}
having 4 complex (8 real ) degrees of freedom, is made of a dotted and an
undotted \emph{4D} Weyl spinors as follows%
\begin{equation}
\begin{tabular}{lllll}
$\Psi_{6D}^{WL}$ & $=\left( \xi^{\alpha+},\bar{\lambda}_{\dot{\alpha}%
}^{-}\right) $ & , & $\alpha$ & $=1,2$%
\end{tabular}%
\end{equation}
Its complex adjoint is 
\begin{equation}
\begin{tabular}{lllll}
$\left( \Psi_{6D}^{WL}\right) ^{c}$ & $=$ & $\left( \lambda^{\alpha+},\bar{%
\xi}_{\dot{\alpha}}^{-}\right) $ & $\sim$ & $4_{+}^{c}$%
\end{tabular}%
\end{equation}
and has the same 6D\ chirality as $4_{+}$. The $\pm$ charges carried by the
fields refer to quantum numbers of $SO_{R}\left( 2\right) \sim U_{R}\left(
1\right) $ resulting from the reduction of the $SO\left( 1,5\right) $
Lorentz symmetry down to $SO\left( 1,3\right) \times SO_{R}\left( 2\right) $%
. The corresponding 6D Weyl right fermion 
\begin{equation*}
\Psi_{6D}^{WR}\sim4_{-}
\end{equation*}
with negative \emph{6D} chirality is given by%
\begin{equation}
\begin{tabular}{lll}
$\Psi_{6D}^{WR}$ & $=$ & $\left( \xi^{\alpha-},\bar{\lambda}_{\dot{\alpha}%
}^{+}\right) $ \\ 
$\left( \Psi_{6D}^{WR}\right) ^{c}$ & $=$ & $\left( \lambda^{\alpha-},\bar{%
\xi}_{\dot{\alpha}}^{+}\right) $%
\end{tabular}%
\end{equation}
The chiral \emph{6D}$\  \mathcal{N}=\left( 1,0\right) $ supersymmetric YM
theory has two types of supermultiplets describing matter and gauge fields
with on shell degrees of freedom as follows:

\  \  \  \ 

$\left( a\right) $ \emph{6D hypermultiplets}\newline
These supermultiplets describe matter; they have 2 complex (4 real) scalars
and a 6D\ Weyl spinor 
\begin{equation}
\begin{tabular}{lll}
$\mathcal{H}_{6D}$ & : & $\left( \frac{1}{2},0^{4}\right) _{6D}$%
\end{tabular}%
\end{equation}
they may belong to any representation of the gauge symmetry including
complex ones; see \textrm{\cite{4R}-\cite{7R}} for other properties.

\  \  \  \  \ 

$\left( b\right) $ \emph{6D vector multiplets} $\mathcal{V}_{6D}$ \newline
These multiplets have a gauge field and a Weyl spinor \newline
\begin{equation}
\begin{tabular}{lll}
$\mathcal{V}_{6D}^{\mathcal{N}=\left( 1,0\right) }$ & : & $\left( 1,\frac{1}{%
2}\right) _{6D}$%
\end{tabular}
\label{V6}
\end{equation}
they transform in the adjoint representation of the gauge symmetry. Below,
we refer to these supermultiplets like%
\begin{equation}
\begin{tabular}{lll}
$\mathcal{V}_{6D}^{\mathcal{N}=1}$ & $=$ & $\left( \mathcal{A}_{M},\psi
^{A}\right) _{6D}$%
\end{tabular}
\label{V66}
\end{equation}
where the field $\mathcal{A}_{M}$ is the 6D hermitian gauge field and $%
\psi^{A}$ the complex 4- dimension Weyl spinor of 
\begin{equation}
SO\left( 1,5\right) \simeq SU^{\ast}\left( 4\right)
\end{equation}
the fields $\mathcal{A}_{M}$ and $\psi^{A}$ are valued in the Lie algebra of
the U$\left( N\right) $ gauge symmetry.

\subsubsection{\emph{Reduction from 6D to 3D and twisting}}

We give two approaches to build the twisted field spectrum of \emph{3D }$%
\mathcal{N}=4$ supersymmetric YM that follows from the reduction of (\ref{V6}%
-\ref{V66}). We also comment on the link between the two methods.

\  \  \ 

$\left( 1\right) $ \emph{first approach}\newline
This approach is a rephrasing of eq(\ref{1U}); it involves two steps: $%
\left( i\right) $ dimension reduction from \emph{6D} to \emph{3D}; $\left(
ii\right) $ twisting the symmetries resulting from the breaking of $SO\left(
1,5\right) $.\ 

\  \  \  \  \newline
Under the reduction of the chiral \emph{6D}$\  \mathcal{N}=\left( 1,0\right) $
supersymmetric YM theory down to the \emph{3D} space, the $SO\left(
1,5\right) $ breaks down to 
\begin{equation*}
SO\left( 1,2\right) \times SO_{R}\left( 3\right)
\end{equation*}
and so the 6 local coordinates $X^{M}$ of $\mathbb{R}^{1,5}$ decompose like%
\begin{equation*}
\left( x^{\mu},y^{m}\right)
\end{equation*}
with $x^{\mu}\in \mathbb{R}^{1,2}$ and $y^{m}\in \mathbb{R}_{R}^{3}$ with
respective isotropy symmetries $SO\left( 1,2\right) $ and $SO_{R}\left(
3\right) $. Similarly, the on shell $4+4$ real degrees of freedom of the 
\emph{6D} chiral $\mathcal{N}=\left( 1,0\right) $ gauge multiplet (\ref{V6}%
), decomposes into a gauge field $A_{\mu}$, three real scalars $\phi_{m}$
and $4$ Majorana spinors $\psi^{\alpha1},...,\psi^{\alpha4}$.\ In the
euclidian version of this theory, the $SO\left( 1,2\right) \times
SO_{R}\left( 3\right) $ isotropy gets mapped to the compact $SO_{E}\left(
3\right) \times SO_{R}\left( 3\right) $ and the 4 Majorana spinors $%
\psi^{\alpha I}$ into 2 complex Dirac spinors $\xi^{\alpha1},\xi^{\alpha2}$
like%
\begin{equation}
\left( 1,\frac{1}{2}_{Dirac}^{2},0^{3}\right) _{3D}=\left(
A_{\mu},\xi^{\alpha i},\phi_{m}\right) _{3D}
\end{equation}
with:

\begin{itemize}
\item the field $A_{\mu}$ being a real \emph{3D} gauge field transforming as 
$\left( 3,1\right) $ under $SO_{E}\left( 3\right) \times SO_{R}\left(
3\right) $,

\item the fields $\xi^{\alpha i}$ are complex fermions of transforming $%
\left( 2,2\right) $ spinors of $SO_{E}\left( 3\right) \times SO_{R}\left(
3\right) $ $\simeq$ $SU_{E}\left( 2\right) \times SU_{R}\left( 2\right) $,

\item the fields $\phi_{m}$ are 3 real scalars transforming as $\left(
1,3\right) $ under $SO_{E}\left( 3\right) \times SO_{R}\left( 3\right) $.
\end{itemize}

\  \  \  \  \  \newline
Notice that in practice these fields should be taken as functions depending
only on the $x$ coordinates;%
\begin{equation}
\begin{tabular}{lll}
$A_{\mu}^{\left( 0\right) }=A_{\mu}\left( x\right) $, & $\xi_{\left(
0\right) }^{\alpha i}=\xi^{\alpha i}\left( x\right) ,$ & $...$%
\end{tabular}
\label{zo}
\end{equation}
but generally speaking they are functions of both coordinates $\left(
x,y\right) ;$%
\begin{equation}
\begin{tabular}{lll}
$A_{\mu}=A_{\mu}\left( x,y\right) $, & $\xi^{\alpha i}=\xi^{\alpha i}\left(
x,y\right) ,$ & $...$ \\ 
&  & 
\end{tabular}%
\end{equation}
By taking $y^{m}$ as the coordinates of a real 3-torus $\mathbb{T}^{3}$ with
large volume 
\begin{equation}
\frac{1}{\left( 2\pi l\right) ^{3}}\int_{\mathbb{T}^{3}}d^{3}y=1,\qquad
vol\left( \mathbb{T}^{3}\right) =\left( 2\pi l\right) ^{3}  \label{vo}
\end{equation}
one may expand these fields into infinite harmonic series like%
\begin{equation}
\begin{tabular}{lll}
$F\left( x,y\right) $ & $=$ & $\dsum \limits_{n_{1},n_{2},n_{3}}e^{in_{m}%
\frac{y^{m}}{2\pi l}}F^{\left( n_{1},n_{2},n_{3}\right) }\left( x\right) $%
\end{tabular}%
\end{equation}
where eqs(\ref{zo}) appear as the zero mode of the expansions and the extra
others as massive modes that break gauge symmetry in the restricted real 3D.

\  \  \  \newline
Under twisting, the quantum numbers of $SO_{E}\left( 3\right) $ and $%
SO_{R}\left( 3\right) $ groups are identified and the $SO_{E}\left( 3\right)
\times SO_{R}\left( 3\right) $ symmetry is reduced down to the diagonal 
\begin{equation}
SO\left( 3\right) =\frac{SO_{E}\left( 3\right) \times SO_{R}\left( 3\right) 
}{SO^{\prime}\left( 3\right) }
\end{equation}
As a consequence of the twisting, the fields of the chiral \emph{6D} $%
\mathcal{N}=\left( 1,0\right) $ gauge multiplet (\ref{V6}) are mapped to the
twisted ones%
\begin{equation}
\begin{tabular}{llll}
fields & : & twisted fields \  \  \  \  \  \  & $SO\left( 3\right) $ repres \\ 
$A_{\mu}$ &  & $\  \ A_{\mu}$ & \  \ $3$ \\ 
$\phi_{m}$ &  & $\  \ B_{\mu}$ & \  \ $3$ \\ 
$\xi^{\alpha \pm}$ &  & $\  \xi,$ $\  \xi^{\mu}$ & $1\oplus3$%
\end{tabular}
\label{tf}
\end{equation}
where now we have:

\begin{itemize}
\item two gauge fields $A_{\mu},$ $B_{\mu}$ that we combine into a complex
gauge field and its adjoint like%
\begin{equation}
\begin{tabular}{lll}
$\mathcal{G}^{\mu}$ & $=$ & $A_{\mu}+iB_{\mu}\  \  \  \  \  \ $ \\ 
$\mathcal{\bar{G}}_{\mu}$ & $=$ & $A_{a}-iB_{\mu}$%
\end{tabular}
\label{ft}
\end{equation}

\item four complex fermionic fields $\xi,$ $\xi^{\mu}$ transforming
respectively as a singlet and triplet of $SO\left( 3\right) $.
\end{itemize}

\  \  \  \  \newline
$\left( 2\right) $ \emph{second approach}\newline
This approach involves one step; and, in some sense, is a direct method. The
idea of this way of doing relies on the fact that since the fields $\mathcal{%
G}^{\mu}$ and $\xi,$ $\xi^{\mu}$ are complex fields, one may be tempted to
use complex groups to deal with them; this extension can be implemented by
considering other breaking modes of the $SO_{E}\left( 6\right) $ isotropy
symmetry of the euclidian space time $\mathbb{R}^{6}$ (following from the
Wick rotation of $\mathbb{R}^{1,5}$); in particular 
\begin{equation}
\begin{tabular}{lll}
&  &  \\ 
$SO_{E}\left( 6\right) $ & $\longrightarrow$ & $SU\left( 3\right) \times
U\left( 1\right) $ \\ 
&  &  \\ 
$\  \  \  \ 6_{v}$ & $\sim$ & $\ 3_{+2q}+\bar{3}_{-2q}$ \\ 
&  &  \\ 
$\  \  \  \ 4_{s}$ & $\sim$ & $3_{+q}+1_{-3q}$ \\ 
&  & 
\end{tabular}
\label{63}
\end{equation}
where $q$ is a unit charge of the abelian $U\left( 1\right) $ factor that
can be fixed to a number $q_{0}$. But here we will keep it free for later
use when considering the singular limit $q=0$.

\  \  \  \newline
Under the breaking mode (\ref{63}), the euclidian space $\mathbb{R}^{6}$,
parameterized by the local coordinates $X^{M}=\left( x^{\mu},y^{\mu}\right) $
with $y^{\mu}=x^{\mu+3}$, get mapped to the complex $\mathbb{C}^{3}$ with
local coordinates 
\begin{equation}
z^{a}=x^{a}+iy^{a}
\end{equation}
where $\left( x^{a}\right) $ the coordinates of the real space $\mathbb{R}%
^{3}$ and $\left( y^{a}\right) $ the coordinates of the internal space $%
\mathbb{R}_{int}^{3}$. Moreover, the fields of the multiplet (\ref{V6})
decompose as follows%
\begin{equation}
\begin{tabular}{lllll}
&  &  &  &  \\ 
$SO_{E}\left( 6\right) $ & $\longrightarrow$ & \multicolumn{3}{l}{$\  \  \  \  \
SU\left( 3\right) \times U\left( 1\right) $} \\ 
$\  \mathcal{A}_{M}$ & : & $\  \  \mathcal{G}^{a\left( -2q\right) }$ & , & $\  \ 
\mathcal{\bar{G}}_{a}^{\left( +2q\right) }$ \\ 
$\  \Psi^{A}$ & : & $\  \  \psi^{a\left( +q\right) }$ & , & $\  \  \psi^{\left(
-3q\right) }$ \\ 
&  &  &  & 
\end{tabular}
\label{36}
\end{equation}
and may be treated in general as functions of $\left( z,\bar{z}\right) $;
this property will be manifested on the lattice side through orientations of
the links; complex p-tensors and their duals are associated with
p-plaquettes with opposite orientations. \newline
\  \  \  \  \newline
Moreover, comparing eqs(\ref{tf}-\ref{ft}) with eq(\ref{63}-\ref{36}), we
end with the following results:

\begin{description}
\item[$\left( a\right) $] the spectrum of twisted fields of the two
approaches are quasi the same; the main difference is that (\ref{63}) depend
on the extra charge $q$ and transform in $SU\left( 3\right) $
representations rather than $SO\left( 3\right) $.

\item[$\left( b\right) $] Eqs(\ref{tf}-\ref{ft}) are recovered from eqs(\ref%
{63}-\ref{36}) by taking the limit 
\begin{equation}
q\rightarrow0
\end{equation}
and restricting the complex $SU\left( 3\right) \times U\left( 1\right) $
symmetry down to the real $SO\left( 3\right) $ which may be thought of as
its "real part". In practice, this corresponds to dropping out the y-
dependence into the fields and using (\ref{vo}) to integrate it out in the
field action.
\end{description}

\section{Twisted \emph{3D} $\mathcal{N}=4$ algebra and superfields}

We first give the basic anticommutators defining this superalgebra; then we
describe the general structure of twisted superspace and superfields.

\subsection{Twisted $\mathcal{N}=4$ supersymmetry in \emph{3D}}

For a generic charge \emph{q}, the twisted \emph{3D} $\mathcal{N}=4$
supersymmetric algebra is generated by 
\begin{equation}
\begin{tabular}{lllll}
&  &  &  &  \\ 
$Q^{{\small (+3q)}}$ & , & $Q_{a}^{{\small (-q)}}$ & , & $P_{a}^{{\small %
(+2q)}}$%
\end{tabular}%
\end{equation}
and 
\begin{equation}
P^{a{\small (-2q)}}
\end{equation}
having no supersymmetric partner; a property that makes \emph{asymmetric}
the formulation of twisted supersymmetric YM.\ 

\subsubsection{\emph{anticommutators}}

These operators transform under $U\left( 1\right) \times SU\left( 3\right) $
as in (\ref{63}); and satisfy the following basic anticommutation relations,%
\begin{equation}
\begin{tabular}{lll}
$\left \{ Q^{{\small (+3q)}},Q_{a}^{{\small (-q)}}\right \} $ & $=$ & $%
2P_{a}^{{\small (+2q)}}$ \\ 
$\left \{ Q_{a}^{{\small (-q)}},Q_{b}^{{\small (-q)}}\right \} $ & $=$ & $0$%
\end{tabular}
\label{CA}
\end{equation}
together with the topological one%
\begin{equation}
\begin{tabular}{lll}
$\left \{ Q^{{\small (+3q)}},Q^{{\small (+3q)}}\right \} $ & $=$ & $0$%
\end{tabular}%
\end{equation}
and%
\begin{equation}
\begin{tabular}{lll}
$\left[ Q^{{\small (+3q)}},P_{a}^{{\small (+2q)}}\right] $ & $=\left[ Q_{a}^{%
{\small (-q)}},P_{b}^{{\small (+2q)}}\right] $ & $=0$ \\ 
$\left[ Q^{{\small (+3q)}},P^{a{\small (-2q)}}\right] $ & $=\left[ Q_{a}^{%
{\small (-q)}},P^{b{\small (-2q)}}\right] $ & $=0$%
\end{tabular}%
\end{equation}
These graded commutation relations preserve the U$\left( 1\right) $ charge
and are covariant under $SU\left( 3\right) $ symmetry.\ 

\subsubsection{\emph{twisted superspace and superderivatives}}

The twisted \emph{3D} $\mathcal{N}=4$ superalgebra may be realized in
superspace by using complex bosonic and fermionic coordinates%
\begin{equation}
\begin{tabular}{lllllll}
$z^{a{\small (-2q)}}$ & $,$ & $z_{a}^{{\small (+2q)}}$ & , & $\theta ^{%
{\small (-3q)}}$ & $,$ & $\vartheta^{a\left( +q\right) }$%
\end{tabular}%
\end{equation}
with%
\begin{equation}
\left( z_{a}^{{\small (+2q)}}\right) ^{\dagger}=z^{a{\small (-2q)}}
\end{equation}
Using the usual supersymmetric covariant derivatives $D^{{\small (+3q)}}$
and $D_{a}^{\left( -q\right) },$ instead of the supercharges $Q^{{\small %
(+3q)}}$ and $Q_{a}^{\left( -q\right) }$, a suitable superspace
representation of the twisted superalgebra (\ref{CA}) is given by 
\begin{equation}
\begin{tabular}{lll}
$D^{{\small (+3q)}}$ & $=$ & $\frac{\partial}{\partial \theta^{{\small (}-%
{\small 3q)}}}$ \\ 
$D_{a}^{\left( -q\right) }$ & $=$ & $\frac{\partial}{\partial \vartheta
^{a\left( +q\right) }}+2\theta^{{\small (-3q)}}\partial_{a}^{\left(
+2q\right) }$ \\ 
&  & 
\end{tabular}%
\end{equation}
with%
\begin{equation}
\begin{tabular}{lllll}
$P^{a{\small (-2q)}}$ & $=$ & $\partial^{a{\small (-2q)}}$ & $=$ & $\frac{%
\partial}{\partial z_{a}^{{\small (+2q)}}}$ \\ 
$\bar{P}_{a}^{{\small (+2q)}}$ & $=$ & $\partial_{a}^{{\small (+2q)}}$ & $=$
& $\frac{\partial}{\partial z^{a{\small (-2q)}}}$ \\ 
&  &  &  & 
\end{tabular}%
\end{equation}
To implement gauge interactions, these superspace derivatives are
covariantized by introducing gauge connexions as follows%
\begin{equation}
\begin{tabular}{lll}
&  &  \\ 
$\mathcal{D}^{\left( +3q\right) }$ & $=$ & $D^{\left( +3q\right)
}+ig_{YM}\Upsilon^{\left( +3q\right) }$ \\ 
$\mathcal{D}_{a}^{\left( -q\right) }$ & $=$ & $D_{a}^{\left( -q\right)
}+ig_{YM}\Upsilon_{a}^{\left( -q\right) }$ \\ 
$\tciLaplace_{a}^{{\small (+2q)}}$ & $=$ & $\partial_{a}^{{\small (+2q)}%
}+ig_{YM}V_{a}^{{\small (+2q)}}$ \\ 
$\tciLaplace^{a\left( -2q\right) }$ & $=$ & $\partial^{a\left( -2q\right)
}+ig_{YM}U^{a\left( -2q\right) }$ \\ 
&  & 
\end{tabular}
\label{der}
\end{equation}
These extended superderivatives are needed for building the gauge covariant
superfields $\Phi_{i}^{\left( q_{i}\right) }$ of the twisted YM theory to be
considered later.

\subsection{Superfields of twisted\emph{\ 3D} $\mathcal{N}=4$ SYM}

To build the field action $\mathcal{S}_{twisted}$ of the twisted \emph{3D} $%
\mathcal{N}=4$ supersymmetric YM theory, we require invariance under the
three following symmetries:

$\left( a\right) $ the gauge symmetry which we take as $U\left( N\right) $,

$\left( b\right) $ the scalar supersymmetric charge $Q^{{\small (+3q)}}$ or
equivalently $D^{{\small (+3q)}}$; and,

$\left( c\right) $ the $U\left( 1\right) \times SU\left( 3\right) $ space
isotropy symmetry

\  \  \  \  \  \newline
First, observe that the gauge invariant action 
\begin{equation}
\mathcal{S}_{twisted}=\frac{1}{l^{3}}\dint \boldsymbol{L}_{twist}
\end{equation}
with the scalar supercharge $Q^{{\small (+3q)}}$ manifestly exhibited reads
in superspace as follows%
\begin{equation}
\begin{tabular}{lll}
$\boldsymbol{L}_{twist}$ & $=$ & $\left( \dint d\theta^{\left( -3q\right) }%
\mathcal{L}^{^{\left( -3q\right) }}\right) _{\vartheta^{a\left( +q\right)
}=0}$%
\end{tabular}
\label{ac}
\end{equation}
The scale factor $\frac{1}{l^{3}}$ is as in eq(\ref{vo}). The superspace
density $\mathcal{L}^{^{\left( -3q\right) }}$ transforms in the $U\left(
1\right) \times SU\left( 3\right) $ representation 1$_{-3}$; that is having $%
-3$ unit charges under $U\left( 1\right) $, and has the form 
\begin{equation}
\begin{tabular}{lll}
$\mathcal{L}^{^{\left( -3q\right) }}$ & $=$ & $Tr\left( \boldsymbol{L}%
_{twist}^{^{\left( -3q\right) }}\right) $ \\ 
&  & 
\end{tabular}%
\end{equation}
with the $N\times N$ superfield matrix 
\begin{equation}
\begin{tabular}{ll}
$\boldsymbol{L}_{twist}^{^{\left( -3q\right) }}$ & $=\boldsymbol{L}%
_{twist}^{^{\left( -3q\right) }}\left( \Phi \right) $ \\ 
& 
\end{tabular}%
\end{equation}
The $\boldsymbol{L}_{twist}^{^{\left( -3q\right) }}$ depends on a set of
superfields 
\begin{equation}
\Phi_{i}^{\left( q_{i}\right) }=\Phi_{i}^{\left( q_{i}\right) }\left(
z,\theta,\vartheta \right)
\end{equation}
that describe the off shell degrees of freedom of twisted \emph{3D} $%
\mathcal{N}=4$ supersymmetric YM theory. Below, we describe this set of
superfields; \textrm{for more details on the explicit derivation see the
analysis given in the appendix}.\  \  \  \  \ 

\subsubsection{\emph{Gauge covariance}}

The $U\left( N\right) $ gauge symmetry of the action (\ref{ac}) acts on the
superfield matrix density $\boldsymbol{L}_{twist}^{^{\left( -3\right) }}$
like%
\begin{equation}
\begin{tabular}{lll}
$\boldsymbol{L}_{twist}^{^{\left( -3q\right) }}$ & $\rightarrow$ & $%
\boldsymbol{g}$ $\boldsymbol{L_{twist}^{^{\left( -3q\right) }}}$ $%
\boldsymbol{g}^{-1}$%
\end{tabular}%
\end{equation}
since 
\begin{equation}
Tr\left( \boldsymbol{L}_{twist}^{^{\left( -3q\right) }}\right) =Tr\left( 
\boldsymbol{g}\text{ }\boldsymbol{L_{twist}^{^{\left( -3q\right) }}}\text{ }%
\boldsymbol{g}^{-1}\right)  \label{pr}
\end{equation}
where, for convenience as described in the appendix, the matrix element $%
\boldsymbol{g}$ is chosen as follows%
\begin{equation}
\begin{tabular}{lll}
$\boldsymbol{g}=\boldsymbol{g}(z,\bar{z},\vartheta^{a\left( +q\right) })$ & ,
& $D^{{\small (+3q)}}\boldsymbol{g=}0$ \\ 
&  & 
\end{tabular}%
\end{equation}
Notice that $\boldsymbol{g}$ depend on $z,$ $\bar{z},$ $\vartheta^{a\left(
+q\right) }$ but has no 
\begin{equation*}
\theta^{\left( -3q\right) }
\end{equation*}
The property (\ref{pr}) is ensured by requiring the superfields $\Phi
_{i}^{\left( q_{i}\right) }$ to be also gauge covariant; this means that
under a generic gauge symmetry transformation $\boldsymbol{g}$, we have%
\begin{equation}
\begin{tabular}{lll}
$\Phi_{i}^{\left( q_{i}\right) }$ & $\rightarrow$ & $\boldsymbol{g}\Phi
_{i}^{\left( q_{i}\right) }\boldsymbol{g}^{-1}$ \\ 
&  & 
\end{tabular}%
\end{equation}
General results on covariant formulation of supersymmetric YM theories in
superspace \textrm{\cite{1G}} applied to our present study lead to the
following set of gauge covariant superfields. 
\begin{equation}
\begin{tabular}{lllllll}
{\small Fermionic sector} & : & $\Psi^{\left( -3q\right) }$ &  & $%
\Psi^{a\left( +q\right) }$ &  & $\Phi_{ab}^{\left( +q\right) }$ \\ 
$SU\left( 3\right) \times U\left( 1\right) $ & : & $1_{-3q}$ &  & $3_{+q}$ & 
& $3_{+q}$ \\ 
scale mass dim &  & 1 &  & 1 &  & 1 \\ 
&  &  &  &  &  &  \\ 
{\small Bosonic sector} & : & $\mathbb{J}^{\left( 0\right) }$ & $,$ & $%
\mathbb{E}^{ab\left( -4q\right) }$ &  & $\mathbb{F}_{ab}^{\left( +4q\right)
} $ \\ 
$SU\left( 3\right) \times U\left( 1\right) $ & : & $1_{0}$ &  & $\bar {3}%
_{-4q}$ &  & $3_{+4q}$ \\ 
scale mass dim &  & $\frac{3}{2}$ &  & $\frac{3}{2}$ &  & $\frac{3}{2}$%
\end{tabular}
\label{gcs}
\end{equation}
built out of commutators of the gauge covariant superderivatives%
\begin{equation}
\begin{tabular}{lll}
&  &  \\ 
$\boldsymbol{\Psi}^{{\small (-3q)}}$ & $=$ & $\frac{1}{ig_{YM}}\left[ 
\mathcal{D}_{a}^{{\small (-q)}},\tciLaplace^{a\left( -2q\right) }\right] $
\\ 
&  &  \\ 
$\boldsymbol{\Psi}^{a{\small (+q)}}$ & $=$ & $\frac{1}{ig_{YM}}\left[ 
\mathcal{D}^{{\small (+3q)}},\tciLaplace^{a\left( -2q\right) }\right] $ \\ 
&  &  \\ 
$\boldsymbol{\Phi}_{ab}^{{\small (+q)}}$ & $=$ & $\frac{1}{ig_{YM}}\left[ 
\mathcal{D}_{a}^{{\small (-q)}},\tciLaplace_{b}^{{\small (+2q)}}\right] $%
\end{tabular}
\label{T}
\end{equation}
and 
\begin{equation}
\begin{tabular}{lll}
&  &  \\ 
$\mathbb{J}^{{\small (0)}}$ & $=$ & $\frac{1}{ig_{YM}}\left[
\tciLaplace_{a}^{{\small (+2q)}},\tciLaplace^{a\left( -2q\right) }\right] $
\\ 
&  &  \\ 
$\mathbb{E}^{ab{\small (-4q)}}$ & $=$ & $\frac{1}{ig_{YM}}\left[
\tciLaplace^{a\left( -2q\right) },\tciLaplace^{b\left( -2q\right) }\right] $
\\ 
&  &  \\ 
$\mathbb{F}_{ab}^{\left( +4q\right) }$ & $=$ & $\frac{1}{ig_{YM}}\left[
\tciLaplace_{a}^{{\small (+2q)}},\tciLaplace_{b}^{{\small (+2q)}}\right] $
\\ 
&  & 
\end{tabular}%
\end{equation}
with gauge coupling constant $g_{YM}$ scaling like $\left( mass\right) ^{%
\frac{1}{2}}$. Notice that as far as superfields with scaling dimension as $%
\left( mass\right) ^{1}$ are concerned, eqs(\ref{gcs}) may also contain the
fermionic superfield%
\begin{equation}
\begin{tabular}{lll}
$\boldsymbol{\Psi}_{a}^{{\small (+5q)}}$ & $=$ & $\frac{1}{ig_{YM}}\left[ 
\mathcal{D}^{{\small (+3q)}},\tciLaplace_{a}^{{\small (+2q)}}\right] $ \\ 
&  & 
\end{tabular}%
\end{equation}
it is constrained to be equal to zero in our construction. The above
fermionic and bosonic gauge covariant superfields obey as well constraint
relations; in particular%
\begin{equation}
\begin{tabular}{lll}
&  &  \\ 
$\mathcal{D}^{{\small (+3q)}}\Psi^{{\small (-3q)}}$ & $=$ & $2\mathbb{J}^{%
{\small (0)}}-\mathcal{D}_{a}^{{\small (-q)}}\Psi^{a{\small (+q)}}$ \\ 
&  &  \\ 
$\mathcal{D}^{{\small (+3q)}}\mathbb{E}^{ab{\small (-4q)}}$ & $=$ & $%
\tciLaplace^{a\left( -2q\right) }\Psi^{b{\small (+q)}}-\tciLaplace^{b\left(
-2q\right) }\Psi^{a{\small (+q)}}$ \\ 
&  &  \\ 
$\mathcal{D}^{{\small (+3q)}}\Phi_{ab}^{{\small (+q)}}$ & $=$ & $\mathbb{F}%
_{ab}^{\left( +4q\right) }$%
\end{tabular}
\label{S}
\end{equation}
and remarkably%
\begin{equation}
\begin{tabular}{lll}
$\mathcal{D}^{{\small (+3q)}}\Psi^{a{\small (+q)}}$ & $=$ & $0$ \\ 
&  &  \\ 
$\mathcal{D}^{{\small (+3q)}}\mathbb{F}_{ab}^{\left( +4q\right) }$ & $=$ & $%
0 $ \\ 
&  &  \\ 
$\tciLaplace_{b}^{{\small (+2q)}}\Psi^{{\small (-3q)}}$ & $=$ & $%
\tciLaplace^{a\left( -2q\right) }\Phi_{ab}^{{\small (+q)}}$ \\ 
&  & 
\end{tabular}
\label{u}
\end{equation}
To deal with these constraint relations, it is helpful to use the following $%
\theta$- expansions%
\begin{equation}
\begin{tabular}{lll}
$\Psi^{a{\small (+q)}}$ & $=$ & $\psi^{a{\small (+q)}}+\theta^{\left(
-3q\right) }f^{a{\small (+4q)}}$ \\ 
$\Psi^{\left( -3q\right) }$ & $=$ & $\psi^{\left( -3q\right)
}+\theta^{\left( -3q\right) }F^{\left( 0\right) }$ \\ 
$\Phi_{ab}^{\left( +q\right) }$ & $=$ & $\phi_{ab}^{\left( +q\right)
}+\theta^{\left( -3q\right) }\mathcal{F}_{ab}^{\left( +4q\right) }$ \\ 
&  &  \\ 
$\mathbb{J}^{\left( 0\right) }$ & $=$ & $\mathcal{F}^{\left( 0\right)
}+\theta^{\left( -3q\right) }\nabla_{a}^{\left( +2q\right) }\psi^{a\left(
+q\right) }$ \\ 
$\mathbb{E}^{ab\left( -4q\right) }$ & $=$ & $\mathcal{F}^{ab\left(
-4q\right) }+\theta^{\left( -3q\right) }\left[ \nabla^{a\left( -2q\right)
}\psi^{b\left( +q\right) }-\nabla^{b\left( -2q\right) }\psi^{a\left(
+q\right) }\right] $ \\ 
$\mathbb{F}_{ab}^{\left( +4q\right) }$ & $=$ & $\mathcal{F}_{ab}^{\left(
+4q\right) }+\theta^{\left( -3q\right) }\varkappa_{ab}^{\left( +7q\right) } $
\\ 
&  & 
\end{tabular}
\label{cgs}
\end{equation}
with the fields 
\begin{equation}
\psi^{\left( -3q\right) },\qquad \psi^{a\left( +q\right) }
\end{equation}
being the twisted fermionic fields of eq\textrm{s(\ref{36})}; the bosonic
fields $f^{a{\small (+4q)}}$ and $F^{\left( 0\right) }$ scaling as $\left(
mas\right) ^{\frac{3}{2}}$ are auxiliary fields; and finally%
\begin{equation}
\begin{tabular}{lllll}
$\mathcal{F}^{\left( 0\right) }$ & $,$ & $\mathcal{F}^{ab\left( -4q\right) }$
& $,$ & $\mathcal{F}_{ab}^{\left( +4q\right) }$ \\ 
&  &  &  & 
\end{tabular}%
\end{equation}
are as follows%
\begin{equation}
\begin{tabular}{lll}
&  &  \\ 
$\mathcal{F}_{ab}^{\left( +4q\right) }$ & $=$ & $\frac{1}{ig_{YM}}\left[
\nabla_{a}^{\left( +2q\right) },\nabla_{b}^{\left( +2q\right) }\right] $ \\ 
&  &  \\ 
$\mathcal{F}^{ab\left( -4q\right) }$ & $=$ & $\frac{1}{ig_{YM}}\left[
\nabla^{a\left( -2q\right) },\nabla^{b\left( -2q\right) }\right] $ \\ 
&  &  \\ 
$\mathcal{F}^{\left( 0\right) }$ & $=$ & $\frac{1}{ig_{YM}}\left[
\nabla_{a}^{\left( +2q\right) },\nabla^{a\left( -2q\right) }\right] $ \\ 
&  & 
\end{tabular}
\label{fff}
\end{equation}
with the gauge covariant $\nabla^{a\left( -2q\right) },$ $\nabla
_{a}^{\left( +2q\right) }$ derivatives given by%
\begin{equation}
\begin{tabular}{lll}
$\nabla^{a\left( -2q\right) }$ & $=$ & $\partial^{a\left( -2q\right)
}+ig_{YM}\mathcal{G}^{a\left( -2q\right) }$ \\ 
$\nabla_{a}^{\left( +2q\right) }$ & $=$ & $\partial_{a}^{\left( +2q\right)
}+ig_{YM}\mathcal{G}_{a}^{\left( +2q\right) }$%
\end{tabular}
\label{ggg}
\end{equation}
with $\mathcal{G}_{a}^{\left( +2q\right) },$ $\mathcal{G}^{a\left( -2\right)
}$ as in (\ref{36}).\newline
We also have the relations 
\begin{equation}
\begin{tabular}{lll}
$\nabla_{a}^{\left( +2q\right) }\psi^{a\left( +q\right) }$ & $=$ & $%
\partial_{a}^{\left( +2q\right) }\psi^{a\left( +q\right) }+ig_{YM}\left[ 
\mathcal{G}_{a}^{\left( +2q\right) },\psi^{a\left( +q\right) }\right] $ \\ 
&  &  \\ 
$\nabla^{a\left( -2q\right) }\psi^{b\left( +q\right) }$ & $=$ & $%
\partial^{a\left( -2q\right) }\psi^{b\left( +q\right) }+ig_{YM}\left[ 
\mathcal{G}^{a\left( -2q\right) },\psi^{b\left( +q\right) }\right] $ \\ 
&  & 
\end{tabular}%
\end{equation}
Notice that the constraint relation $\mathcal{D}^{{\small (+3q)}}\Psi^{a%
{\small (+q)}}=0$, which by using the gauge fixing described in the appendix
reads also like 
\begin{equation}
D^{{\small (+3q)}}\Psi^{a{\small (+q)}}=0
\end{equation}
is solved as follows 
\begin{equation}
\begin{tabular}{lllllll}
$\Psi^{a{\small (+q)}}$ & $=$ & $\psi^{a{\small (+q)}}$ & , & $f^{a{\small %
(+4q)}}$ & $=$ & $0$ \\ 
&  &  &  &  &  & 
\end{tabular}%
\end{equation}
This solution shows that $\psi^{a{\small (+q)}}$ is a supersymmetric
invariant field in agreement with the $\theta$- expansion of the the gauge
superfield $V^{a{\small (-2q)}}$ involved in (\ref{der}), 
\begin{equation}
\begin{tabular}{lll}
$U^{a{\small (-2q)}}$ & $=$ & $\mathcal{G}^{a{\small (-2q)}}+\theta^{\left(
-3q\right) }\psi^{a{\small (+q)}}$%
\end{tabular}%
\end{equation}
Similarly, we have for the constraint $D^{{\small (+3q)}}\mathbb{F}%
_{ab}^{\left( +q\right) }=0$, the following%
\begin{equation}
\begin{tabular}{lllllll}
$\mathbb{F}_{ab}^{\left( +4q\right) }$ & $=$ & $\mathcal{F}_{ab}^{\left(
+4q\right) }$ & , & $\varkappa_{ab}^{\left( +7q\right) }$ & $=$ & $0$ \\ 
&  &  &  &  &  & 
\end{tabular}
\label{sg}
\end{equation}
showing that $\mathcal{F}_{ab}^{\left( +q\right) }$ is a supersymmetric
invariant field in agreement with the $\theta$- expansion of the the gauge
superfield $\Upsilon_{a}^{{\small (-q)}}$ involved in eqs(\ref{der}) namely%
\begin{equation}
\begin{tabular}{lll}
$\Upsilon_{a}^{{\small (-q)}}$ & $=$ & $\gamma_{a}^{{\small (-q)}%
}+\theta^{\left( -3q\right) }\mathcal{G}_{a}^{{\small (+2q)}}$%
\end{tabular}%
\end{equation}
From this relation, we also learn that $\mathcal{G}_{a}^{{\small (+2q)}}$ is
supersymmetric invariant and so the superfield $V_{a}^{{\small (+2q)}}$
appearing in (\ref{der}) has no $\theta^{\left( -3q\right) }$\ dependence
and then should be as%
\begin{equation}
V_{a}^{{\small (+2q)}}=\mathcal{G}_{a}^{{\small (+2q)}}
\end{equation}
Regarding the constraint relation $\tciLaplace_{b}^{{\small (+2q)}}\Psi^{%
{\small (-3q)}}=\tciLaplace^{a\left( -2q\right) }\Phi_{ab}^{{\small (+q)}}$,
we use the $\theta$- expansions of the superfields to end, on one hand, with 
\begin{equation}
\begin{tabular}{lll}
$\tciLaplace_{b}^{{\small (+2q)}}\Psi^{{\small (-3q)}}$ & $=$ & $%
\partial_{b}^{{\small (+2q)}}\Psi^{{\small (-3q)}}+ig_{YM}\left[ \mathcal{G}%
_{a}^{{\small (+2q)}},\Psi^{{\small (-3q)}}\right] $ \\ 
&  &  \\ 
& = & $\nabla_{a}^{{\small (+2q)}}\psi^{\left( -3q\right)
}+ig_{YM}\theta^{\left( -3q\right) }\nabla_{a}^{{\small (+2q)}}F^{\left(
0\right) }$ \\ 
&  & 
\end{tabular}%
\end{equation}
and, on the other hand, with%
\begin{equation*}
\begin{tabular}{lll}
$\tciLaplace^{a\left( -2q\right) }\Phi_{ab}^{{\small (+q)}}$ & $=$ & $%
\partial^{a\left( -2q\right) }\Phi_{ab}^{{\small (+q)}}+ig_{YM}\left[
U^{a\left( -2q\right) },\Phi_{ab}^{{\small (+q)}}\right] $ \\ 
&  &  \\ 
& = & $\nabla^{a\left( -2q\right) }\phi_{ab}^{\left( +q\right)
}+ig_{YM}\theta^{\left( -3q\right) }\left( \left[ \mathcal{G}^{a\left(
-2q\right) },\mathcal{F}_{ab}^{\left( +4q\right) }\right] +\left \{
\psi^{a\left( +q\right) },\phi_{ab}^{\left( +q\right) }\right \} \right) $
\\ 
&  & 
\end{tabular}%
\end{equation*}
By equating, we obtain%
\begin{equation}
\begin{tabular}{lll}
$\nabla_{b}^{{\small (+2q)}}\psi^{\left( -3q\right) }$ & $=$ & $%
\nabla^{a\left( -2q\right) }\phi_{ab}^{\left( +q\right) }$ \\ 
&  &  \\ 
$\left[ \mathcal{G}_{b}^{{\small (+2q)}},F^{\left( 0\right) }\right] $ & $=$
& $\left[ \mathcal{G}^{a\left( -2q\right) },\mathcal{F}_{ab}^{\left(
+4q\right) }\right] +\left \{ \psi^{a\left( +q\right) },\phi_{ab}^{\left(
+q\right) }\right \} $ \\ 
&  & 
\end{tabular}
\label{uu}
\end{equation}
Under gauge transformations with matrix element $\boldsymbol{g}$ chosen, for
simplicity, as%
\begin{equation}
\begin{tabular}{lll}
$\boldsymbol{g}=\boldsymbol{g}\left( z,\bar{z},\vartheta^{a\left( +q\right)
}\right) $ & , & $D^{{\small (+3q)}}\boldsymbol{g=}0$ \\ 
&  & 
\end{tabular}
\label{gz}
\end{equation}
the superfields (\ref{gcs}) satisfy%
\begin{equation}
\begin{tabular}{lll}
$\Psi^{{\small (-3q)}}$ & $\rightarrow$ & $\boldsymbol{g\Psi}^{{\small (-3q)}%
}\boldsymbol{g}^{-1}$ \\ 
&  &  \\ 
$\Phi_{ab}^{\left( +q\right) }$ & $\rightarrow$ & $\boldsymbol{g\Phi
_{ab}^{\left( +q\right) }g}^{-1}$ \\ 
&  & 
\end{tabular}%
\end{equation}
and%
\begin{equation}
\begin{tabular}{lll}
$\mathbb{E}^{ab{\small (-4q)}}$ & $\rightarrow$ & $\boldsymbol{g}\mathbb{E}%
^{ab{\small (-4q)}}\boldsymbol{g}^{-1}$ \\ 
&  &  \\ 
$\mathbb{J}^{{\small (0)}}$ & $\rightarrow$ & $\boldsymbol{g}\mathbb{J}^{%
{\small (0)}}\boldsymbol{g}^{-1}$ \\ 
&  & 
\end{tabular}%
\end{equation}

\subsubsection{\emph{Supersymmetric transformations}}

First, we give the supersymmetric transformations of the on shell degrees of
freedom of eq(\ref{36}); then we consider the transformations of a
particular set of off shell ones.

\begin{itemize}
\item \emph{On shell multiplet}\newline
Using the equations of motion of the on shell twisted fields; in particular $%
\nabla_{a}^{\left( +2q\right) }\psi^{a\left( +q\right) }=0$ and $%
\nabla_{a}^{\left( +2q\right) }\psi^{\left( -3q\right) }=0$, we can write
down the supersymmetric transformations generated by the scalar operator $Q^{%
{\small (+3q)}}$; they are given by%
\begin{equation}
\begin{tabular}{lll}
&  &  \\ 
$Q^{{\small (+3q)}}\mathcal{G}^{a\left( -2q\right) }$ & $=$ & $\psi
^{a\left( +q\right) }$ \\ 
$Q^{{\small (+3q)}}\psi^{a\left( +q\right) }$ & $=$ & $0$ \\ 
&  &  \\ 
$Q^{{\small (+3q)}}\mathcal{G}_{a}^{\left( +2q\right) }$ & $=$ & $0$ \\ 
&  &  \\ 
$Q^{{\small (+3q)}}\psi^{\left( -3q\right) }$ & $=$ & $\mathcal{F}^{\left(
0\right) }$ \\ 
$Q^{{\small (+3q)}}\mathcal{F}^{\left( 0\right) }$ & $=$ & $\nabla
_{a}^{\left( +2q\right) }\psi^{a\left( +q\right) }=0$ \\ 
&  & 
\end{tabular}%
\end{equation}
where we used 
\begin{align}
\mathcal{F}^{\left( 0\right) } & =\partial_{a}^{\left( +2q\right) }\mathcal{G%
}^{a\left( -2q\right) }-\partial^{a\left( -2q\right) }\mathcal{G}%
_{a}^{\left( +2q\right) }+ig_{YM}\left[ \mathcal{G}_{a}^{\left( +2q\right) },%
\mathcal{G}^{a\left( -2q\right) }\right] \\
&  \notag
\end{align}

\item \emph{Off shell case}\newline
A set of off shell degrees of freedom is as in eqs(\ref{gcs}-\ref{cgs}); the
supersymmetric transformations of the fields are therefore given by%
\begin{equation}
\begin{tabular}{lll}
&  &  \\ 
$Q^{{\small (+3q)}}\psi^{{\small (-3q)}}$ & $=$ & $F^{\left( 0\right) }$ \\ 
$Q^{{\small (+3q)}}F^{\left( 0\right) }$ & $=$ & $0$ \\ 
$Q^{{\small (+3q)}}\phi_{ab}^{\left( +q\right) }$ & $=$ & $\mathcal{F}%
_{ab}^{\left( +4q\right) }$ \\ 
$Q^{{\small (+3q)}}\mathcal{F}_{ab}^{\left( +4q\right) }$ & $=$ & $0$ \\ 
&  & 
\end{tabular}%
\end{equation}
where $F^{\left( 0\right) }$\ is an auxiliary field; and%
\begin{equation}
\begin{tabular}{lll}
&  &  \\ 
$Q^{{\small (+3q)}}\mathcal{F}^{ab\left( -4q\right) }$ & $=$ & $%
\psi^{ab\left( -q\right) }$ \\ 
$Q^{{\small (+3q)}}\psi^{ab\left( -q\right) }$ & $=$ & $0$ \\ 
$Q^{{\small (+3q)}}\mathcal{F}^{\left( 0\right) }$ & $=$ & $\nabla
_{a}^{\left( +2q\right) }\psi^{a\left( +q\right) }$ \\ 
$Q^{{\small (+3q)}}\nabla_{a}^{\left( +2q\right) }\psi^{a\left( +q\right) } $
& $=$ & $0$ \\ 
&  & 
\end{tabular}%
\end{equation}
where we have set%
\begin{equation}
\psi^{ab\left( -q\right) }=\nabla^{a\left( -2q\right) }\psi^{b\left(
+q\right) }-\nabla^{b\left( -2q\right) }\psi^{a\left( +q\right) }
\end{equation}
\end{itemize}

\section{Action in superspace}

The action of twisted fields of chiral \emph{3D} $\mathcal{N}=4$
supersymmetry, exhibiting manifestly the scalar supercharge $Q^{{\small (+3q)%
}}$, reads in superspace like%
\begin{equation}
\begin{tabular}{lll}
$\boldsymbol{L}_{twist}$ & $=$ & $\left( \dint d\theta^{\left( -3q\right) }%
\mathcal{L}^{^{\left( -3q\right) }}\right) _{\vartheta=0},\qquad$ \\ 
&  & 
\end{tabular}%
\end{equation}
with lagrangian superdensity $\mathcal{L}^{^{\left( -3q\right) }}$ depending
on the Grassman variable $\theta^{\left( -3q\right) }$; but also on%
\begin{equation}
\begin{tabular}{ll}
$\vartheta^{a{\small (+q)}}$ & ,%
\end{tabular}%
\end{equation}
Because of the role played by the supersymmetric generator $\mathcal{D}_{a}^{%
{\small (-q)}}$ in our construction; eg (\ref{T}-\ref{S}) and appendix, the
dependence into the $\vartheta^{a{\small (+q)}}$ is implicit; and is killed
at the end after performing integration with respect to $\theta ^{{\small %
(-3q)}}$.

\subsection{Lagrangian superdensity}

The general form of the fermionic superdensity $\mathcal{L}^{^{\left(
-3q\right) }}$ scaling as $\left( mass\right) ^{\frac{5}{2}}$ one may
construct out of the\ set of gauge covariant superfields (\ref{gcs}) is as
follows%
\begin{equation}
\begin{tabular}{lll}
&  &  \\ 
$\mathcal{L}^{^{\left( -3q\right) }}$ & $=$ & $\alpha_{1}Tr\left[ \Psi^{%
{\small (-3q)}}D^{{\small (+3q)}}\Psi^{{\small (-3q)}}\right] +\alpha_{2}Tr%
\left[ \Psi^{{\small (-3q)}}\mathbb{J}^{{\small (0)}}\right] +$ \\ 
&  &  \\ 
&  & $\alpha_{3}Tr\left[ \varepsilon_{abc}\Psi^{a{\small (+q)}}\mathbb{E}^{bc%
{\small (-4q)}}\right] +$ \\ 
&  &  \\ 
&  & $\alpha_{4}Tr\left[ \Phi_{ab}^{{\small (+q)}}\mathbb{E}^{ab{\small (-4q)%
}}\right] +$ \\ 
&  &  \\ 
&  & $\nu_{_{FI}}Tr\left[ \Psi^{{\small (-3q)}}\right] $ \\ 
&  & 
\end{tabular}%
\end{equation}
where the $\alpha_{i}$'s are normalization numbers and the coupling scaling
as $\left( mass\right) ^{\frac{3}{2}}$ 
\begin{equation}
\nu_{_{FI}}
\end{equation}
is a Fayet-Iliopoulous coupling constant. This term breaks scalar
supersymmetry; it will be dropped out below. \newline
Notice that the integration of with respect to the Grassamn variable of%
\begin{equation}
\nu_{_{FI}}\dint d\theta^{\left( -3q\right) }\text{ }Tr\left[ \Psi^{{\small %
(-3q)}}\right]
\end{equation}
leads in general to 
\begin{equation}
\nu_{_{FI}}Tr\left( F^{\left( 0\right) }\right) =\nu_{_{FI}}\dsum
\limits_{A=1}^{\dim U(N)}F_{A}^{\left( 0\right) }Tr\left( \mathcal{T}%
^{A}\right)
\end{equation}
which does'nt vanish due to the abelian gauge subsymmetry 
\begin{equation}
U\left( 1\right) =\frac{U\left( N\right) }{SU\left( N\right) }
\end{equation}
In the above relation, the matrices $\mathcal{T}^{A}$ stand for the
generators of $U\left( N\right) $.

\subsection{Component field action}

Using the $\theta$- expansions (\ref{cgs}-\ref{sg}) and integrating with
respect to the Grassman variable $\theta^{\left( -3q\right) }$; we obtain%
\begin{equation}
\begin{tabular}{lll}
&  &  \\ 
$\boldsymbol{L}_{twist}$ & $=$ & $\alpha_{1}Tr\left[ F^{\left( 0\right)
}F^{\left( 0\right) }\right] +\alpha_{2}Tr\left[ F^{\left( 0\right) }%
\mathcal{F}^{\left( 0\right) }\right] $ \\ 
&  &  \\ 
&  & $-\alpha_{2}Tr\left[ \psi^{{\small (-3q)}}\nabla_{a}^{\left( +2q\right)
}\psi^{a\left( +q\right) }\right] $ \\ 
&  &  \\ 
&  & $+2\alpha_{3}Tr\left[ \varepsilon_{abc}\psi^{a{\small (+q)}%
}\nabla^{b\left( -2q\right) }\psi^{c\left( +q\right) }\right] $ \\ 
&  &  \\ 
&  & $+\alpha_{4}Tr\left[ \mathcal{F}_{ab}^{\left( +4q\right) }\mathcal{F}%
^{ab{\small (-4q)}}\right] $ \\ 
&  &  \\ 
&  & $-\alpha_{4}Tr\left[ \phi_{ab}^{{\small (+q)}}\left[ \nabla^{a\left(
-2q\right) }\psi^{b\left( +q\right) }-\nabla^{b\left( -2q\right)
}\psi^{a\left( +q\right) }\right] \right] $ \\ 
&  & 
\end{tabular}
\label{fac}
\end{equation}
Notice that the terms with coefficients $\alpha_{1},$ $\alpha_{2},$ $%
\alpha_{4}$ are manifestly invariant with respect to the scalar
supersymmetric transformations. However the variation of the term%
\begin{equation}
Tr\left[ \varepsilon_{abc}\psi^{a{\small (+q)}}\nabla^{b\left( -2q\right)
}\psi^{c\left( +q\right) }\right]
\end{equation}
leads to%
\begin{equation}
Tr\left[ \varepsilon_{abc}\left \{ \psi^{a{\small (+q)}},\left \{
\psi^{b\left( +q\right) },\psi^{c\left( +q\right) }\right \} \right \} %
\right]
\end{equation}
or equivalently%
\begin{equation*}
\frac{1}{3}\varepsilon_{abc}\psi_{A}^{a{\small (+q)}}\left( \nabla^{b\left(
-2q\right) }\psi^{c\left( +q\right) }\right) _{B}\psi_{C}^{c\left( +q\right)
}Tr\left( \left[ \mathcal{T}^{A},\left[ \mathcal{T}^{B},\mathcal{T}^{C}%
\right] \right] +\text{ cyclic perm}\right)
\end{equation*}
which vanishes identically due to Jacobi-Identity.

\  \  \newline
Notice that the component field action (\ref{fac}) can be rewritten in a
more convenient form by eliminating $\phi_{ab}^{\left( +q\right) }$ through
the constraint relation%
\begin{equation}
\begin{tabular}{lll}
$\nabla_{b}^{{\small (+2q)}}\psi^{{\small (-3q)}}$ & $=$ & $\nabla^{a\left(
-2q\right) }\phi_{ab}^{{\small (+q)}}$ \\ 
&  & 
\end{tabular}%
\end{equation}
following from the superfield constraint eqs(\ref{u}-\ref{uu}) namely%
\begin{equation}
\begin{tabular}{lll}
$\tciLaplace_{b}^{{\small (+2q)}}\Psi^{{\small (-3q)}}$ & $=$ & $%
\tciLaplace^{a\left( -2q\right) }\Phi_{ab}^{{\small (+q)}}$ \\ 
&  & 
\end{tabular}%
\end{equation}
Indeed, starting from 
\begin{equation}
\begin{tabular}{lll}
$Tr\left( \phi_{ab}^{{\small (+q)}}\nabla^{a\left( -2q\right) }\psi^{b\left(
+q\right) }\right) $ & $=$ & $Tr\left( \phi_{ab}^{{\small (+q)}%
}\partial^{a\left( -2q\right) }\psi^{b\left( +q\right) }\right) $ \\ 
&  & $ig_{YM}Tr\left( \phi_{ab}^{{\small (+q)}}\left[ \mathcal{G}^{a\left(
-2q\right) },\psi^{b\left( +q\right) }\right] \right) $ \\ 
&  & 
\end{tabular}%
\end{equation}
and integrating by part, we get up to a total divergence, 
\begin{equation}
\begin{tabular}{lll}
$Tr\left( \phi_{ab}^{{\small (+q)}}\nabla^{a\left( -2q\right) }\psi^{b\left(
+q\right) }\right) $ & $=$ & $-Tr\left[ \left( \nabla^{a\left( -2q\right)
}\phi_{ab}^{{\small (+q)}}\right) \psi^{b\left( +q\right) }\right] $ \\ 
&  &  \\ 
& = & $-Tr\left[ \psi^{b\left( +q\right) }\nabla_{b}^{{\small (+2q)}}\psi^{%
{\small (-3q)}}\right] $ \\ 
&  & 
\end{tabular}%
\end{equation}
By substituting back into (\ref{fac}), we end with the lagrangian density%
\begin{equation}
\begin{tabular}{lll}
&  &  \\ 
$\boldsymbol{L}_{twist}$ & $=$ & $\alpha_{1}Tr\left[ F^{\left( 0\right)
}F^{\left( 0\right) }\right] +\alpha_{2}Tr\left[ F^{\left( 0\right) }%
\mathcal{F}^{\left( 0\right) }\right] $ \\ 
&  &  \\ 
&  & $+\alpha_{4}Tr\left[ \mathcal{F}_{ab}^{\left( +4q\right) }\mathcal{F}%
^{ab{\small (-4q)}}\right] $ \\ 
&  &  \\ 
&  & $+\left( \alpha_{2}+2\alpha_{4}\right) Tr\left[ \psi^{a\left( +q\right)
}\nabla_{a}^{\left( +2q\right) }\psi^{{\small (-3q)}}\right] $ \\ 
&  &  \\ 
&  & $+2\alpha_{3}Tr\left[ \varepsilon_{abc}\psi^{a{\small (+q)}%
}\nabla^{b\left( -2q\right) }\psi^{c\left( +q\right) }\right] $ \\ 
&  & 
\end{tabular}%
\end{equation}
Eliminating the auxiliary field $F^{\left( 0\right) }$ through its equation
of motion 
\begin{equation}
F^{\left( 0\right) }=-\frac{\alpha_{2}}{2\alpha_{1}}\mathcal{F}^{\left(
0\right) }
\end{equation}
and putting back into the lagrangian density, we end with

\begin{equation}
\begin{tabular}{lll}
&  &  \\ 
$\boldsymbol{L}_{twist}$ & $=$ & $\alpha_{4}Tr\left[ \mathcal{F}%
_{ab}^{\left( +4q\right) }\mathcal{F}^{ab{\small (-4q)}}\right] $ \\ 
&  &  \\ 
&  & $-\frac{\left( \alpha_{2}\right) ^{2}}{4\alpha_{1}}Tr\left[ \mathcal{F}%
^{\left( 0\right) }\mathcal{F}^{\left( 0\right) }\right] $ \\ 
&  &  \\ 
&  & $+\left( \alpha_{2}+2\alpha_{4}\right) Tr\left[ \psi^{a\left( +q\right)
}\nabla_{a}^{\left( +2q\right) }\psi^{{\small (-3q)}}\right] $ \\ 
&  &  \\ 
&  & $+2\alpha_{3}Tr\left[ \varepsilon_{abc}\psi^{a{\small (+q)}%
}\nabla^{b\left( -2q\right) }\psi^{c\left( +q\right) }\right] $ \\ 
&  & 
\end{tabular}
\label{caf}
\end{equation}
with scaling mass dimension $\left( mass\right) ^{3}$. Notice that the YM
coupling constant g$_{YM}$ is within the gauge covariant derivatives $%
\nabla_{a}^{\left( +2q\right) }$, $\nabla^{a\left( -2q\right) }$ and the
field strengths $\mathcal{F}_{ab}^{\left( +4q\right) }$, $\mathcal{F}^{ab%
{\small (-4q)}}$ as shown on eqs(\ref{fff}-\ref{ggg}).

\section{Twisted 3D$\  \mathcal{N}=4$ SYM theory on lattice}

In all what follows, we focuss on the study of twisted \emph{3D} $\mathcal{N}%
=4$ supersymmetric YM on particular 3- dimensional lattice $\mathcal{L}%
_{3D}^{su_{3}\times u_{1}}$. This crystal is given by the fibration 
\begin{equation}
\begin{tabular}{lll}
&  &  \\ 
$\mathcal{L}_{1D}^{u_{1}}$ & $\rightarrow$ & $\  \  \mathcal{L}%
_{3D}^{su_{3}\times u_{1}}$ \\ 
&  & $\  \  \  \  \downarrow$ \\ 
&  & $\  \  \mathcal{L}_{2D}^{su_{3}}=\mathbb{A}_{2}^{\ast}$%
\end{tabular}
\label{3D}
\end{equation}
with the two following components:

\begin{description}
\item[$\left( i\right) $] the base sublattice $\mathcal{L}_{2D}^{su_{3}}$
given by $\mathbb{A}_{2}^{\ast}$, the dual of the 2-dimensional root lattice 
$\mathbb{A}_{2}$ that is associated with the $SU\left( 3\right) $ symmetry 
\textrm{\cite{7S,8S}}; and

\item[$\left( ii\right) $] the fiber $\mathcal{L}_{1D}^{u_{1}}$ associated
with the $U\left( 1\right) $ factor of the symmetry $SU\left( 3\right)
\times U\left( 1\right) $, it is a 1-dimensional lattice with direction
normal to $\mathbb{A}_{2}^{\ast}$.
\end{description}

\  \  \  \newline
To fix the ideas, $\mathcal{L}_{3D}^{su_{3}\times u_{1}}$ will be realized
as \emph{a twist} of the 3- dimensional lattice $\mathbb{A}_{3}^{\ast}$; the
dual to the 3D root lattice $\mathbb{A}_{3}$ generated by the 3 simple roots
of $SU\left( 4\right) $; that is:%
\begin{equation}
\begin{tabular}{llll}
&  &  &  \\ 
$\mathcal{L}_{3D}^{su_{3}\times u_{1}}$ & $\sim$ & twist of $\mathbb{A}%
_{3}^{\ast}$ &  \\ 
&  &  & 
\end{tabular}%
\end{equation}
Notice that the $\mathbb{A}_{3}^{\ast}$ crystal is generated by the 3
fundamental weight of $SO\left( 6\right) \simeq SU\left( 4\right) $; and the
twist of $\mathbb{A}_{3}^{\ast}$ we are looking for is the one induced by
the breaking mode 
\begin{equation}
\begin{tabular}{lll}
$SU\left( 4\right) $ & $\rightarrow$ & $SU\left( 3\right) \times U\left(
1\right) $%
\end{tabular}
\label{31}
\end{equation}
For the explicit engineering of $\mathcal{L}_{3D}^{su_{3}\times u_{1}}$; see
next section; in due time let us focus on building the lattice analogue of
the twist field of continuum.

\subsection{Tensor fields on lattice $\mathcal{L}_{3D}^{su_{3}\times u_{1}}$}

In this subsection, we study the discretization of the twisted \emph{3D}$\ 
\mathcal{N}=4$ SYM theory to the lattice $\mathcal{L}_{3D}^{su_{3}\times
u_{1}}$ of (\ref{3D}) by first focussing on the projection of this gauge
theory on the base sublattice 
\begin{equation*}
\mathcal{L}_{2D}^{su_{3}}=\mathbb{A}_{2}^{\ast}
\end{equation*}
The implementation of the effect of the fiber $\mathcal{L}_{1D}^{u_{1}}$
will be considered later on.

\subsubsection{\emph{Discretizing continuum}}

We begin by considering real scalar fields and then \textrm{real} tensor
ones living on 3D space. After that, we give the extension to complex space
and complex fields appearing in the formulation 3D$\  \mathcal{N}=4$ SYM
theory given in previous sections.

\  \  \ 

1) \emph{Discretizing space}\newline
In the discretization of the real 3D continuum space, generic points $%
\boldsymbol{P}$ with local coordinates $\left( x^{\mu}\right) =\left(
x,y,z\right) $ get mapped to lattice nodes 
\begin{equation*}
\boldsymbol{N}=\boldsymbol{N}\left( n_{1},n_{2},n_{3}\right)
\end{equation*}
described by 3- dimensional integral position vectors 
\begin{equation*}
\vec{R}_{n}
\end{equation*}
In the example of a \emph{simple cubic} lattice with spacing parameter $L$,
the nodes $\boldsymbol{N}$ are represented by 
\begin{equation}
\begin{tabular}{lll}
$\vec{R}_{n}$ & $=$ & $x_{n}\vec{e}_{1}+y_{n}\vec{e}_{2}+z_{n}\vec{e}_{3}$%
\end{tabular}%
\end{equation}
with $\vec{e}_{i}$ the usual canonical basis obeying%
\begin{equation*}
\vec{e}_{i}.\vec{e}_{j}=\delta_{ij}
\end{equation*}
Each site $\vec{R}_{n}$ in this simple lattice has 6 first nearest neighbors
located at%
\begin{equation}
\begin{tabular}{lllll}
$\vec{R}_{n}\pm L\vec{e}_{1}$ & , & $\vec{R}_{n}\pm L\vec{e}_{2}$ & , & $%
\vec{R}_{n}\pm L\vec{e}_{3}$%
\end{tabular}%
\end{equation}
In the case of the lattice $\mathcal{L}_{3D}^{su_{3}\times u_{1}}$ with an $%
SU\left( 3\right) \times U\left( 1\right) $ symmetry, the site positions $%
\vec{R}_{n}$ are given by 
\begin{equation}
\begin{tabular}{lll}
$\vec{R}_{n}$ & $=$ & $n_{1}\vec{L}_{1}+n_{2}\vec{L}_{2}+n_{3}\vec{L}_{3}$
\\ 
$n$ & $=$ & $\left( n_{1},n_{2},n_{3}\right) $%
\end{tabular}%
\end{equation}
with the basis $\vec{L}_{i}$ satisfying a non trivial $3\times3$
intersection matrix 
\begin{equation}
\begin{tabular}{lll}
&  &  \\ 
$\mathcal{J}_{ij}^{su_{3}\times u_{1}}$ & $=$ & $\vec{L}_{i}.\vec{L}_{j}$ \\ 
&  &  \\ 
$\vec{R}_{n}.\vec{R}_{m}$ & $=$ & $n_{i}\mathcal{J}_{ij}^{su_{3}\times
u_{1}}m_{j}$ \\ 
&  & 
\end{tabular}%
\end{equation}
capturing the shape of the crystal $\mathcal{L}_{3D}^{su_{3}\times u_{1}}$. 
\newline
The intersection matrix $\mathcal{J}_{ij}^{su_{3}\times u_{1}}$ has a set of
features; in particular the 2 following useful ones.

\begin{description}
\item[$\left( a\right) $] the matrix $\mathcal{J}_{ij}^{su_{3}\times u_{1}}$
is exactly given by%
\begin{equation}
\mathcal{J}_{ij}^{su_{3}\times u_{1}}=\left( 
\begin{array}{ccc}
\frac{2}{3}+q^{2} & \frac{1}{3}+2q^{2} & 3q^{2} \\ 
\frac{1}{3}+2q^{2} & \frac{2}{3}+4q^{2} & 6q^{2} \\ 
3q^{2} & 6q^{2} & 9q^{2}%
\end{array}
\right)  \label{fbb}
\end{equation}
It depends on the number $q$ that encodes the charges of the twisted
supersymmetric YM fields under the abelian $U\left( 1\right) $ symmetry of (%
\ref{31}) and moreover defines the fibration (\ref{bff}).

\item[$\left( b\right) $] In the particular and remarkable case $q=0$, the
intersection matrix $\mathcal{J}_{ij}^{su_{3}\times u_{1}}$ reduces to the
singular matrix 
\begin{equation}
\left( \mathcal{J}_{ij}^{su_{3}\times u_{1}}\right) _{q=0}=\left( 
\begin{array}{cc}
\mathcal{J}_{ij}^{su_{3}} & 0 \\ 
0 & 0%
\end{array}
\right)
\end{equation}
with%
\begin{equation}
\mathcal{J}_{ij}^{su_{3}}=\frac{1}{3}\left( 
\begin{array}{cc}
2 & 1 \\ 
1 & 2%
\end{array}
\right)
\end{equation}
This singular case corresponds exactly to the projection of sites 
\begin{equation}
\vec{R}_{\left( n_{1},n_{2},n_{3}\right) }
\end{equation}
in the 3- dimensional $\mathcal{L}_{3D}^{su_{3}\times u_{1}}$ onto sites 
\begin{equation}
\vec{r}_{\left( n_{1},n_{2}\right) }
\end{equation}
in the base sublattice 
\begin{equation*}
\mathbb{A}_{2}^{\ast}
\end{equation*}
This means that site positions in $\mathcal{L}_{3D}^{su_{3}\times u_{1}}$
depend on the parameter $q$; and so can be parameterized like%
\begin{equation}
\begin{tabular}{llll}
$\vec{R}_{\left( n_{1},n_{2},n_{3}\right) }^{\left( q\right) }$ & $=$ & $%
\left( 
\begin{array}{c}
\vec{r}_{\left( n_{1},n_{2}\right) } \\ 
qZ_{n_{3}}%
\end{array}
\right) $ & 
\end{tabular}%
\end{equation}
with third component belonging to the fiber,%
\begin{equation}
\begin{tabular}{lllll}
&  &  &  &  \\ 
$Z_{n_{3}}\in \mathcal{L}_{1D}^{u_{1}}$ & $,$ & $\mathcal{L}_{1D}^{u_{1}}$ & 
$\simeq$ & $q\mathbb{Z}$ \\ 
&  &  &  & 
\end{tabular}%
\end{equation}
The same property is valid for the $\vec{L}_{i}$ basis generators; they
depend on the charge $q$ and may be decomposed as well like 
\begin{equation}
\begin{tabular}{lll}
&  &  \\ 
$\vec{L}_{1}^{\left( q\right) }=\left( 
\begin{array}{c}
\vec{l}_{1} \\ 
q%
\end{array}
\right) ,$ & $\vec{L}_{2}^{\left( q\right) }=\left( 
\begin{array}{c}
\vec{l}_{2} \\ 
2q%
\end{array}
\right) ,$ & $\vec{L}_{3}^{\left( q\right) }=\left( 
\begin{array}{c}
\vec{0} \\ 
3q%
\end{array}
\right) $ \\ 
&  & 
\end{tabular}%
\end{equation}
with the 2-dimensional vectors $\vec{r}_{n}$ giving the sites in the base
sublattice $\mathbb{A}_{2}^{\ast}$ 
\begin{equation}
\begin{tabular}{lll}
$\vec{r}_{n}$ & $=$ & $n_{1}\vec{l}_{1}+n_{2}\vec{l}_{2}$ \\ 
$n$ & $=$ & $\left( n_{1},n_{2}\right) $%
\end{tabular}%
\end{equation}
So the case $q=0$ define a projection from $\mathcal{L}_{3D}^{su_{3}\times
u_{1}}$ down to the base $\mathbb{A}_{2}^{\ast}$; we have%
\begin{equation}
\begin{tabular}{llll}
$\vec{R}_{n}^{\left( 0\right) }$ & $=$ & $\left( 
\begin{array}{c}
\vec{r}_{n} \\ 
0%
\end{array}
\right) $ & 
\end{tabular}%
\end{equation}
\end{description}

\  \  \  \  \ 

2) \emph{Discrete field variables}

\begin{itemize}
\item \emph{scalar fields}\newline
Under discretization of the real \emph{3D} continuum space into $\mathcal{L}%
_{3D}^{su_{3}\times u_{1}}$, local \emph{scalar} fields $\Phi \left(
x\right) $ of the continuum get mapped to an infinite set of discrete
variables 
\begin{equation}
\begin{tabular}{lll}
$\Phi \left( R_{n}\right) $ & $=$ & $\Phi_{\left( n_{1},n_{2},n_{3}\right)
}^{\left( q\right) }$ \\ 
&  &  \\ 
& $\equiv$ & $\Phi_{n}^{\left( q\right) }$%
\end{tabular}%
\end{equation}
living at the lattice nodes 
\begin{equation*}
R_{n},\qquad n=\left( n_{1},n_{2},n_{3}\right) \in \mathbb{Z}^{3}
\end{equation*}
The variables $\Phi_{n}^{\left( q\right) }$ may have either an even
statistics or an odd one depending on whether $\Phi \left( x\right) $ is
bosonic or fermionic. In the case of twisted \emph{3D} $\mathcal{N}=4$
supersymmetric YM, odd variables are given by the twisted fermion $%
\psi^{\left( -3q\right) }$ and the Grassman variable $\theta^{\left(
-3q\right) }$.

\item \emph{antisymmetric p-tensors}\newline
Real p- form fields $\mathcal{T}_{\left[ p\right] }\left( x\right) $ in
continuum 
\begin{equation}
\begin{tabular}{lll}
$\mathcal{T}_{\left[ p\right] }$ & $=$ & $\frac{1}{p!}dx^{\mu_{1}}\wedge
dx^{\mu_{2}}...\wedge dx^{\mu_{p}}$ $T_{\mu_{1}...\mu_{p}}$%
\end{tabular}%
\end{equation}
are associated with p- dimensional plaquettes in the lattice. In the case of 
$\mathcal{L}_{3D}^{su_{3}\times u_{1}}$, we have vectors and and their duals
namely the rank 2 antisymmetric tensors; rank 3 antisymmetric tensors are
dual to scalars. So we have%
\begin{equation}
\begin{tabular}{lllll}
fields & $\rightarrow$ \  \  \  \  \  \  \  \  & \multicolumn{3}{l}{\  \ p-plaquettes
} \\ 
$\ T$ &  & \  \ sites &  &  \\ 
$\ T_{\mu}$ &  & \  \ 1d- \emph{links} &  &  \\ 
$\ T_{\mu \nu}$ &  & \ 2d- plaquettes & $\sim$ & 1d- \emph{links} \\ 
&  &  &  & 
\end{tabular}%
\end{equation}
Let us illustrate the construction on the particular case of the gradient $%
\partial_{\mu}\Phi \left( x\right) .$ \newline
To get the discrete expression representing $\partial_{\mu}\Phi \left(
x\right) $ on the lattice $\mathcal{L}_{3D}^{su_{3}\times u_{1}}$, it is\
useful to consider%
\begin{equation}
\begin{tabular}{lll}
$d\Phi \left( x\right) $ & $=$ & $dx^{\mu}\partial_{\mu}\Phi \left( x\right) 
$ \\ 
&  & 
\end{tabular}
\label{df}
\end{equation}
which is a particular \emph{1-form }in 3D. This differential, which involves
the operator $d=dx^{\mu}\partial_{\mu}$, behaves as a scalar under $SO\left(
3\right) $ and is related to $\Phi \left( x\right) $ by 
\begin{equation}
\begin{tabular}{lll}
$d\Phi \left( \vec{x}\right) $ & $=$ & $\underset{d\vec{x}\rightarrow \vec{0}%
}{\lim}\left[ \Phi \left( \vec{x}+d\vec{x}\right) -\Phi \left( \vec {x}%
\right) \right] $%
\end{tabular}%
\end{equation}
In the standard case of a simple cubic lattice with spacing parameter $L$,
the arbitrary elementary variations $d\vec{x}$ of the 3-dimensional
continuum are given by the \emph{6} first nearest neighbors namely 
\begin{equation}
\begin{tabular}{lllll}
$\pm L\vec{e}_{1}$ & , & $\pm L\vec{e}_{2}$ & , & $\pm L\vec{e}_{3}$%
\end{tabular}%
\end{equation}
In the case of discretization of space to the $\mathcal{L}%
_{3D}^{su_{3}\times u_{1}}$ of eq(\ref{31}), vectors $\vec{x}$ in continuum
are mapped to $\vec {R}_{n}^{\left( q\right) }$ and the elementary
variations $d\vec{x}$ are mapped to first nearest neighbors of $\vec{R}%
_{n}^{\left( q\right) }$; that is 
\begin{equation}
\begin{tabular}{lll}
$\vec{x}+d\vec{x}$ & $\rightarrow$ & $\vec{R}_{n}^{\left( q\right) }+\vec {V}%
_{I}^{\left( q\right) }$%
\end{tabular}%
\end{equation}
with 
\begin{equation}
\vec{V}_{I}^{\left( q\right) }=\left( 
\begin{array}{c}
\vec{\upsilon}_{I} \\ 
qZ_{I}%
\end{array}
\right)  \label{vi}
\end{equation}
and the 6 non zero $\vec{\upsilon}_{I}$'s as 
\begin{equation}
\vec{\upsilon}_{i}=\left( 
\begin{array}{c}
\upsilon_{i}^{1} \\ 
\upsilon_{i}^{2}%
\end{array}
\right) ,\qquad i=1,...,6
\end{equation}
referring to the first nearest neighbors of the base sublattice $\mathbb{A}%
_{2}^{\ast}$. So the 3D crystal analogue of $d\Phi \left( x\right) $ is
given by%
\begin{equation}
\begin{tabular}{lll}
&  &  \\ 
$\Phi_{I}\left( R_{n}\right) $ & $=$ & $\Phi(R_{n}+V_{I}^{\left( q\right)
})-\Phi \left( R_{n}\right) $ \\ 
&  & 
\end{tabular}
\label{MA}
\end{equation}
with projection on the $\mathbb{A}_{2}^{\ast}$ base sublattice corresponding
to $q=0$ as follows%
\begin{equation}
\begin{tabular}{lll}
$\phi_{I}\left( \vec{r}_{n}\right) $ & $=$ & $\phi \left( \vec{r}_{n}+\vec{%
\upsilon}_{I}\right) -\phi \left( \vec{r}_{n}\right) $ \\ 
&  & 
\end{tabular}
\label{AM}
\end{equation}
where the $\vec{\upsilon}_{I}$'s are as in (\ref{vi}). For later use, it is
convenient to denote $\phi_{I}\left( \vec{r}_{n}\right) $ like 
\begin{equation}
\begin{tabular}{lllll}
$\phi_{I}\left( \vec{r}_{n}\right) $ & $=$ & $\phi_{\mathbf{r}%
_{n}\rightarrow \left( \mathbf{r}_{n}+\mathbf{\upsilon}_{I}\right) }$ & $%
\equiv$ & $\phi_{n,I}$%
\end{tabular}%
\end{equation}
\end{itemize}

\subsubsection{\emph{First result}}

From eqs(\ref{MA}-\ref{AM}), we learn a set of useful features that we
collect below:

\begin{itemize}
\item the field $\partial_{\mu}\Phi \left( x\right) $ is mapped to link
variables 
\begin{equation}
\Phi_{n,I}^{\left( q\right) }
\end{equation}
living on edges of the 3D crystal $\mathcal{L}_{3D}^{su_{3}\times u_{1}}$. 
\newline
For $q=0$, the link variables $\Phi_{n,I}^{\left( q\right) }$ are projected
down to 
\begin{equation}
\phi_{n,I}
\end{equation}
living on the $\mathbb{A}_{2}^{\ast}$ the edge 
\begin{equation}
\begin{tabular}{lll}
$\overrightarrow{P_{\mathbf{r}_{n}}P}_{\mathbf{r}_{n}+\mathbf{\upsilon}_{I}} 
$ & $\sim$ & $\mathbf{\vec{\upsilon}}_{I}$%
\end{tabular}
\label{lk}
\end{equation}

\item the analogue of the gauge field $A_{\mu}\left( x\right) $ on the
lattice $\mathcal{L}_{3D}^{su_{3}\times u_{1}}$ is given by the link
variables 
\begin{equation}
\begin{tabular}{lll}
$\mathcal{U}_{I}(R_{n}^{\left( q\right) })$ & $=$ & $\mathcal{U}%
_{n,I}^{\left( q\right) }$%
\end{tabular}%
\end{equation}
with projection on the $\mathbb{A}_{2}^{\ast}$ base sublattice as%
\begin{equation}
\begin{tabular}{lll}
$\mathcal{U}_{I}\left( \vec{r}_{n}\right) $ & $=$ & $U_{n,I}$%
\end{tabular}%
\end{equation}
living on the links (\ref{lk}). \newline
The usual gauge transformation in continuum with $G$ a generic gauge group
element 
\begin{equation}
\partial_{\mu}+ig_{YM}A_{\mu}\left( x\right) \rightarrow G\left( x\right) 
\left[ \partial_{\mu}+ig_{YM}A_{\mu}\left( x\right) \right] G^{\dagger
}\left( x\right)
\end{equation}
is mapped to%
\begin{equation}
\begin{tabular}{lll}
$\mathcal{U}_{I}(R_{n}^{\left( q\right) })$ & $\rightarrow$ & $%
G(R_{n}^{\left( q\right) })\mathcal{U}_{I}(R_{n}^{\left( q\right)
})G^{\dagger }(R_{n}^{\left( q\right) }+V_{I})$ \\ 
&  & 
\end{tabular}%
\end{equation}
On the $\mathbb{A}_{2}^{\ast}$ base sublattice, these transformations reduce
to 
\begin{equation}
\begin{tabular}{lll}
$U_{I}\left( \vec{r}_{n}\right) $ & $\rightarrow$ & $G\left( \vec{r}%
_{n}\right) U_{I}\left( \vec{r}_{n}\right) G^{\dagger}\left( \vec{r}_{n}+%
\vec{\upsilon}_{I}\right) $ \\ 
&  & 
\end{tabular}%
\end{equation}

\item the discrete analogue of the field strength $F_{\mu \nu}\left(
x\right) $ is given by the plaquette variables%
\begin{equation}
\begin{tabular}{lll}
$\mathcal{W}_{\left[ IJ\right] }\left( R_{n}^{\left( q\right) }\right) $ & $%
= $ & $\mathcal{W}_{n,\left[ IJ\right] }^{\left( q\right) },\qquad I\neq J$
\\ 
&  & 
\end{tabular}%
\end{equation}
which is dual to 1-dimensional link. On the $\mathbb{A}_{2}^{\ast}$ base
sublattice, these variables are projected to the 2- dimensional plaquette
variables 
\begin{equation}
\begin{tabular}{lll}
$W_{\left[ IJ\right] }\left( \mathbf{r}_{n}\right) $ & $=$ & $W_{n,\left[ IJ%
\right] },\qquad I\neq J$%
\end{tabular}%
\end{equation}
associated with 
\begin{equation}
\begin{tabular}{lll}
$\overrightarrow{P_{\mathbf{r}_{n}}P}_{\mathbf{r}_{n}+\mathbf{\upsilon}%
_{I}}\wedge \overrightarrow{P_{\mathbf{r}_{n}}P}_{\mathbf{r}_{n}+\mathbf{%
\upsilon }_{J}}$ & $\sim$ & $\mathbf{\vec{\upsilon}}_{I}\wedge \mathbf{\vec{%
\upsilon}}_{J}$ \\ 
&  & 
\end{tabular}
\label{PP}
\end{equation}
This plaquette has 4 vertices located at 
\begin{equation}
\begin{tabular}{lllllll}
&  &  &  &  &  &  \\ 
$P_{\mathbf{r}_{n}}$ & $\sim$ & $\vec{r}_{n}$ & $,$ & $P_{\mathbf{r}_{n}+%
\mathbf{\upsilon}_{I}}$ & $\sim$ & $\vec{r}_{n}+\vec{\upsilon}_{I}$ \\ 
&  &  &  &  &  &  \\ 
$P_{\mathbf{r}_{n}+\mathbf{\upsilon}_{J}}$ & $\sim$ & $\vec{r}_{n}+\vec{%
\upsilon}_{J}$ & $,$ & $P_{\mathbf{r}_{n}+\mathbf{\upsilon}_{I}+\mathbf{%
\upsilon}_{J}}$ & $\sim$ & $\vec{r}_{n}+\vec{\upsilon}_{I}+\vec{\upsilon}%
_{J} $ \\ 
&  &  &  &  &  & 
\end{tabular}%
\end{equation}
and has an interpretation in terms of the vector surface $\vec{s}_{IJ}=\vec{%
\upsilon}_{I}\wedge \vec{\upsilon}_{J}$ with components 
\begin{equation}
s_{IJ}^{\mu}=\frac{1}{2}\varepsilon_{\mu \nu \rho}\upsilon_{I}^{\mu}\upsilon
_{J}^{\rho}
\end{equation}

\item Non abelian $U\left( N\right) $ gauge fields\newline
In the case of YM theory with non abelian $U\left( N\right) $\ gauge
symmetry, the fields are valued in the adjoint representation of the Lie
algebra of the gauge symmetry; so the gradient $\partial_{\mu}\Phi \left(
x\right) $ and the field strength $F_{\mu \nu}$ involve gauge covariant
derivatives 
\begin{equation}
\begin{tabular}{lll}
&  &  \\ 
$D_{\mu}\Phi$ & $=$ & $\partial_{\mu}\Phi+\left[ A_{\mu},\Phi \right] $ \\ 
$F_{\mu \nu}$ & $=$ & $\left[ D_{\mu},D_{\nu}\right] $ \\ 
&  & 
\end{tabular}%
\end{equation}
By discretization, $D_{\mu}\Phi \left( x\right) $ and $F_{\mu \nu}\left(
x\right) $\ are respectively mapped to discrete $N\times N$ matrix variables%
\begin{equation}
\begin{tabular}{lll}
$\Phi_{I}^{\left( q\right) }\left( R_{n}\right) $ & , & $\mathcal{W}%
_{IJ}^{\left( p\right) }\left( \vec{r}_{n}\right) $ \\ 
&  & 
\end{tabular}%
\end{equation}
carrying moreover charges under $U\left( 1\right) $. On the base sublattice $%
\mathbb{A}_{2}^{\ast}$ where $q=0$, these quantities become 
\begin{equation}
\begin{tabular}{lll}
$\Phi_{I}\left( \vec{r}_{n}\right) $ & $=$ & $U_{I}\left( \vec{r}_{n}\right)
\Phi \left( \vec{r}_{n}+\vec{\upsilon}_{I}\right) -\Phi \left( \vec{r}%
_{n}\right) U_{I}\left( \vec{r}_{n}\right) $ \\ 
&  &  \\ 
$W_{IJ}\left( \vec{r}_{n}\right) $ & $=$ & $U_{I}\left( \vec{r}_{n}\right)
U_{J}\left( \vec{r}_{n}+\vec{\upsilon}_{I}\right) -U_{J}\left( \vec{r}%
_{n}\right) U_{I}\left( \vec{r}_{n}+\vec{\upsilon}_{J}\right) $ \\ 
&  & 
\end{tabular}%
\end{equation}
\end{itemize}

\subsection{Complex extension and orientation}

\begin{figure}[ptbh]
\begin{center}
\hspace{0cm} \includegraphics[width=7cm]{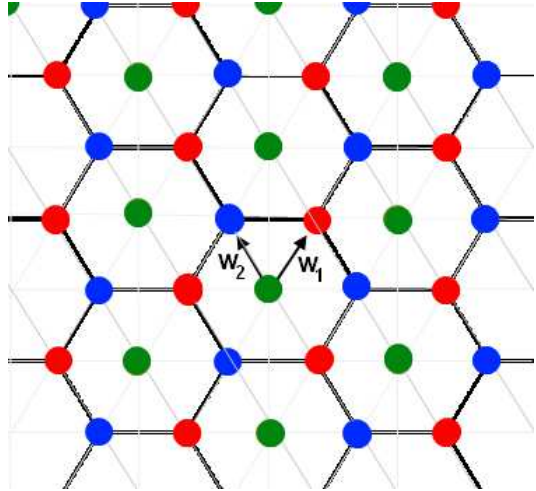}
\end{center}
\par
\vspace{0 cm}
\caption{the lattice $\mathbb{A}_{2}^{\ast}$ generated by the 2 basic weight
vectors $\vec{\protect \omega}_{1},$ $\vec{\protect \omega}_{2}$ of $SU\left(
3\right) $. Green nodes are associated with the lattice variable $\protect%
\psi _{n}^{\left( -3q\right) }$; red nodes with $\protect \psi_{n}^{I\left(
+q\right) }$ and blue nodes with the lattice gauge variables $%
(U_{n}^{I\left( -2q\right) },U_{In}^{\left( +2q\right) })$.\ More precisely $%
\protect \psi_{n}^{I\left( +q\right) }$ transforms into the representation $%
\mathbf{3}$; it is given by the link from Green to red nodes. Similarly, $%
U_{n}^{I\left( -2q\right) }$; it transforms in the $\mathbf{3}$
representation and is given by links from the blue to the green nodes while $%
U_{In}^{\left( +2q\right) }$ transforms in the adjoint $\mathbf{\bar{3}}$
and is given by links from the green to the blue sites.}
\label{1}
\end{figure}

In the case of the complex 3D space, on which the 3-dimensional twisted $%
\mathcal{N}=4$ supersymmetric YM has been formulated, one distinguishes two
kinds of quantities:

\begin{itemize}
\item complex antisymmetric tensor fields of type $B^{\left( q\right)
a_{1}...a_{p}}$ transforming in some complex representation $\boldsymbol{R}%
_{q}$ of $SU\left( 3\right) \times U\left( 1\right) ,$

\item the adjoint fields $B_{a_{1}...a_{p}}^{\left( -q\right) }$
transforming in the adjoint conjugate $\boldsymbol{\bar{R}}_{-q}$.
\end{itemize}

\  \  \  \newline
On the lattice $\mathcal{L}_{3D}^{su_{3}\times u_{1}}$, these objects are
interpreted in terms of \emph{oriented} p-simplex. To that purpose, recall
the objects appearing in the field action (\ref{fac}-\ref{caf}); there, we
have bosons and fermions: In the bosonic sector, each object have an adjoint
as shown below 
\begin{equation}
\begin{tabular}{lll}
\ Object \  & \  \  \  \  \  & Adjoint \\ 
$\  \ z^{a\left( -2q\right) }$ &  & $\  \  \bar{z}_{a}^{\left( +2q\right) }$ \\ 
$\  \ dz^{a\left( -2q\right) }$ &  & $\  \ d\bar{z}_{a}^{\left( +2q\right) }$
\\ 
$\  \  \mathcal{G}^{a\left( -2q\right) }$ &  & $\  \  \mathcal{\bar{G}}%
_{a}^{\left( +2q\right) }$ \\ 
$\  \  \nabla^{a\left( -2q\right) }$ &  & $\  \  \nabla_{a}^{\left( +2q\right) }$
\\ 
$\  \  \mathcal{F}^{ab\left( -4q\right) }$ &  & $\  \  \mathcal{\bar{F}}%
_{ab}^{\left( +4q\right) }$%
\end{tabular}%
\end{equation}
and so both orientations of bosonic link variables are involved; contrary to
the chiral fermionic sector%
\begin{equation}
\begin{tabular}{lll}
\  \ Object \  & \  \  & Adjoint \\ 
$\  \  \theta^{\left( -3q\right) }$ &  & \ - \\ 
$\  \  \vartheta^{a\left( +q\right) }$ &  & \ - \\ 
$\  \  \psi^{\left( -3q\right) }$ &  & \ - \\ 
$\  \  \psi^{a\left( +\right) }$ &  & \ - \\ 
$\  \  \nabla^{a\left( -2q\right) }\psi^{b\left( +\right) }$ &  & \ - \\ 
$\  \nabla_{a}^{\left( +2q\right) }\psi^{a\left( +\right) }$ &  & \ -%
\end{tabular}%
\end{equation}
where we have only one orientation for fermionic lattice link variables.

\  \  \  \  \  \  \newline
To study the discrete version of (\ref{fac}-\ref{caf}), we proceed in two
steps:

\emph{a) step 1}: we describe lattice theory living on the 2-dimensional $%
\mathbb{A}_{2}^{\ast}$ given by fig \ref{1}.

\begin{figure}[ptbh]
\begin{center}
\hspace{0cm} \includegraphics[width=14cm]{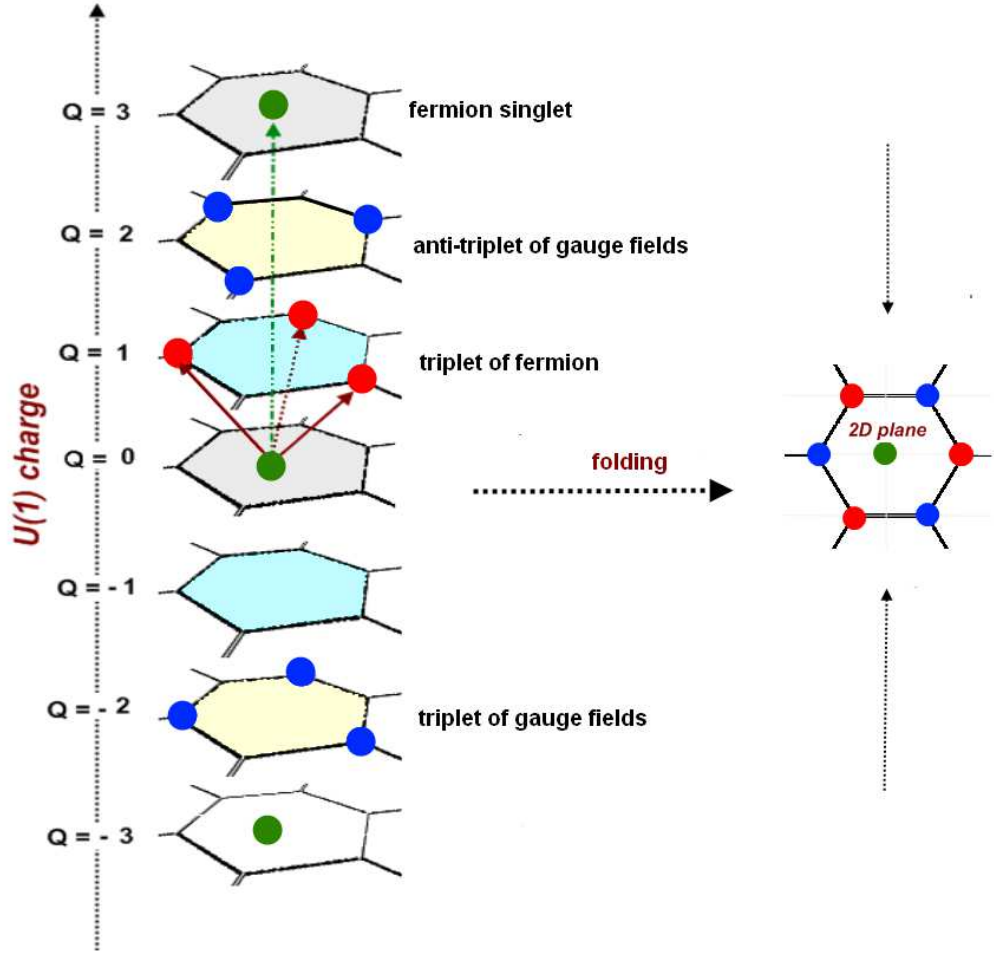}
\end{center}
\par
\vspace{0 cm}
\caption{the 3D lattice $\mathcal{L}_{3D}^{su_{3}\times u_{1}}$ given by a $%
\mathcal{L}_{2D}^{su_{3}}\times \mathcal{L}_{1D}^{u_{1}}$ fibration with 2D
base sublattice given by $\mathbb{A}_{2}^{\ast}$ and 1D fiber $\mathcal{L}%
_{1D}^{u_{1}}$ isomorphic to $\mathbb{Z}$, the set of integers. The
superposition of the 3 sublattices $\mathfrak{G}$, $\mathfrak{R}$ and $%
\mathfrak{B}$ making $\mathbb{A}_{2}^{\ast}$ as given by eq(\protect \ref{GB}%
) is lifted. Sheets with $Q=3q$ corresponds to the sublattice $\mathfrak{G}$%
, sheets with $Q=q$ to the sublattice $\mathfrak{R}$ and those with $Q=2q$
to the sublattice $\mathfrak{B}$.}
\label{2}
\end{figure}

\emph{b) step 2}: we extend the construction from $\mathbb{A}_{2}^{\ast}$ to
the 3-dimensional lattice $\mathcal{L}_{3D}^{su_{3}\times u_{1}}$ by
implementing the fiber direction $\mathcal{L}_{1D}^{u_{1}}$. This
corresponds to unfolding the normal direction\ to $\mathbb{A}_{2}^{\ast}$ as
illustrated on fig \ref{2}. 

\  \  \  \  \ 

\subsubsection{\emph{twisted supersymmetric YM on} $\mathbb{A}_{2}^{\ast}$}

The lattice $\mathbb{A}_{2}^{\ast}$ is generated by the 2 fundamental weight
vectors $\vec{\omega}_{1}$, $\vec{\omega}_{2}$ of the Lie algebra of the
simple $SU\left( 3\right) $ symmetry. These are non canonical vectors obeying%
\begin{equation}
\begin{tabular}{lllll}
$\vec{\omega}_{i}.\vec{\omega}_{i}$ & $=\frac{2}{3}$ & , & $\left( \widehat{%
\vec{\omega}_{1},\vec{\omega}_{2}}\right) $ & $=\frac{\pi}{6}$ \\ 
&  &  &  & 
\end{tabular}%
\end{equation}
Each site $\vec{r}_{n}$ in $\mathbb{A}_{2}^{\ast}$ ( say a green site of fig %
\ref{1} ) has $6$ first nearest neighbors located at $\vec{r}_{n}+\vec{%
\upsilon}_{i}$ and which can be organized into 2 subsets, each having $3$
elements like%
\begin{equation}
\begin{tabular}{lll}
$\vec{\upsilon}_{1}$ & $=$ & $L\sqrt{\frac{3}{2}}\vec{\omega}_{1}$ \\ 
$\vec{\upsilon}_{2}$ & $=$ & $L\sqrt{\frac{3}{2}}\left( \vec{\omega}_{2}-%
\vec{\omega}_{1}\right) $ \\ 
$\vec{\upsilon}_{3}$ & $=$ & $-L\sqrt{\frac{3}{2}}\vec{\omega}_{2}$%
\end{tabular}
\label{VI}
\end{equation}
and%
\begin{equation}
\begin{tabular}{lll}
$\vec{\upsilon}_{1}^{\prime}$ & $=$ & $-\vec{\upsilon}_{1}$ \\ 
$\vec{\upsilon}_{2}^{\prime}$ & $=$ & $-\vec{\upsilon}_{2}$ \\ 
$\vec{\upsilon}_{3}^{\prime}$ & $=$ & $-\vec{\upsilon}_{3}$%
\end{tabular}%
\end{equation}
Notice that each triplet obeys a traceless property%
\begin{equation}
\begin{tabular}{lll}
$\vec{\upsilon}_{1}+\vec{\upsilon}_{2}+\vec{\upsilon}_{3}$ & $=$ & $\vec{0}$
\\ 
$\vec{\upsilon}_{1}^{\prime}+\vec{\upsilon}_{2}^{\prime}+\vec{\upsilon}%
_{3}^{\prime}$ & $=$ & $\vec{0}$%
\end{tabular}
\label{49}
\end{equation}
Notice also that the set $\mathfrak{G}$ of (green) sites $\vec{r}_{n}$ at
the centre of the hexagons of fig \ref{1} form a sublattice of $\mathbb{A}%
_{2}^{\ast}$ generated by the particular vectors%
\begin{equation}
\begin{tabular}{lll}
$\vec{\alpha}_{1}$ & $=$ & $2\vec{\omega}_{1}-\vec{\omega}_{2}$ \\ 
$\vec{\alpha}_{2}$ & $=$ & $\vec{\omega}_{1}-2\vec{\omega}_{2}$%
\end{tabular}%
\end{equation}
This means that green sites are related amongst others as%
\begin{equation}
\begin{tabular}{lll}
$\vec{r}_{m}$ & $=\dsum \limits_{m_{1},m_{2}}m_{1}\vec{\alpha}_{1}+m_{2}\vec{%
\alpha}_{2}$ & , \\ 
&  & 
\end{tabular}%
\end{equation}
and form a sublattice%
\begin{equation*}
\begin{tabular}{lll}
$\mathfrak{G}$ & $\equiv$ & $\left \{ \vec{r}_{m}\right \} _{green\text{ }%
sites}$%
\end{tabular}%
\end{equation*}
which is nothing but $\mathbb{A}_{2},$ the \emph{root lattice}\ of $SU\left(
3\right) $. So we have the isomorphism%
\begin{equation}
\mathfrak{G}\simeq \mathbb{A}_{2}
\end{equation}
Notice moreover that, because of the symmetric role played by the \emph{3}
types of nodes (green, red, blue in fig \ref{1}), the same thing may be said
about the set $\mathfrak{R}$ of red sites and the set $\mathfrak{B}$ of blue
ones. In other words the set $\mathfrak{R}$ is isomorphic to a root lattice
of SU$\left( 3\right) $ and similarly the set $\mathfrak{B}$ is isomorphic
as well to a root lattice of SU$\left( 3\right) $. Thus we have the
isomorphisms%
\begin{equation}
\begin{tabular}{lll}
$\mathfrak{R}\simeq \mathbb{A}_{2}$ & , & $\mathfrak{B}\simeq \mathbb{A}_{2}$%
\end{tabular}%
\end{equation}
and formally%
\begin{equation}
\begin{tabular}{lll}
$\mathfrak{R}\simeq \mathfrak{G}+\vec{\omega}_{1}$ & , & $\mathfrak{B}\simeq 
\mathfrak{G}+\vec{\omega}_{2}$%
\end{tabular}%
\end{equation}
From this representation, it follows that the lattice $\mathbb{A}_{2}^{\ast} 
$ is made by the superposition of the 3 sublattices $\mathfrak{G}$, $%
\mathfrak{R}$ and $\mathfrak{B}$ or equivalently 
\begin{equation}
\mathbb{A}_{2}^{\ast}=\mathfrak{G}\cup \mathfrak{R}\cup \mathfrak{B}
\label{GB}
\end{equation}
For more details on the matrix describing the shape of the 2- dimensional
base lattice $\mathbb{A}_{2}^{\ast}$; \textrm{see subsection 5.}\newline
Under discretization of continuum, we have the following correspondence%
\begin{equation}
\begin{tabular}{lll}
continuum & \  \  \  \  \  \  \  \  \  \  \  \  \  & crystal $\mathbb{A}_{2}^{\ast}$ \\ 
$\left( z,\bar{z}\right) $ &  & $\vec{r}_{n}$ \\ 
$z+dz$ &  & $\vec{r}_{n}+\vec{\upsilon}_{I}$ \\ 
$\bar{z}+d\bar{z}$ &  & $\vec{r}_{n}-\vec{\upsilon}_{I}$ \\ 
&  & 
\end{tabular}%
\end{equation}
with $n=\left( n_{1},n_{2}\right) $ arbitrary integers; and where $\vec{%
\upsilon}_{I}$ stand for the 3 oriented first nearest neighbors given by (%
\ref{VI}). Observe that complex conjugation is captured by the change of the
orientation of $\vec{\upsilon}_{I}$.\newline
Regarding the lattice analogue of the twist fields in continuum, we have the
following dictionary:

\begin{description}
\item[$\left( i\right) $ \emph{bosonic fields}] : 
\begin{equation}
\begin{tabular}{lll}
continuum & \  \  \  \  \  \  \  \  \  \  \  \  & crystal $A_{2}^{\ast}$ \\ 
$\  \mathcal{G}^{a}\left( z,\bar{z}\right) $ &  & $\ U_{n}^{I}$ \\ 
$\  \mathcal{\bar{G}}_{a}\left( z,\bar{z}\right) $ &  & $\ U_{n,I}^{\dagger} $
\\ 
$\  \mathcal{F}^{ab}\left( z,\bar{z}\right) $ &  & $\ W_{n}^{IJ}$ \\ 
$\  \mathcal{\bar{F}}_{ab}\left( z,\bar{z}\right) $ &  & $\
W_{n,IJ}^{\dagger} $ \\ 
$\  \mathcal{F}_{a}^{b}\left( z,\bar{z}\right) $ &  & $\ W_{n,I}^{J}$ \\ 
&  & 
\end{tabular}%
\end{equation}
with%
\begin{equation}
\begin{tabular}{lll}
$U_{n}^{I}$ & $=$ & $U^{I}\left( \vec{r}_{n}\right) $ \\ 
&  &  \\ 
$U_{n,I}^{\dagger}$ & $=$ & $U_{I}^{\dagger}\left( \vec{r}_{n}\right) $ \\ 
&  &  \\ 
$W_{n}^{IJ}$ & $=$ & $U^{I}\left( \vec{r}_{n}\right) U^{J}\left( \vec {r}%
_{n}+\vec{\upsilon}_{I}\right) -U^{J}\left( \vec{r}_{n}\right) U^{I}\left( 
\vec{r}_{n}+\vec{\upsilon}_{J}\right) $ \\ 
&  &  \\ 
$W_{n,IJ}^{\dagger}$ & $=$ & $U_{J}^{\dagger}\left( \vec{r}_{n}+\vec {%
\upsilon}_{I}\right) U_{I}^{\dagger}\left( \vec{r}_{n}\right)
-U_{I}^{\dagger}\left( \vec{r}_{n}+\vec{\upsilon}_{J}\right) U_{J}^{\dagger
}\left( \vec{r}_{n}\right) $ \\ 
&  & 
\end{tabular}%
\end{equation}
and%
\begin{equation}
\begin{tabular}{lll}
$W_{n,I}^{I}$ & $=$ & $U^{I}\left( \vec{r}_{n}\right) U_{I}^{\dagger}\left( 
\vec{r}_{n}\right) -U_{I}^{\dagger}\left( \vec{r}_{n}-\vec{\upsilon}%
_{I}\right) U^{I}\left( \vec{r}_{n}-\vec{\upsilon}_{I}\right) $ \\ 
&  & 
\end{tabular}%
\end{equation}

\item[$\left( ii\right) $ \emph{fermionic fields}] :%
\begin{equation}
\begin{tabular}{lll}
continuum & \  \  \  \  \  \  \  \  \  \  \  \  & crystal $A_{2}^{\ast}$ \\ 
$\  \  \psi \left( z,\bar{z}\right) $ &  & $\  \  \psi_{n}$ \\ 
$\  \  \psi^{a}\left( z,\bar{z}\right) $ &  & $\  \  \psi_{n}^{I}$ \\ 
$\  \  \nabla^{a}\psi^{b}$ &  & $\  \  \psi_{n}^{IJ}$ \\ 
$\  \  \bar{\nabla}_{a}\psi^{a}$ &  & $\  \  \psi_{n,I}^{I}$%
\end{tabular}%
\end{equation}
with%
\begin{equation}
\begin{tabular}{lll}
$\  \  \psi_{n}$ & $=$ & $\psi \left( \vec{r}_{n}\right) $ \\ 
$\  \  \psi_{n}^{I}$ & $=$ & $\psi^{I}\left( \vec{r}_{n}\right) $ \\ 
$\  \  \psi_{n}^{IJ}$ & $=$ & $\psi^{IJ}\left( \vec{r}_{n}\right) $ \\ 
$\  \  \psi_{n,I}^{I}$ & $=$ & $\psi_{I}^{I}\left( \vec{r}_{n}\right) $%
\end{tabular}%
\end{equation}
and%
\begin{equation}
\begin{tabular}{lll}
&  &  \\ 
$\psi^{IJ}\left( \vec{r}_{n}\right) $ & $=$ & $U^{I}\left( \vec{r}%
_{n}\right) \psi^{J}\left( \vec{r}_{n}+\vec{\upsilon}_{I}\right) -\psi
^{J}\left( \vec{r}_{n}\right) U^{I}\left( \vec{r}_{n}+\vec{\upsilon}%
_{J}\right) $ \\ 
&  &  \\ 
$\psi_{I}^{I}\left( \vec{r}_{n}\right) $ & $=$ & $\psi^{I}\left( \vec {r}%
_{n}\right) U_{I}^{\dagger}\left( \vec{r}_{n}\right) -U_{I}^{\dagger }\left( 
\vec{r}_{n}-\vec{\upsilon}_{I}\right) \psi^{I}\left( \vec{r}_{n}-\vec{%
\upsilon}_{I}\right) $%
\end{tabular}
\label{IJ}
\end{equation}
\end{description}

\  \  \  \  \  \newline
The $U\left( N\right) $ gauge transformations with generic unitary $N\times
N $ matrix $G\left( \vec{r}_{n}\right) $ act on the lattice fields as follows%
\begin{align}
& 
\begin{tabular}{lll}
field $\  \ $ & $\rightarrow$ & $\  \ $gauge transform \\ \hline
&  &  \\ 
$\  \ U_{n}^{I}$ &  & $\  \ G\left( \vec{r}_{n}\right) U^{I}\left( \vec {r}%
_{n}\right) G^{\dagger}\left( \vec{r}_{n}+\vec{\upsilon}_{I}\right) $ \\ 
&  &  \\ 
$\  \ U_{n,I}^{\dagger}$ &  & $\  \ G\left( \vec{r}_{n}+\vec{\upsilon}%
_{I}\right) U_{I}^{\dagger}\left( \vec{r}_{n}\right) G^{\dagger}\left( \vec{r%
}_{n}\right) $ \\ 
&  &  \\ 
$\  \ W_{n}^{IJ}$ &  & $\  \ G\left( \vec{r}_{n}\right) W_{n}^{IJ}G^{\dagger
}\left( \vec{r}_{n}+\vec{\upsilon}_{I}+\vec{\upsilon}_{J}\right) $ \\ 
&  &  \\ 
$\  \ W_{n,IJ}^{\dagger}$ &  & $\  \ G\left( \vec{r}_{n}+\vec{\upsilon}_{I}+%
\vec{\upsilon}_{J}\right) W_{n}^{IJ}G^{\dagger}\left( \vec{r}_{n}\right) $
\\ 
&  &  \\ 
$\  \  \psi \left( \vec{r}_{n}\right) $ $\  \ $ &  & $\  \ G\left( \vec{r}%
_{n}\right) \psi \left( \vec{r}_{n}\right) G^{\dagger}\left( \vec{r}%
_{n}\right) $ \\ 
&  &  \\ 
$\  \  \psi^{I}\left( \vec{r}_{n}\right) $ &  & $\  \ G\left( \vec{r}%
_{n}\right) \psi^{I}\left( \vec{r}_{n}\right) G^{\dagger}\left( \vec {r}_{n}+%
\vec{\upsilon}_{I}\right) $ \\ 
&  &  \\ 
$\  \  \psi^{IJ}\left( \vec{r}_{n}\right) $ $\  \ $ &  & $\  \ G\left( \vec {r}%
_{n}\right) \psi^{IJ}\left( \vec{r}_{n}\right) G^{\dagger}\left( \vec{r}_{n}+%
\vec{\upsilon}_{I}+\vec{\upsilon}_{J}\right) $ \\ 
&  &  \\ 
$\  \  \psi_{I}^{I}\left( \vec{r}_{n}\right) $ &  & $\  \ G\left( \vec{r}%
_{n}\right) \psi_{I}^{I}\left( \vec{r}_{n}\right) G^{\dagger}\left( \vec{r}%
_{n}\right) $ \\ 
&  &  \\ \hline
\end{tabular}
\\
&
\end{align}
Using these gauge transformations, one can check that the following
couplings 
\begin{equation}
\begin{tabular}{lll}
$\left( i\right) $ & : & $Tr\left[ \psi \left( \vec{r}_{n}\right) \psi
_{I}^{I}\left( \vec{r}_{n}\right) \right] $ \\ 
&  &  \\ 
$\left( ii\right) $ & : & $-\frac{1}{2}\varepsilon_{IJK}Tr\left[ \psi
^{K}\left( \vec{r}_{n}-\vec{\upsilon}_{K}\right) \psi^{IJ}\left( \vec {r}%
_{n}\right) \right] $ \\ 
&  & 
\end{tabular}%
\end{equation}
are gauge invariant provided we have%
\begin{equation}
G^{\dagger}\left( \vec{r}_{n}+\vec{\upsilon}_{I}+\vec{\upsilon}_{J}\right)
G\left( \vec{r}_{n}-\vec{\upsilon}_{K}\right) =I
\end{equation}
But this constraint equation requires%
\begin{equation}
\vec{\upsilon}_{I}+\vec{\upsilon}_{J}+\vec{\upsilon}_{K}=\vec{0}  \label{vvv}
\end{equation}
which, up on using the antisymmetry property of the tensor $%
\varepsilon_{IJK} $, can be also written as 
\begin{equation}
\vec{\upsilon}_{1}+\vec{\upsilon}_{2}+\vec{\upsilon}_{3}=\vec{0}
\end{equation}
This constraint relation is identically satisfied for the lattice $\mathbb{A}%
_{2}^{\ast}$ as shown by eq(\ref{49}).

\subsubsection{\emph{\ action of twisted supersymmetric YM on} $\mathbb{A}%
_{2}^{\ast}$}

Using the above dictionary giving the analogue of fields in continuum to
lattice variables on $\mathcal{L}_{3D}^{su_{3}\times u_{1}}$, we can work
out the action of twisted supersymmetric YM on the base sublattice $\mathbb{A%
}_{2}^{\ast}$. This action may decomposed in 2 parts as%
\begin{equation}
\begin{tabular}{lll}
$\mathcal{S}_{lattice}$ & $=$ & $\mathcal{S}_{lattice}^{bose}+\mathcal{S}%
_{lattice}^{fermi}$ \\ 
&  & 
\end{tabular}%
\end{equation}
where $\mathcal{S}_{lattice}^{bose}$ involving bosonic degrees of freedom
and $\mathcal{S}_{lattice}^{fermi}$ describing lattice fermionic variables
coupled to the gauge link variables.

\  \  \  \  \  \  \ 

$\left( \alpha \right) $ \emph{Bosonic term}\newline
Under discretization, the bosonic part of the field action in continuum
namely 
\begin{equation}
\begin{tabular}{lll}
$\mathcal{S}_{cont}^{bose}$ & $=$ & $\alpha_{4}\dint Tr\left[ \mathcal{F}%
_{ab}^{\left( +4q\right) }\mathcal{F}^{ab{\small (-4q)}}\right] $ \\ 
&  &  \\ 
&  & $-\frac{\left( \alpha_{2}\right) ^{2}}{4\alpha_{1}}\dint Tr\left[ 
\mathcal{F}^{\left( 0\right) }\mathcal{F}^{\left( 0\right) }\right] $ \\ 
&  & 
\end{tabular}%
\end{equation}
gets mapped to the following gauge invariant lattice one%
\begin{equation}
\begin{tabular}{lll}
&  &  \\ 
$\mathcal{S}_{latt}^{bose}$ & $=$ & $\alpha_{4}\dsum \limits_{\mathbb{A}%
_{2}^{\ast}}Tr\left( W_{n}^{IJ}W_{n,IJ}^{\dagger}\right) -\frac{\left(
\alpha _{2}\right) ^{2}}{4\alpha_{1}}\dsum \limits_{\mathbb{A}%
_{2}^{\ast}}Tr\left( W_{n}^{\left( 0\right) }W_{n}^{\dagger \left( 0\right)
}\right) $%
\end{tabular}%
\end{equation}
with%
\begin{equation}
\begin{tabular}{lll}
$W_{n}^{IJ}W_{n,IJ}^{\dagger}$ & $=$ & $\mathcal{P}_{1}+\mathcal{P}_{2}-%
\mathcal{P}_{3}-\mathcal{P}_{4}$ \\ 
&  &  \\ 
$W_{n}^{\left( 0\right) }W_{n}^{\dagger \left( 0\right) }$ & $=$ & $\mathcal{%
R}_{1}+\mathcal{R}_{2}-\mathcal{R}_{3}-\mathcal{R}_{4}$%
\end{tabular}%
\end{equation}
and%
\begin{equation}
\begin{tabular}{lll}
$\mathcal{P}_{1}$ & $=$ & $U^{I}\left( \vec{r}_{n}\right) U^{J}\left( \vec{r}%
_{n}+\vec{\upsilon}_{I}\right) U_{J}^{\dagger}\left( \vec{r}_{n}+\vec{%
\upsilon}_{I}\right) U_{I}^{\dagger}\left( \vec{r}_{n}\right) $ \\ 
&  &  \\ 
$\mathcal{P}_{2}$ & $=$ & $U^{J}\left( \vec{r}_{n}\right) U^{I}\left( \vec{r}%
_{n}+\vec{\upsilon}_{J}\right) U_{I}^{\dagger}\left( \vec{r}_{n}+\vec{%
\upsilon}_{J}\right) U_{J}^{\dagger}\left( \vec{r}_{n}\right) $ \\ 
&  &  \\ 
$\mathcal{P}_{3}$ & $=$ & $U^{J}\left( \vec{r}_{n}\right) U^{I}\left( \vec{r}%
_{n}+\vec{\upsilon}_{J}\right) U_{J}^{\dagger}\left( \vec{r}_{n}+\vec{%
\upsilon}_{I}\right) U_{I}^{\dagger}\left( \vec{r}_{n}\right) $ \\ 
&  &  \\ 
$\mathcal{P}_{4}$ & $=$ & $U^{I}\left( \vec{r}_{n}\right) U^{J}\left( \vec{r}%
_{n}+\vec{\upsilon}_{I}\right) U_{I}^{\dagger}\left( \vec{r}_{n}+\vec{%
\upsilon}_{J}\right) U_{J}^{\dagger}\left( \vec{r}_{n}\right) $ \\ 
&  & 
\end{tabular}%
\end{equation}
as well as%
\begin{equation}
\begin{tabular}{lll}
$\mathcal{R}_{1}$ & $=$ & $U^{I}\left( \vec{r}_{n}\right) U_{I}^{\dagger
}\left( \vec{r}_{n}\right) U^{J}\left( \vec{r}_{n}\right) U_{J}^{\dagger
}\left( \vec{r}_{n}\right) $ \\ 
&  &  \\ 
$\mathcal{R}_{2}$ & $=$ & $U_{I}^{\dagger}\left( \vec{r}_{n}-\vec{\upsilon }%
_{I}\right) U^{I}\left( \vec{r}_{n}-\vec{\upsilon}_{I}\right)
U_{J}^{\dagger}\left( \vec{r}_{n}-\vec{\upsilon}_{J}\right) U^{J}\left( \vec{%
r}_{n}-\vec{\upsilon}_{J}\right) $ \\ 
&  &  \\ 
$\mathcal{R}_{3}$ & $=$ & $U^{I}\left( \vec{r}_{n}\right) U_{I}^{\dagger
}\left( \vec{r}_{n}\right) U_{J}^{\dagger}\left( \vec{r}_{n}-\vec{\upsilon }%
_{J}\right) U^{J}\left( \vec{r}_{n}-\vec{\upsilon}_{J}\right) $ \\ 
&  &  \\ 
$\mathcal{R}_{4}$ & $=$ & $U_{I}^{\dagger}\left( \vec{r}_{n}-\vec{\upsilon }%
_{I}\right) U^{I}\left( \vec{r}_{n}-\vec{\upsilon}_{I}\right) U^{J}\left( 
\vec{r}_{n}\right) U_{J}^{\dagger}\left( \vec{r}_{n}\right) $ \\ 
&  & 
\end{tabular}%
\end{equation}

$\left( \beta \right) $ \emph{fermionic term }\newline
For the fermionic terms, the analogue of%
\begin{equation}
\begin{tabular}{lll}
$\mathcal{S}_{cont}^{fermi}$ &  & $\left( \alpha_{2}+2\alpha_{4}\right)
\dint Tr\left[ \psi^{a\left( +q\right) }\nabla_{a}^{\left( +2q\right) }\psi^{%
{\small (-3q)}}\right] $ \\ 
&  &  \\ 
&  & $+2\alpha_{3}\dint Tr\left[ \varepsilon_{abc}\psi^{a{\small (+q)}%
}\nabla^{b\left( -2q\right) }\psi^{c\left( +q\right) }\right] $ \\ 
&  & 
\end{tabular}%
\end{equation}
is given by the following gauge invariant expression%
\begin{equation}
\begin{tabular}{lll}
&  &  \\ 
$\mathcal{S}_{latt}^{fermi}$ & $=$ & $\left( \alpha_{2}+2\alpha_{4}\right)
\dsum \limits_{\mathbb{A}_{2}^{\ast}}Tr\left[ \psi \left( \vec{r}_{n}\right)
\psi^{I}\left( \vec{r}_{n}\right) U_{I}^{\dagger}\left( \vec{r}_{n}\right) %
\right] +$ \\ 
&  &  \\ 
&  & $\left( \alpha_{2}+2\alpha_{4}\right) \dsum \limits_{\mathbb{A}%
_{2}^{\ast}}Tr\left[ \psi \left( \vec{r}_{n}\right) U_{I}^{\dagger}\left( 
\vec{r}_{n}-\vec{\upsilon}_{I}\right) \psi^{I}\left( \vec{r}_{n}-\vec{%
\upsilon}_{I}\right) \right] $ \\ 
&  &  \\ 
&  & $+2\alpha_{3}\dsum \limits_{\mathbb{A}_{2}^{\ast}}\varepsilon_{IJK}Tr%
\left[ \psi^{K}\left( \vec{r}_{n}-\vec{\upsilon}_{K}\right) \psi^{IJ}\left( 
\vec{r}_{n}\right) \right] $ \\ 
&  & 
\end{tabular}%
\end{equation}
with $\psi^{IJ}\left( \vec{r}_{n}\right) $ as in eq(\ref{IJ}).

\section{Twisted theory on the 3D lattice}

First, we construct the \emph{3D} lattice by giving further details on the
base sublattice $\mathbb{A}_{2}^{\ast}$; and the fiber $\mathcal{L}%
_{1D}^{u_{1}}$. Then we turn to derive the gauge invariant lattice action of
twisted supersymmetric YM theory on $\mathcal{L}_{3D}^{su_{3}\times u_{1}}$.

\subsection{More on the base sublattice $\mathbb{A}_{2}^{\ast}$}

The base sublattice $\mathbb{A}_{2}^{\ast}$ is generated by the 2
fundamental weight vectors $\vec{\omega}_{1},$ $\vec{\omega}_{2}$ of $%
SU\left( 3\right) $. These fundamental weight vectors, having the length $%
\frac{2}{3}$ and angle ($\widehat{\omega_{1},\omega_{2}})=\frac{\pi}{6}$,
are the dual of the 2 simple roots $\vec{\alpha}_{1},$ $\vec{\alpha}_{2}$ of 
$SU\left( 3\right) $,%
\begin{equation}
\vec{\omega}_{i}.\vec{\alpha}_{j}=\delta_{ij}
\end{equation}
Below, we take $\vec{\omega}_{1},$ $\vec{\omega}_{2}$ and $\vec{\alpha}_{1},$
$\vec{\alpha}_{2}$ in the real plane as follows%
\begin{equation}
\begin{tabular}{lll}
$\vec{\omega}_{1}=\left( \frac{\sqrt{2}}{2},\frac{\sqrt{6}}{6}\right) $ & ,
& $\vec{\omega}_{2}=\left( 0,\frac{\sqrt{6}}{3}\right) $ \\ 
&  &  \\ 
$\vec{\alpha}_{1}=\left( \sqrt{2},0\right) $ & $,$ & $\vec{\alpha}%
_{2}=\left( -\frac{\sqrt{2}}{2},\frac{\sqrt{6}}{2}\right) $%
\end{tabular}%
\end{equation}
They are related to each other like 
\begin{equation}
\begin{tabular}{lll}
$\vec{\alpha}_{1}$ & $=$ & $2\vec{\omega}_{1}-\vec{\omega}_{2}$ \\ 
$\vec{\alpha}_{2}$ & $=$ & $2\vec{\omega}_{2}-\vec{\omega}_{1}$%
\end{tabular}%
\end{equation}
and%
\begin{equation}
\begin{tabular}{lll}
$\vec{\omega}_{1}$ & $=$ & $\frac{1}{3}\left( 2\vec{\alpha}_{1}+\vec{\alpha }%
_{2}\right) $ \\ 
$\vec{\omega}_{2}$ & $=$ & $\frac{1}{3}\left( \vec{\alpha}_{1}+2\vec{\alpha }%
_{2}\right) $ \\ 
&  & 
\end{tabular}
\label{al}
\end{equation}
The vectors $\vec{\alpha}_{1},$ $\vec{\alpha}_{2}$ generate $\mathbb{A}_{2}$%
; the root lattice of $SU\left( 3\right) $. 
\begin{equation*}
\begin{tabular}{lllll}
&  &  &  &  \\ 
$\vec{r}_{n}\in \mathbb{A}_{2}$ & $\Leftrightarrow$ & $\vec{r}_{n}$ & $=$ & $%
\dsum \limits_{n,m}n\vec{\alpha}_{1}+m\vec{\alpha}_{2}$ \\ 
&  &  & $=$ & $\dsum \limits_{n,m}\left( 2n-m\right) \vec{\omega}_{1}+\left(
2m-n\right) \vec{\omega}_{2}$%
\end{tabular}%
\end{equation*}
Using $\vec{\omega}_{1},$ $\vec{\omega}_{2}$, position vectors $\vec{r}_{n}$
of sites in the wight lattice $\mathbb{A}_{2}^{\ast}$ are expanded like%
\begin{equation}
\vec{r}_{n}=\sqrt{\frac{3}{2}}L\text{ }\vec{\omega}_{\left(
n_{1},n_{2}\right) }
\end{equation}
with 
\begin{equation}
\vec{\omega}_{\left( n_{1},n_{2}\right) }=n_{1}\vec{\omega}_{1}+n_{2}\vec{%
\omega}_{2}
\end{equation}
and where $L$ stands for the spacing parameter of the lattice and $n=\left(
n_{1},n_{2}\right) $ are arbitrary integers. Using (\ref{al}), we also have 
\begin{equation}
\vec{r}_{n}=\sqrt{\frac{3}{2}}L\text{ }\frac{\left( 2n_{1}+n_{2}\right) }{3}%
\vec{\alpha}_{1}+\sqrt{\frac{3}{2}}L\frac{\left( n_{1}+2n_{2}\right) }{3}%
\vec{\alpha}_{2}
\end{equation}
The architecture of the sites of the $\mathbb{A}_{2}^{\ast}$ crystal is
encoded into the intersection matrix 
\begin{equation}
J_{ij}^{su_{3}}=\vec{\omega}_{i}.\vec{\omega}_{j}
\end{equation}
given by%
\begin{equation}
J_{ij}^{su_{3}}=\frac{1}{3}\left( 
\begin{array}{cc}
2 & 1 \\ 
1 & 2%
\end{array}
\right)
\end{equation}
with inverse%
\begin{equation}
\begin{tabular}{lll}
$K_{ij}^{su_{3}}=\left( 
\begin{array}{cc}
2 & -1 \\ 
-1 & 2%
\end{array}
\right) $ & , & $K_{ij}^{su_{3}}=\vec{\alpha}_{i}.\vec{\alpha}_{j}$%
\end{tabular}%
\end{equation}

\  \  \  \  \  \ 

\emph{classifying closest neighbors}\newline
Each site $\vec{r}_{n}$ in the base lattice $\mathbb{A}_{2}^{\ast}$ has $%
\left( 3+3\right) $ first nearest neighbors and \emph{6} second nearest
ones; they are as follows:

\begin{itemize}
\item \emph{6 first nearests}\newline
Up to a scaling factor L, the first nearest neighbors are given by%
\begin{equation}
\begin{tabular}{ll}
$\vec{\lambda}_{1}$ & $=\vec{\omega}_{1}$ \\ 
$\vec{\lambda}_{2}$ & $=\vec{\omega}_{2}-\vec{\omega}_{1}$ \\ 
$\vec{\lambda}_{3}$ & $=-\vec{\omega}_{2}$%
\end{tabular}
\label{LZ}
\end{equation}
and%
\begin{equation}
\begin{tabular}{ll}
$\vec{\zeta}_{1}$ & $=\vec{\omega}_{2}$ \\ 
$\vec{\zeta}_{2}$ & $=\vec{\omega}_{1}-\vec{\omega}_{2}$ \\ 
$\vec{\zeta}_{3}$ & $=-\vec{\omega}_{1}$%
\end{tabular}
\label{ZL}
\end{equation}
obeying the identities%
\begin{equation}
\begin{tabular}{lll}
$\vec{\lambda}_{1}+\vec{\lambda}_{2}+\vec{\lambda}_{3}$ & $=$ & $\vec{0}$ \\ 
$\vec{\zeta}_{1}+\vec{\zeta}_{2}+\vec{\zeta}_{3}$ & $=$ & $\vec{0}$ \\ 
&  & 
\end{tabular}%
\end{equation}
As an illustration on fig \ref{1}, choose as a node the green site at $\vec {%
r}_{n}$, its first nearest neighbors (\ref{LZ}) are given by the red sites;
and those of associated with (\ref{ZL}) are given by the blue ones.

\item \emph{6 second nearest}\newline
The 6 second nearest neighbors are given by%
\begin{equation}
\begin{tabular}{lll}
$\vec{\omega}_{\left( 2,-1\right) }$ & $=$ & $+2\vec{\omega}_{1}-\vec {\omega%
}_{2}$ \\ 
$\vec{\omega}_{\left( 1,1\right) }$ & $=$ & $+\vec{\omega}_{1}+\vec{\omega }%
_{2}$ \\ 
$\vec{\omega}_{\left( -1,2\right) }$ & $=$ & $-\vec{\omega}_{1}+2\vec {\omega%
}_{2}$ \\ 
$\vec{\omega}_{\left( -2,1\right) }$ & $=$ & $-2\vec{\omega}_{1}+\vec {\omega%
}_{2}$ \\ 
$\vec{\omega}_{\left( -1,-1\right) }$ & $=$ & $-\vec{\omega}_{1}-\vec {\omega%
}_{2}$ \\ 
$\vec{\omega}_{\left( 1,-2\right) }$ & $=$ & $+\vec{\omega}_{1}-2\vec {\omega%
}_{2}$ \\ 
&  & 
\end{tabular}
\label{WL}
\end{equation}
and are nothing but the six roots of the $SU\left( 3\right) $ namely%
\begin{equation}
\begin{tabular}{lllll}
$\pm \vec{\alpha}_{1}$ & $,$ & $\pm \vec{\alpha}_{2}$ & , & $\pm \left( \vec{%
\alpha}_{1}+\vec{\alpha}_{2}\right) $ \\ 
&  &  &  & 
\end{tabular}%
\end{equation}
As an illustration on fig \ref{1}, each green site $\vec{r}_{n}$ has 6
nearest neighbors located at%
\begin{equation}
\begin{tabular}{lllll}
$\vec{r}_{n}\pm L\vec{\alpha}_{1}$ & $,$ & $\vec{r}_{n}\pm L\vec{\alpha}_{2}$
& , & $\vec{r}_{n}\pm L\left( \vec{\alpha}_{1}+\vec{\alpha}_{2}\right) $ \\ 
&  &  &  & 
\end{tabular}%
\end{equation}
Observe also that a generic red site located at%
\begin{equation}
\begin{tabular}{lll}
$\vec{r}_{n}+L\vec{\lambda}_{i}$ & $,$ & $i=1,2,3$%
\end{tabular}%
\end{equation}
has 6 first nearest neighbors given by the red ones located 
\begin{equation}
\begin{tabular}{lllll}
&  &  &  &  \\ 
$\vec{r}_{n}+L\vec{\lambda}_{i}\pm L\vec{\alpha}_{1}$ & $,$ & $\vec{r}_{n}+L%
\vec{\lambda}_{i}\pm L\vec{\alpha}_{2}$ & , & $\vec{r}_{n}+L\vec {\lambda}%
_{i}\pm L\left( \vec{\alpha}_{1}+\vec{\alpha}_{2}\right) $ \\ 
&  &  &  & 
\end{tabular}%
\end{equation}
This result may be also checked by computing the relative vector $\vec{V}%
_{ij}$ between two nearest sites located at $\vec{r}_{n}+L\vec{\lambda}_{i}$
and $\vec{r}_{n}+L\vec{\lambda}_{j}$. We have 
\begin{equation}
\begin{tabular}{lll}
$\vec{V}_{ij}$ & $=$ & $\left( \vec{r}_{n}+L\vec{\lambda}_{i}\right) -\left( 
\vec{r}_{n}+L\vec{\lambda}_{j}\right) $ \\ 
& $=$ & $L\left( \vec{\lambda}_{i}-\vec{\lambda}_{j}\right) $ \\ 
&  & 
\end{tabular}%
\end{equation}
which, by using eqs(\ref{ZL}-\ref{WL}), the 6 vectors $\vec{\lambda}_{i}-%
\vec{\lambda}_{j}$ are precisely the 6 roots of $SU\left( 3\right) $.
\end{itemize}

\subsection{Building the lattice\emph{\ }$\mathcal{L}_{3D}^{su_{3}\times
u_{1}}$}

The lattice $\mathcal{L}_{3D}^{su_{3}\times u_{1}}$ is a 3-dimensional
crystal that may be thought of as given by a twisting of weight lattice $%
\mathbb{A}_{3}^{\ast}$ of the Lie algebra of the symmetry%
\begin{equation}
SO\left( 6\right) \simeq SU\left( 4\right)
\end{equation}
To build this 3D crystal, we begin by describing the lattice $\mathbb{A}%
_{3}^{\ast}$; after what we turn to construct $\mathcal{L}%
_{3D}^{su_{3}\times u_{1}}$.

\subsubsection{\emph{Construction of} \emph{the lattice }$\mathbb{A}%
_{3}^{\ast}$}

\begin{figure}[ptbh]
\begin{center}
\hspace{0cm} \includegraphics[width=4cm]{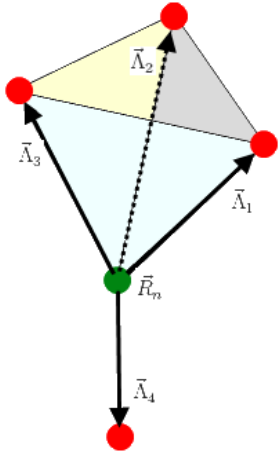}
\end{center}
\par
\vspace{0 cm}
\caption{Each site $\vec{R}_{n}$ in the lattice $\mathbb{A}_{3}^{\ast}$ has $%
8=4+4$ first nearest neighbors associated with the weight vectors $\vec{%
\Lambda}_{i}$ and $\vec{\Lambda}_{i}^{\prime}$. In the figure, 4 sites of
the 8 ones are represented and are related to the $\Omega_{i}$ generators as
in eqs(\protect \ref{wv}).}
\label{4}
\end{figure}

The 3- dimensional lattice $\mathbb{A}_{3}^{\ast}$ is the dual of the root
lattice of $SU\left( 4\right) $; it is generated by the 3 \emph{fundamental
weight} vectors of $SU\left( 4\right) $%
\begin{equation}
\begin{tabular}{lllll}
$\vec{\Omega}_{1}$ & $,$ & $\vec{\Omega}_{2}$ & $,$ & $\vec{\Omega}_{3}$%
\end{tabular}%
\end{equation}
Using the lattice spacing parameter $L_{su_{4}}$ of the crystal $\mathbb{A}%
_{3}^{\ast}$, we can express the positions $\vec{R}_{n}$ of sites in this
lattice as follows%
\begin{equation}
\vec{R}_{n}=L_{su_{4}}\text{ }\sqrt{\frac{4}{3}}\text{ }\vec{\Omega}_{\left(
n_{1},n_{2},n_{3}\right) }  \label{RR}
\end{equation}
with 
\begin{equation}
\text{ }\vec{\Omega}_{\left( n_{1},n_{2},n_{3}\right) }=n_{1}\vec{\Omega }%
_{1}+n_{2}\vec{\Omega}_{2}+n_{3}\vec{\Omega}_{3}
\end{equation}
where $n_{i}$ are arbitrary integers. \newline
The shape of $\mathbb{A}_{3}^{\ast}$ is encoded in the intersection matrix%
\begin{equation*}
\mathcal{J}_{ij}^{su_{4}}=\vec{\Omega}_{i}.\vec{\Omega}_{j}
\end{equation*}
given by%
\begin{equation}
\mathcal{J}_{ij}^{su_{4}}=\left( 
\begin{array}{ccc}
\frac{3}{4} & \frac{1}{2} & \frac{1}{4} \\ 
\frac{1}{2} & 1 & \frac{1}{2} \\ 
\frac{1}{4} & \frac{1}{2} & \frac{3}{4}%
\end{array}
\right)
\end{equation}
The fundamental weight vectors $\vec{\Omega}_{i}$ are the dual of the 3
simple roots $\vec{a}_{1},$ $\vec{a}_{2},$ $\vec{a}_{3}$ of $SU\left(
4\right) $ obeying%
\begin{equation}
\vec{\Omega}_{i}.\vec{a}_{j}=\delta_{ij}
\end{equation}
The vectors $\vec{\Omega}_{i}$ are also the highest weight vectors of the
complex $\mathbf{4}$, the adjoint conjugate $\mathbf{\bar{4}}$ and the real $%
6$ dimensional representations of $SU\left( 4\right) $.%
\begin{equation}
\begin{tabular}{lllll}
representation & : & $\mathbf{4}$ & $6$ & $\mathbf{\bar{4}}$ \\ 
highest weights & : & $\vec{\Omega}_{1}$ & $\vec{\Omega}_{2}$ & $\vec{\Omega 
}_{3}$%
\end{tabular}%
\end{equation}
The set of the weight vectors $\vec{\Lambda}_{i}$ defining the states of the
complex 4- dimensional highest weight representations $\mathbf{4}$ and its
conjugate $\mathbf{\bar{4}}$ are as follows 
\begin{equation}
\begin{tabular}{lll|lll}
\multicolumn{3}{l|}{weight vectors of $\mathbf{4}$ \  \  \  \  \  \ } & 
\multicolumn{3}{|l}{weight vectors of $\mathbf{\bar{4}}$} \\ \hline \hline
&  &  &  &  &  \\ 
$\vec{\Lambda}_{1}$ & $=$ & $\vec{\Omega}_{1}$ & $\vec{\Lambda}_{1}^{\prime} 
$ & $=$ & $\vec{\Omega}_{3}$ \\ 
$\vec{\Lambda}_{2}$ & $=$ & $\vec{\Omega}_{2}-\vec{\Omega}_{1}$ & $\vec{%
\Lambda}_{2}^{\prime}$ & $=$ & $\vec{\Omega}_{2}-\vec{\Omega}_{3}$ \\ 
$\vec{\Lambda}_{3}$ & $=$ & $\vec{\Omega}_{3}-\vec{\Omega}_{2}$ & $\vec{%
\Lambda}_{3}^{\prime}$ & $=$ & $\vec{\Omega}_{1}-\vec{\Omega}_{2}$ \\ 
$\vec{\Lambda}_{4}$ & $=$ & $-\vec{\Omega}_{3}$ & $\vec{\Lambda}%
_{4}^{\prime} $ & $=$ & $-\vec{\Omega}_{1}$ \\ 
&  &  &  &  &  \\ \hline
\end{tabular}
\label{wv}
\end{equation}
These quartets obey the following traceless properties%
\begin{align}
& 
\begin{tabular}{lll}
$\vec{\Lambda}_{1}+\vec{\Lambda}_{2}+\vec{\Lambda}_{3}+\vec{\Lambda}_{4}$ & $%
=$ & $\vec{0}$ \\ 
$\vec{\Lambda}_{1}^{\prime}+\vec{\Lambda}_{2}^{\prime}+\vec{\Lambda}%
_{3}^{\prime}+\vec{\Lambda}_{4}^{\prime}$ & $=$ & $\vec{0}$%
\end{tabular}
\\
&  \notag
\end{align}
indicating that the sum of the relative positions of the first nearest
neighbors should be equal to zero as schematized on fig \ref{4}. 
Similarly, the set of weight vectors $\vec{\Lambda}_{i}^{\prime \prime}$ of
the real 6- dimensional representation of $SU\left( 4\right) \simeq SO\left(
6\right) $ are given by%
\begin{align}
& 
\begin{tabular}{lll}
$\vec{\Lambda}_{1}^{\prime \prime}$ & $=$ & $\vec{\Omega}_{2}$ \\ 
$\vec{\Lambda}_{2}^{\prime \prime}$ & $=$ & $\vec{\Omega}_{1}+\vec{\Omega}%
_{3}-\vec{\Omega}_{2}$ \\ 
$\vec{\Lambda}_{3}^{\prime \prime}$ & $=$ & $\vec{\Omega}_{3}-\vec{\Omega}%
_{1}$ \\ 
$\vec{\Lambda}_{4}^{\prime \prime}$ & $=$ & $\vec{\Omega}_{1}-\vec{\Omega}%
_{3}$ \\ 
$\vec{\Lambda}_{5}^{\prime \prime}$ & $=$ & $\vec{\Omega}_{2}-\vec{\Omega}%
_{1}-\vec{\Omega}_{3}$ \\ 
$\vec{\Lambda}_{6}^{\prime \prime}$ & $=$ & $-\vec{\Omega}_{2}$%
\end{tabular}
\\
&
\end{align}
satisfying also a traceless property $\vec{\Lambda}_{1}^{\prime \prime }+...+%
\vec{\Lambda}_{6}^{\prime \prime}=0$ and having an interpretation in terms
of second nearest neighbors.

\subsubsection{\emph{The lattice }$\mathcal{L}_{3D}^{su_{3}\times u_{1}}$ 
\emph{as a twist of} $\mathbb{A}_{3}^{\ast}$}

Under the breaking of $SU\left( 4\right) $ down to $U\left( 1\right) \times
SU\left( 3\right) $, the representations $\mathbf{4}$ and $\mathbf{\bar{4}}$
break down to \textrm{\cite{1K}}%
\begin{equation}
\begin{tabular}{lll}
$\mathbf{4}$ & $\rightarrow$ & $\mathbf{3}_{+q}$ $\mathbf{\oplus}$ $\mathbf{1%
}_{-3q}$ \\ 
$\mathbf{\bar{4}}$ & $\rightarrow$ & $\mathbf{\bar{3}}_{-q}$ $\mathbf{\oplus}
$ $\mathbf{\bar{1}}_{+3q}$%
\end{tabular}%
\end{equation}
and the weight vectors (\ref{wv}) become%
\begin{equation}
\begin{tabular}{lll|lll}
\multicolumn{3}{l|}{weight vectors of $\mathbf{3}_{+q}\mathbf{\oplus1}_{-3q} 
$} & \multicolumn{3}{|l}{weight vectors of $\mathbf{\bar{3}}_{-q}\mathbf{%
\oplus1}_{+3q}$} \\ \hline
&  &  &  &  &  \\ 
$\vec{\Lambda}_{1}$ & $=$ & $\left( \vec{\lambda}_{1},+q\right) $ & $\vec{%
\Lambda}_{0}^{\prime}$ & $=$ & $\left( \vec{0},+3q\right) $ \\ 
$\vec{\Lambda}_{2}$ & $=$ & $\left( \vec{\lambda}_{2},+q\right) $ & $\vec{%
\Lambda}_{1}^{\prime}$ & $=$ & $\left( \vec{\zeta}_{1},-q\right) $ \\ 
$\vec{\Lambda}_{3}$ & $=$ & $\left( \vec{\lambda}_{3},+q\right) $ & $\vec{%
\Lambda}_{2}^{\prime}$ & $=$ & $\left( \vec{\zeta}_{2},-q\right) $ \\ 
$\vec{\Lambda}_{0}$ & $=$ & $\left( 0,0,-3q\right) $ & $\vec{\Lambda}%
_{3}^{\prime}$ & $=$ & $\left( \vec{\zeta}_{3},-q\right) $ \\ 
&  &  &  &  &  \\ \hline
\end{tabular}%
\end{equation}
where where $\vec{\lambda}_{i},$ $\vec{\zeta}_{i}$ are as in eqs(\ref{LZ}). 
\newline
Similarly, we have for the 6-dimensional representation the following
decomposition%
\begin{equation}
\begin{tabular}{lll}
$\mathbf{6}$ & $\rightarrow$ & $\mathbf{3}_{-2q}$ $\mathbf{\oplus}$ $\mathbf{%
\bar{3}}_{+2q}$%
\end{tabular}%
\end{equation}
with weight vectors $\vec{\Lambda}_{i}^{\prime \prime}$ as follows 
\begin{equation}
\begin{tabular}{lll}
$\vec{\Lambda}_{1}^{\prime \prime}$ & $=$ & $\left( \vec{\zeta}%
_{1},+2q\right) $ \\ 
$\vec{\Lambda}_{2}^{\prime \prime}$ & $=$ & $\left( \vec{\zeta}%
_{2},+2q\right) $ \\ 
$\vec{\Lambda}_{3}^{\prime \prime}$ & $=$ & $\left( \vec{\zeta}%
_{3},+2q\right) $ \\ 
$\vec{\Lambda}_{4}^{\prime \prime}$ & $=$ & $\left( \vec{\lambda}%
_{1},-2q\right) $ \\ 
$\vec{\Lambda}_{5}^{\prime \prime}$ & $=$ & $\left( \vec{\lambda}%
_{2},-2q\right) $ \\ 
$\vec{\Lambda}_{6}^{\prime \prime}$ & $=$ & $\left( \vec{\lambda}%
_{3},-2q\right) $ \\ 
&  & 
\end{tabular}%
\end{equation}
From these decompositions, we deduce the expressions of the fundamental
weight vectors of $U\left( 1\right) \times SU\left( 3\right) $ that we
denote as $\vec{\Gamma}_{1},$ $\vec{\Gamma}_{2},$ $\vec{\Gamma}_{3}$. These
vectors are given by%
\begin{equation}
\begin{tabular}{lll}
$\vec{\Gamma}_{1}$ & $=$ & $\left( \vec{\omega}_{1},q\right) $ \\ 
$\vec{\Gamma}_{2}$ & $=$ & $\left( \vec{\omega}_{2},2q\right) $ \\ 
$\vec{\Gamma}_{0}$ & $=$ & $\left( 0,0,3q\right) $%
\end{tabular}%
\end{equation}
with unit charge $q$ to be fixed later. Computing the intersection matrix of
these weights 
\begin{equation}
\mathcal{J}_{ij}^{su_{3}\times u_{1}}=\vec{\Gamma}_{i}.\vec{\Gamma}_{j}
\end{equation}
we find 
\begin{equation}
\mathcal{J}_{ij}^{su_{3}\times u_{1}}=\left( 
\begin{array}{ccc}
\frac{2}{3}+q^{2} & \frac{1}{3}+2q^{2} & 3q^{2} \\ 
\frac{1}{3}+2q^{2} & \frac{2}{3}+4q^{2} & 6q^{2} \\ 
3q^{2} & 6q^{2} & 9q^{2}%
\end{array}
\right)
\end{equation}
with 
\begin{equation}
\det \mathcal{J}_{ij}^{su_{3}\times u_{1}}=3q^{2}
\end{equation}
and inverse given by%
\begin{equation}
\mathcal{K}_{ij}^{su_{3}\times u_{1}}=\left( 
\begin{array}{ccc}
2 & -1 & 0 \\ 
-1 & 2 & -1 \\ 
0 & -1 & \frac{6q^{2}+1}{9q^{2}}%
\end{array}
\right)
\end{equation}
By requiring the following condition on the self intersection of $\vec{%
\Gamma }_{3}$ 
\begin{equation}
\frac{6q^{2}+1}{9q^{2}}=1
\end{equation}
it results 
\begin{equation}
q^{2}=\frac{1}{3}  \label{3Q}
\end{equation}
Putting this value back into the above intersection matrices, we end with 
\begin{equation}
\mathcal{J}_{ij}^{su_{3}\times u_{1}}=\left( 
\begin{array}{ccc}
1 & 1 & 1 \\ 
1 & 2 & 2 \\ 
1 & 2 & 3%
\end{array}
\right)
\end{equation}
and 
\begin{equation}
\mathcal{K}_{ij}^{su_{3}\times u_{1}}=\left( 
\begin{array}{ccc}
2 & -1 & 0 \\ 
-1 & 2 & -1 \\ 
0 & -1 & 1%
\end{array}
\right)
\end{equation}
From these intersection matrices, we determine the following relations%
\begin{equation}
\begin{tabular}{lll}
$\vec{a}_{1}$ & $=$ & $2\vec{\Gamma}_{1}-\vec{\Gamma}_{2}$ \\ 
$\vec{a}_{2}$ & $=$ & $2\vec{\Gamma}_{2}-\vec{\Gamma}_{1}-\vec{\Gamma}_{0}$
\\ 
$\vec{a}_{0}$ & $=$ & $\vec{\Gamma}_{0}-\vec{\Gamma}_{2}$%
\end{tabular}%
\end{equation}
and 
\begin{equation}
\begin{tabular}{lll}
$\vec{\Gamma}_{1}$ & $=$ & $\vec{a}_{1}+\vec{a}_{2}+\vec{a}_{0}$ \\ 
$\vec{\Gamma}_{2}$ & $=$ & $\vec{a}_{1}+2\vec{a}_{2}+2\vec{a}_{0}$ \\ 
$\vec{\Gamma}_{0}$ & $=$ & $\vec{a}_{1}+2\vec{a}_{2}+3\vec{a}_{0}$%
\end{tabular}%
\end{equation}
Sites $\vec{R}_{n}$ in $\mathcal{L}_{3D}^{su_{3}\times u_{1}}$ are therefore
expanded like%
\begin{equation}
\vec{R}_{n}=L\text{ }\vec{\Gamma}_{n},\qquad L=L_{su_{4}}\sqrt{\frac{4}{3}}
\end{equation}
with 
\begin{equation}
\vec{\Gamma}_{n}=n_{1}\vec{\Gamma}_{1}+n_{2}\vec{\Gamma}_{2}+n_{0}\vec{%
\Gamma }_{0}
\end{equation}
and%
\begin{equation}
\begin{tabular}{lllll}
$\vec{\Gamma}_{1}=\left( 
\begin{array}{c}
\vec{\omega}_{1} \\ 
q%
\end{array}
\right) $ & $,$ & $\vec{\Gamma}_{2}=\left( 
\begin{array}{c}
\vec{\omega}_{2} \\ 
2q%
\end{array}
\right) $ & , & $\vec{\Gamma}_{0}=\left( 
\begin{array}{c}
\vec{0} \\ 
3q%
\end{array}
\right) $%
\end{tabular}
\label{ga}
\end{equation}
obeying the duality relation 
\begin{equation}
\vec{\Gamma}_{i}.\vec{a}_{j}=\delta_{ij}
\end{equation}
We also have%
\begin{equation}
\begin{tabular}{lllll}
$\vec{a}_{1}=\left( 
\begin{array}{c}
2\vec{\omega}_{1}-\vec{\omega}_{2} \\ 
0%
\end{array}
\right) $ & $,$ & $\vec{a}_{2}=\left( 
\begin{array}{c}
2\vec{\omega}_{2}-\vec{\omega}_{1} \\ 
0%
\end{array}
\right) $ & , & $\vec{a}_{0}=\left( 
\begin{array}{c}
-\vec{\omega}_{2} \\ 
q%
\end{array}
\right) $%
\end{tabular}%
\end{equation}

\subsection{Lattice interpretation of BRST symmetry}

Here we use the weight vectors of the $SU\left( 3\right) \times U\left(
1\right) $ representations carried by the fermionic and the bosonic fields
of the spectrum of twisted supersymmetric YM theory on $\mathcal{L}%
_{3D}^{su_{3}\times u_{1}}$ to give a crystal interpretation of the BRST
symmetry. To that purpose, we first derive the closed nearest neighbors in $%
\mathcal{L}_{3D}^{su_{3}\times u_{1}}$; then, we give the lattice
realization of BRST symmetry.

\subsubsection{\emph{Closest neighbors in }$\mathcal{L}_{3D}^{su_{3}\times
u_{1}}$}

Given a generic site $\vec{R}_{n}$ in $\mathcal{L}_{3D}^{su_{3}\times u_{1}}$%
, its closest neighbors are as follows:

\  \ 

\emph{a) first nearest neighbors and the links }$\psi_{n}^{I\left( +q\right)
}$: \newline
Each site in $\mathcal{L}_{3D}^{su_{3}\times u_{1}}$ has 6 first nearest
neighbors that split into $3+3$ ones respectively located at%
\begin{equation}
\vec{R}_{n}+L\vec{\Lambda}_{I}=\vec{R}_{n}+\vec{V}_{I}^{\left( +q\right) }
\end{equation}
and%
\begin{equation}
\begin{tabular}{lll}
$\vec{R}_{n}+L\vec{\Lambda}_{I}^{\prime}$ & $=$ & $\vec{R}_{n}-\vec{V}%
_{I}^{\left( +q\right) }$%
\end{tabular}%
\end{equation}
with%
\begin{equation}
\begin{tabular}{lll}
$\vec{V}_{1}^{\left( +q\right) }$ & $=$ & $L\vec{\Gamma}_{1}$ \\ 
$\vec{V}_{2}^{\left( +q\right) }$ & $=$ & $L\left( \vec{\Gamma}_{2}-\vec{%
\Gamma}_{1}\right) $ \\ 
$\vec{V}_{3}^{\left( +q\right) }$ & $=$ & $L\left( \vec{\Gamma}_{0}-\vec{%
\Gamma}_{2}\right) $%
\end{tabular}%
\end{equation}
satisfying%
\begin{equation}
\begin{tabular}{lll}
$\vec{V}_{1}^{\left( +q\right) }+\vec{V}_{2}^{\left( +q\right) }+\vec {V}%
_{3}^{\left( +q\right) }$ & $=$ & $L\vec{\Gamma}_{0}$ \\ 
&  & 
\end{tabular}%
\end{equation}
Using (\ref{ga}), these relations read also like%
\begin{equation}
\begin{tabular}{lll}
$\vec{V}_{1}^{\left( +q\right) }$ & $=$ & $L\left( \vec{\lambda}%
_{1},q\right) $ \\ 
$\vec{V}_{2}^{\left( +q\right) }$ & $=$ & $L\left( \vec{\lambda}%
_{2},q\right) $ \\ 
$\vec{V}_{3}^{\left( +q\right) }$ & $=$ & $L\left( \vec{\lambda}%
_{3},q\right) $%
\end{tabular}%
\end{equation}
with%
\begin{equation}
\begin{tabular}{lll}
$\vec{V}_{1}^{\left( +q\right) }+\vec{V}_{2}^{\left( +q\right) }+\vec {V}%
_{3}^{\left( +q\right) }$ & $=$ & $\left( 0,0,3Lq\right) $ \\ 
&  & 
\end{tabular}%
\end{equation}
and length%
\begin{equation}
\begin{tabular}{lllll}
$\left( \vec{V}_{I}^{\left( +q\right) }\right) ^{2}$ & $=$ & $\left( -\vec{V}%
_{I}^{\left( +q\right) }\right) ^{2}$ &  &  \\ 
& $=$ & $\left( \frac{2}{3}+q^{2}\right) L^{2}$ & $=$ & $L^{2}$%
\end{tabular}%
\end{equation}
since $3q^{2}=1$. \newline
Therefore the oriented links 
\begin{equation}
\begin{tabular}{lll}
$\overrightarrow{P_{R_{n}}P}_{R_{n}+V_{I}^{\left( +q\right) }}$ & $\sim$ & $%
V_{I}^{\left( +q\right) }$ \\ 
&  & 
\end{tabular}%
\end{equation}
are associated with the fermionic lattice variables 
\begin{equation}
\psi_{n}^{I\left( +q\right) }=\psi^{I\left( +q\right) }\left( R_{n}\right)
\end{equation}

\  \  \  \  \ 

\emph{b) second nearest neighbors and gauge field variables}\newline
Each site $R_{n}$ in $\mathcal{L}_{3D}^{su_{3}\times u_{1}}$ has 6 second
nearest neighbors that also split into $3+3$ ones respectively located at 
\begin{equation}
\vec{R}_{n}+\vec{V}_{I}^{\left( -2q\right) }
\end{equation}
and at%
\begin{equation}
\begin{tabular}{lll}
$\vec{R}_{n}+\vec{V}_{I}^{\left( +2q\right) }$ & $=$ & $\vec{R}_{n}-\vec {V}%
_{I}^{\left( -2q\right) }$ \\ 
&  &  \\ 
& $\sim$ & $-\vec{V}_{I}^{\left( -2q\right) }$%
\end{tabular}%
\end{equation}
Upon the normalization $L=1$, we have 
\begin{equation}
\begin{tabular}{lll}
$\vec{V}_{1}^{\left( +2q\right) }$ & $=$ & $\vec{\Gamma}_{2}$ \\ 
$\vec{V}_{2}^{\left( +2q\right) }$ & $=$ & $\vec{\Gamma}_{1}+\vec{\Gamma }%
_{0}-\vec{\Gamma}_{2}$ \\ 
$\vec{V}_{3}^{\left( +2q\right) }$ & $=$ & $\vec{\Gamma}_{0}-\vec{\Gamma }%
_{1} $ \\ 
$\vec{V}_{1}^{\left( -2q\right) }$ & $=$ & $\vec{\Gamma}_{1}-\vec{\Gamma }%
_{0} $ \\ 
$\vec{V}_{2}^{\left( -2q\right) }$ & $=$ & $\vec{\Gamma}_{2}-\vec{\Gamma }%
_{1}-\vec{\Gamma}_{0}$ \\ 
$\vec{V}_{3}^{\left( -2q\right) }$ & $=$ & $-\vec{\Gamma}_{2}$%
\end{tabular}%
\end{equation}
or equivalently%
\begin{equation}
\begin{tabular}{lll}
$\vec{V}_{1}^{\left( +2q\right) }$ & $=$ & $\left( -\vec{\lambda}%
_{1},2q\right) $ \\ 
$\vec{V}_{2}^{\left( +2q\right) }$ & $=$ & $\left( -\vec{\lambda}%
_{2},2q\right) $ \\ 
$\vec{V}_{3}^{\left( +2q\right) }$ & $=$ & $\left( -\vec{\lambda}%
_{3},2q\right) $ \\ 
$\vec{V}_{1}^{\left( -2q\right) }$ & $=$ & $\left( \vec{\lambda}%
_{1},-2q\right) $ \\ 
$\vec{V}_{2}^{\left( -2q\right) }$ & $=$ & $\left( \vec{\lambda}%
_{2},-2q\right) $ \\ 
$\vec{V}_{3}^{\left( -2q\right) }$ & $=$ & $\left( \vec{\lambda}%
_{3},-2q\right) $%
\end{tabular}
\label{IV}
\end{equation}
with length%
\begin{equation}
\begin{tabular}{lll}
$\left( \vec{V}_{a}^{\left( +2q\right) }\right) ^{2}$ & $=$ & $\left( \frac{2%
}{3}+4q^{2}\right) L^{2}=2L^{2}$ \\ 
$\left( \vec{V}_{a}^{\left( -2q\right) }\right) $ & $=$ & $\left( \frac{2}{3}%
+4q^{2}\right) L^{2}=2L^{2}$%
\end{tabular}%
\end{equation}
The 3 oriented links 
\begin{equation}
\begin{tabular}{lll}
$\overrightarrow{P_{R_{n}}P}_{R_{n}+V_{I}^{\left( +2q\right) }}$ & $\sim$ & $%
-\vec{V}_{I}^{\left( -2q\right) }$ \\ 
&  & 
\end{tabular}%
\end{equation}
are associated with the gauge field variables 
\begin{equation*}
U_{n,I}^{\left( +2q\right) }=U_{I}^{\left( +2q\right) }\left( R_{n}\right)
\end{equation*}
while the opposite ones 
\begin{equation}
\begin{tabular}{lll}
$\overrightarrow{P_{R_{n}}P}_{R_{n}-V_{I}^{\left( +2q\right) }}$ & $\equiv$
& $P_{R_{n}+V_{I}^{\left( -2q\right) }}P_{R_{n}}$ \\ 
&  &  \\ 
& $\sim$ & $V_{I}^{\left( -2q\right) }$%
\end{tabular}%
\end{equation}
with the complex conjugate fields 
\begin{equation}
U_{n}^{I\left( -2q\right) }=U^{I\left( -2q\right) }\left( R_{n}\right)
\end{equation}

\emph{c) third nearest neighbors}\newline
The site $\vec{R}_{n}$ of $\mathcal{L}_{3D}^{su_{3}\times u_{1}}$ has 2
third nearest neighbors that also split into $1+1$ respectively located at 
\begin{equation}
\begin{tabular}{lll}
$\vec{R}_{n}+L\vec{\Gamma}_{0}$ & $=$ & $\vec{R}_{n}+\vec{V}^{\left(
+3q\right) }$ \\ 
$\vec{R}_{n}-L\vec{\Gamma}_{0}$ & $=$ & $\vec{R}_{n}+\vec{V}^{\left(
-3q\right) }$%
\end{tabular}%
\end{equation}
with%
\begin{equation}
\vec{\Gamma}_{0}=\left( 
\begin{array}{c}
\vec{0} \\ 
3q%
\end{array}
\right)
\end{equation}
and%
\begin{equation}
\vec{\Gamma}_{0}.\vec{\Gamma}_{0}=3
\end{equation}
The oriented links 
\begin{equation}
\begin{tabular}{lll}
$\overrightarrow{P_{R_{n}}P}_{R_{n}+V^{\left( -3q\right) }}$ & $\sim$ & $%
V^{\left( -3q\right) }$ \\ 
&  & 
\end{tabular}%
\end{equation}
are associated with the fermionic singlets $\psi_{n}^{\left( -3q\right) }$
and the Grassman variable 
\begin{equation}
\begin{tabular}{lllll}
&  &  &  &  \\ 
$\psi_{n}^{\left( -3q\right) }$ & $=$ & $\psi^{\left( -3q\right) }\left(
R_{n}\right) $ & , & $\theta^{\left( -3q\right) }$ \\ 
&  &  &  & 
\end{tabular}%
\end{equation}
while the opposite ones 
\begin{equation}
\begin{tabular}{lll}
$\overrightarrow{P_{R_{n}}P}_{R_{n}-V^{\left( -3q\right) }}$ & $\equiv$ & $%
P_{R_{n}+V^{\left( +3q\right) }}P_{R_{n}}$ \\ 
&  &  \\ 
& $\sim$ & $-V^{\left( -3q\right) }$%
\end{tabular}%
\end{equation}
with objects carrying $+3$ unit charges of U$\left( 1\right) $ like the
discrete analogue of $\nabla_{a}^{\left( +2q\right) }\psi^{a\left( +\right)
} $.

\subsubsection{\emph{BRST symmetry on lattice}}

Being a scalar objet under $SU\left( 3\right) $ but having non trivial
charges of U$\left( 1\right) $, the scalar supersymmetric $Q^{{\small (+3q)}%
} $ may be interpreted as a link operator living on the direction 
\begin{equation}
\vec{\Gamma}=\left( 0,0,3q\right)
\end{equation}
Therefore supersymmetric transformations generated by $Q^{{\small (+3q)}}$
can be interpreted as particular shifts on the lattice $\mathcal{L}%
_{3D}^{su_{3}\times u_{1}}$; more precisely shifts in the $\mathcal{L}%
_{1D}^{u_{1}}$ sublattice of the fibration 
\begin{equation*}
\begin{tabular}{lll}
$\mathcal{L}_{1D}^{u_{1}}$ & $\rightarrow$ & $\  \  \mathcal{L}%
_{3D}^{su_{3}\times u_{1}}$ \\ 
&  & $\  \  \  \  \downarrow$ \\ 
&  & $\mathcal{\  \  \ }\mathbb{A}_{2}^{\ast}$%
\end{tabular}%
\end{equation*}
For example, the supersymmetric transformation%
\begin{equation}
\begin{tabular}{lll}
$Q^{{\small (+3q)}}U_{n}^{I\left( -2q\right) }$ & $=$ & $\psi_{n}^{I\left(
+q\right) }$ \\ 
&  & 
\end{tabular}%
\end{equation}
corresponds to shifting $\vec{V}_{I}^{\left( -2q\right) }$ as follows%
\begin{equation}
\begin{tabular}{lll}
$\vec{V}_{I}^{\left( -2q\right) }+\vec{V}_{0}^{\left( +3q\right) }$ & $=$ & $%
\vec{V}_{I}^{\left( +q\right) }$%
\end{tabular}%
\end{equation}
In general, the operator $Q^{{\small (+3q)}}$ acts on the various lattice
variables like%
\begin{equation}
\begin{tabular}{lll}
$Q^{{\small (+3q)}}U_{n}^{I\left( -2q\right) }$ & $=$ & $\psi_{n}^{I\left(
+q\right) }$ \\ 
$Q^{{\small (+3q)}}\psi_{n}^{I\left( +q\right) }$ & $=$ & $0$ \\ 
&  &  \\ 
$Q^{{\small (+3q)}}U_{I,n}^{\left( +2q\right) }$ & $=$ & $0$ \\ 
$Q^{{\small (+3q)}}W_{IJ,n}^{\left( +4q\right) }$ & $=$ & $0$ \\ 
&  &  \\ 
$Q^{{\small (+3q)}}\psi_{n}^{\left( -3q\right) }$ & $=$ & $F_{n}^{\left(
0\right) }$ \\ 
$Q^{{\small (+3q)}}F_{n}^{\left( 0\right) }$ & $=$ & $0$ \\ 
&  &  \\ 
$Q^{{\small (+3q)}}W_{n}^{IJ\left( -4q\right) }$ & $=$ & $\Upsilon
_{n}^{IJ(-q)}$ \\ 
$Q^{{\small (+3q)}}\Upsilon_{n}^{IJ(-q)}$ & $=$ & $0$ \\ 
$Q^{{\small (+3q)}}W_{n}^{\left( 0\right) }$ & $=$ & $\Upsilon_{n}^{(+3q)}$
\\ 
$Q^{{\small (+3q)}}\Upsilon_{n}^{(+3q)}$ & $=$ & $0$ \\ 
&  & 
\end{tabular}%
\end{equation}
with%
\begin{equation}
\begin{tabular}{lll}
&  &  \\ 
$\Upsilon_{n}^{IJ\left( -q\right) }$ & $=$ & $U^{I\left( -2q\right) }\left(
R_{n}\right) \psi^{J\left( +q\right) }(R_{n}+\vec{V}_{I}^{\left( -2q\right)
})-$ \\ 
&  & $\psi^{J\left( +q\right) }\left( R_{n}\right) U^{I\left( -2q\right)
}(R_{n}+\vec{V}_{J}^{\left( -2q\right) })$ \\ 
&  &  \\ 
$\Upsilon_{n}^{(0)}$ & $=$ & $\psi^{I\left( +q\right) }\left( R_{n}\right)
U_{I}^{\left( +2q\right) }\left( R_{n}\right) -$ \\ 
&  & $U_{I}^{\left( +2q\right) }(R_{n}-\vec{V}_{I}^{\left( -2q\right)
})\psi^{I\left( +q\right) }(R_{n}-\vec{V}_{I}^{\left( -2q\right) })$ \\ 
&  & 
\end{tabular}%
\end{equation}

\section{Action of 3D $\mathcal{N}=4$ on Lattice $\mathcal{L}%
_{3D}^{su_{3}\times u_{1}}$}

First we give the expression of the fields on the lattice $\mathcal{L}%
_{3D}^{su_{3}\times u_{1}}$; then we build the gauge invariant action on $%
\mathcal{L}_{3D}^{su_{3}\times u_{1}}$.

\subsection{Fields on $\mathcal{L}_{3D}^{su_{3}\times u_{1}}$: a dictionary}

Under discretization of the the complex 3D continuum into the crystal $%
\mathcal{L}_{3D}^{su_{3}\times u_{1}}$, the analogue of coordinate variables
and fields are given by the following dictionary

\begin{itemize}
\item \emph{Coordinates variables}%
\begin{equation}
\begin{tabular}{lll}
continuum & : \  \  \  \  \  \  \  \  & crystal $\mathcal{L}_{3D}^{su_{3}\times
u_{1}}$ \\ 
$(z^{a\left( -2q\right) },\bar{z}_{a}^{\left( +2q\right) })$ &  & $\vec {R}%
_{n}$ \\ 
$z^{a\left( -2q\right) }+dz^{a\left( -2q\right) }$ \  \  \  \  \  \  \  \  &  & $%
\vec{R}_{n}+\vec{V}_{I}^{\left( -2q\right) }$ \\ 
$\bar{z}_{a}^{\left( +2q\right) }+d\bar{z}_{a}^{\left( +2q\right) }$ &  & $%
\vec{R}_{n}+\vec{V}_{I}^{\left( +2q\right) }=\vec{R}_{n}-\vec{V}_{I}^{\left(
-2q\right) }$ \\ 
&  & 
\end{tabular}
\label{dic}
\end{equation}
with, by setting $L=1,$ 
\begin{equation}
\begin{tabular}{lll}
$\vec{V}_{I}^{\left( +2q\right) }$ & $=$ & $\left( -\vec{\lambda}%
_{a},+2q\right) $ \\ 
$\vec{V}_{I}^{\left( -2q\right) }$ & $=$ & $\left( +\vec{\lambda}%
_{a},-2q\right) $%
\end{tabular}%
\end{equation}

\item \emph{Bosonic fields}

\begin{equation}
\begin{tabular}{lll}
continuum & : \  \  \  \  \  \  \  \  & crystal $\mathcal{L}_{3D}^{su_{3}\times
u_{1}}$ \\ 
$\mathcal{G}^{a\left( -2q\right) }\left( z,\bar{z}\right) $ & \  \  \  \  & $%
U_{n}^{I\left( -2q\right) }$ \\ 
$\mathcal{G}_{a}^{\left( +2q\right) }\left( z,\bar{z}\right) $ &  & $%
U_{n,I}^{\left( +2q\right) }$ \\ 
$\mathcal{F}^{ab\left( -4q\right) }\left( z,\bar{z}\right) $ \  \  \  \  \  \  \  \ 
&  & $W_{n}^{IJ\left( -4q\right) }$ \\ 
$\mathcal{\bar{F}}_{ab}^{\left( +4q\right) }\left( z,\bar{z}\right) $ &  & $%
W_{n,IJ}^{\left( +4q\right) }$ \\ 
$\mathcal{F}^{\left( 0\right) }\left( z,\bar{z}\right) $ &  & $W_{n}^{\left(
0\right) }$ \\ 
&  & 
\end{tabular}%
\end{equation}
with%
\begin{equation}
\begin{tabular}{lll}
$U_{n}^{I\left( -2q\right) }$ & $=$ & $U^{I\left( -2q\right) }\left(
R_{n}\right) $ \\ 
$U_{n,I}^{\left( +2q\right) }$ & $=$ & $U_{I}^{\left( +2q\right) }\left(
R_{n}\right) $ \\ 
&  &  \\ 
$W_{n}^{IJ\left( -4q\right) }$ & $=$ & $U^{I\left( -2q\right) }\left(
R_{n}\right) U^{J\left( -2q\right) }(R_{n}+V_{I}^{\left( -2q\right) })-$ \\ 
&  & $U^{J\left( -2q\right) }\left( R_{n}\right) U^{I\left( -2q\right)
}(R_{n}+V_{J}^{\left( -2q\right) })$ \\ 
&  &  \\ 
$W_{n,IJ}^{\left( +4q\right) }$ & $=$ & $U_{J}^{\left( +2q\right)
}(R_{n}+V_{I}^{\left( -2q\right) })U_{I}^{\left( +2q\right) }\left(
R_{n}\right) -$ \\ 
&  & $U_{I}^{\left( +2q\right) }(R_{n}+V_{J}^{\left( -2q\right)
})U_{J}^{\left( +2q\right) }\left( R_{n}\right) $ \\ 
&  &  \\ 
$W_{n}^{\left( 0\right) }$ & $=$ & $U^{I\left( -2q\right) }\left(
R_{n}\right) U_{I}^{\left( +2q\right) }\left( R_{n}\right) -$ \\ 
&  & $U_{I}^{\left( +2q\right) }(R_{n}-V_{I}^{\left( -2q\right) })U^{I\left(
-2q\right) }(R_{n}-V_{I}^{\left( -2q\right) })$%
\end{tabular}%
\end{equation}%
\begin{equation}
\end{equation}

\item \emph{Fermionic fields}%
\begin{equation*}
\end{equation*}%
\begin{equation}
\begin{tabular}{lll}
continuum & : \  \  \  \  \  \  \  \  & crystal $\mathcal{L}_{3D}^{su_{3}\times
u_{1}}$ \\ 
$\psi^{\left( -3q\right) }\left( z,\bar{z}\right) $ &  & $\psi _{n}^{\left(
-3q\right) }$ \\ 
$\psi^{I\left( +q\right) }\left( z,\bar{z}\right) $ &  & $\psi _{n}^{I\left(
+q\right) }$ \\ 
$\nabla^{a\left( -2q\right) }\psi^{b\left( +\right) }$ \  \  \  \  \  \  \  \  &  & 
$\psi_{n}^{IJ\left( -q\right) }$ \\ 
$\bar{\nabla}_{a}^{\left( +2q\right) }\psi^{a\left( -\right) }$ &  & $%
\psi_{n,I}^{I\left( +\right) }$ \\ 
&  & 
\end{tabular}%
\end{equation}
with%
\begin{equation}
\begin{tabular}{lll}
$\psi_{n}^{\left( -3q\right) }$ & $=$ & $\psi^{\left( -3q\right) }\left(
R_{n}\right) $ \\ 
$\psi_{n}^{I\left( +q\right) }$ & $=$ & $\psi^{I\left( +q\right) }(R_{n})$
\\ 
&  &  \\ 
$\psi_{n}^{IJ\left( -q\right) }$ & $=$ & $U^{I\left( -2q\right)
}(R_{n}+V_{J}^{\left( -q\right) })\psi^{J\left( +q\right)
}(R_{n}+V_{I}^{\left( -2q\right) })-$ \\ 
&  & $\psi^{J\left( +q\right) }\left( R_{n}\right) U^{I\left( -2q\right)
}(R_{n})$ \\ 
&  &  \\ 
$\psi_{n,I}^{I\left( +3q\right) }$ & $=$ & $\psi^{I\left( +q\right)
}(R_{n}+V_{I}^{\left( -2q\right) })U_{I}^{\left( +2q\right) }\left(
R_{n}\right) -$ \\ 
&  & $U_{I}^{\left( +2q\right) }(R_{n}-V_{I}^{\left( -q\right)
})\psi^{I\left( +q\right) }(R_{n})$ \\ 
&  & 
\end{tabular}
\label{JK}
\end{equation}

\item \emph{Gauge symmetry}\ 
\begin{equation*}
\end{equation*}%
\begin{equation}
\begin{tabular}{lll}
{\small lattice variable} & ${\small \rightarrow}$ & $\  \ ${\small gauge
transform } \\ \hline
&  &  \\ 
$\  \ U_{n}^{I\left( -2q\right) }$ &  & $\  \ G\left( R_{n}\right) U^{I\left(
-2q\right) }\left( R_{n}\right) G^{\dagger}(R_{n}+V_{I}^{\left( -2q\right)
}) $ \\ 
&  &  \\ 
$\  \ U_{n,I}^{\left( +2q\right) }$ &  & $\  \ G(R_{n}+V_{I}^{\left(
-2q\right) })U_{I}^{\left( +2q\right) }\left( R_{n}\right) G^{\dagger
}\left( R_{n}\right) $ \\ 
&  &  \\ 
$\  \ W_{n}^{IJ\left( -4q\right) }$ &  & $\  \ G\left( R_{n}\right)
W_{n}^{IJ\left( -4q\right) }G^{\dagger}(R_{n}+V_{I}^{\left( -2q\right)
}+V_{J}^{\left( -2q\right) })$ \\ 
&  &  \\ 
$\  \ W_{n,IJ}^{\left( +4q\right) }$ &  & $\  \ G(R_{n}+V_{I}^{\left(
-2q\right) }+V_{J}^{\left( -2q\right) })W_{n}^{IJ\left( +4q\right)
}G^{\dagger}\left( R_{n}\right) $ \\ 
&  &  \\ 
$\  \  \psi_{n}^{\left( -3q\right) }$ $\  \ $ &  & $\  \ G\left( R_{n}\right)
\psi^{\left( -3q\right) }\left( R_{n}\right) G^{\dagger}(R_{n}+V_{0}^{\left(
-3q\right) })$ \\ 
&  &  \\ 
$\  \  \psi_{n}^{I\left( +q\right) }$ &  & $\  \ G\left( R_{n}+V_{I}^{\left(
-q\right) }\right) \psi^{I\left( +q\right) }\left( R_{n}\right)
G^{\dagger}(R_{n})$ \\ 
&  &  \\ 
$\  \  \psi_{n}^{IJ\left( -q\right) }$ $\  \ $ &  & $\  \ G(R_{n}+V_{J}^{\left(
-q\right) })\psi^{IJ\left( -q\right) }\left( R_{n}\right) G^{\dagger
}(R_{n}+V_{I}^{\left( -2q\right) })$ \\ 
&  &  \\ 
$\  \  \psi_{n,I}^{I\left( +3q\right) }$ &  & $\  \ G(R_{n}+V_{I}^{\left(
-q\right) }+V_{I}^{\left( -2q\right) })\psi_{I}^{I\left( +3q\right) }\left(
R_{n}\right) G^{\dagger}(R_{n})$ \\ 
&  &  \\ \hline
\end{tabular}
\label{cid}
\end{equation}%
\begin{equation*}
\end{equation*}
\end{itemize}

\  \ 

\subsection{Useful identities}

Here we collect some relations useful for checking gauge invariance of the
lattice field action $\mathcal{L}_{3D}^{su_{3}\times u_{1}}$

\begin{equation}
\begin{tabular}{llllll}
$\vec{V}_{1}^{\left( +q\right) }$ & $=$ & $\left( \vec{\lambda}_{1},q\right) 
$ &  &  &  \\ 
$\vec{V}_{2}^{\left( +q\right) }$ & $=$ & $\left( \vec{\lambda}_{2},q\right) 
$ & $,\qquad$ & $\vec{V}_{I}^{\left( -q\right) }$ & $=-\vec{V}_{I}^{\left(
+q\right) }$ \\ 
$\vec{V}_{3}^{\left( +q\right) }$ & $=$ & $\left( \vec{\lambda}_{3},q\right) 
$ &  &  &  \\ 
&  &  &  &  &  \\ 
$\vec{V}_{1}^{\left( +2q\right) }$ & $=$ & $\left( -\vec{\lambda}%
_{1},2q\right) $ &  &  &  \\ 
$\vec{V}_{2}^{\left( +2q\right) }$ & $=$ & $\left( -\vec{\lambda}%
_{2},2q\right) $ & , & $\vec{V}_{I}^{\left( -2q\right) }$ & $=-\vec{V}%
_{I}^{\left( +2q\right) }$ \\ 
$\vec{V}_{3}^{\left( +2q\right) }$ & $=$ & $\left( -\vec{\lambda}%
_{3},2q\right) $ &  &  &  \\ 
&  &  &  &  &  \\ 
$\vec{V}_{0}^{\left( +3q\right) }$ & $=$ & $\left( \vec{0},3q\right) $ & , & 
$\vec{V}_{0}^{\left( -3q\right) }$ & $=-\vec{V}_{0}^{\left( +3q\right) }$ \\ 
&  &  &  &  & 
\end{tabular}
\label{agr}
\end{equation}
with 
\begin{equation}
\vec{\lambda}_{3}=-\vec{\lambda}_{1}-\vec{\lambda}_{2}
\end{equation}
From these relations, we learn a set of identities; in particular 
\begin{equation}
\begin{tabular}{lll}
&  &  \\ 
$V_{1}^{\left( +q\right) }+V_{2}^{\left( +q\right) }+V_{3}^{\left(
-2q\right) }$ & $=$ & $\vec{0}$ \\ 
$V_{1}^{\left( +q\right) }+V_{2}^{\left( -2q\right) }+V_{3}^{\left(
+q\right) }$ & $=$ & $\vec{0}$ \\ 
$V_{1}^{\left( -2q\right) }+V_{2}^{\left( +q\right) }+V_{3}^{\left(
+q\right) }$ & $=$ & $\vec{0}$%
\end{tabular}%
\end{equation}
and%
\begin{equation}
\begin{tabular}{lll}
&  &  \\ 
$\vec{V}_{1}^{\left( +q\right) }+\vec{V}_{2}^{\left( +q\right) }+\vec {V}%
_{3}^{\left( +q\right) }$ & $=$ & $\vec{V}_{0}^{\left( +3q\right) }$ \\ 
$\vec{V}_{1}^{\left( -q\right) }+\vec{V}_{2}^{\left( -q\right) }+\vec {V}%
_{3}^{\left( -q\right) }$ & $=$ & $\vec{V}_{0}^{\left( -3q\right) }$ \\ 
&  &  \\ 
$\vec{V}_{I}^{\left( +q\right) }+\vec{V}_{I}^{\left( +2q\right) }$ & $=$ & $%
\vec{V}_{0}^{\left( +3q\right) }$ \\ 
$\vec{V}_{I}^{\left( -q\right) }+\vec{V}_{I}^{\left( -2q\right) }$ & $=$ & $%
\vec{V}_{0}^{\left( -3q\right) }$ \\ 
&  & 
\end{tabular}%
\end{equation}
These identities are important for establishing gauge symmetry of monomials
like%
\begin{equation}
\begin{tabular}{llll}
$\left( i\right) $ & : &  & $Tr\left[ \psi^{\left( -3q\right) }\left(
R_{n}\right) \psi_{I}^{I\left( +3q\right) }\left( R_{n}\right) \right] $ \\ 
&  &  &  \\ 
$\left( ii\right) $ & : &  & $\varepsilon_{IJK}Tr\left[ \psi^{I\left(
+q\right) }(R_{n}-V_{I}^{\left( -q\right) })\psi^{JK\left( -q\right) }\left(
R_{n}\right) \right] $ \\ 
&  &  & 
\end{tabular}
\label{fte}
\end{equation}

\begin{itemize}
\item \emph{checking gauge invariance of the term }$\left( i\right) $\newline
Under a gauge transformation $\boldsymbol{G}\left( R_{n}\right) $, the
lattice variables $\psi_{n}^{\left( -3q\right) }$ and the divergence $%
\psi_{n,I}^{I\left( +3q\right) }$ transform as%
\begin{equation}
\begin{tabular}{lll}
&  &  \\ 
$\psi_{n}^{\left( -3q\right) }$ $\  \ $ & $\rightarrow$ & $\  \ G\left(
R_{n}\right) \psi^{\left( -3q\right) }\left( R_{n}\right) G^{\dagger
}(R_{n}+V_{0}^{\left( -3q\right) })$ \\ 
&  &  \\ 
$\psi_{n,I}^{I\left( +3q\right) }$ & $\rightarrow$ & $\  \
G(R_{n}+V_{I}^{\left( -q\right) }+V_{I}^{\left( -2q\right)
})\psi_{I}^{I\left( +3q\right) }\left( R_{n}\right) G^{\dagger}(R_{n})$ \\ 
&  & 
\end{tabular}
\label{tfe}
\end{equation}
Using the identity%
\begin{equation}
V_{I}^{\left( -q\right) }+V_{I}^{\left( -2q\right) }=V_{0}^{\left(
-3q\right) }
\end{equation}
the second term of (\ref{tfe}) reads also like%
\begin{equation}
\begin{tabular}{lll}
&  &  \\ 
$\psi_{n,I}^{I\left( +3q\right) }$ & $\rightarrow$ & $\  \
G(R_{n}+V_{0}^{\left( -3q\right) })\psi_{I}^{I\left( +3q\right) }\left(
R_{n}\right) G^{\dagger}(R_{n})$ \\ 
&  & 
\end{tabular}%
\end{equation}
Therefore the gauge transformation of the term $\psi^{\left( -3q\right)
}\left( R_{n}\right) \psi_{I}^{I\left( +3q\right) }\left( R_{n}\right) $ is
given by%
\begin{align*}
& \\
& Tr\left[ G{\small (R}_{n}{\small )}\psi^{\left( -3q\right) }{\small (R}_{n}%
{\small )}G^{\dagger}{\small (R}_{n}{\small +V}_{{\small 0}}^{\left( {\small %
-3q}\right) }{\small )}G{\small (R}_{n}{\small +V}_{{\small 0}}^{\left( 
{\small -3q}\right) }{\small )}\psi_{I}^{I\left( +3q\right) }{\small (R}_{n}%
{\small )}G^{\dagger}{\small (R}_{n}{\small )}\right] \\
&
\end{align*}
and leads then to%
\begin{align}
& Tr\left[ G{\small (R}_{n}{\small )}\psi^{\left( -3q\right) }{\small (R}_{n}%
{\small )}\psi_{I}^{I\left( +3q\right) }{\small (R}_{n}{\small )}G^{\dagger}%
{\small (R}_{n}{\small )}\right] \\
&  \notag
\end{align}
which, due to the cyclic property of the trace, reduce further to 
\begin{equation}
\begin{tabular}{ll}
& $Tr\left[ \psi^{\left( -3q\right) }{\small (R}_{n}{\small )}\psi
_{I}^{I\left( +3q\right) }{\small (R}_{n}{\small )}\right] $ \\ 
& 
\end{tabular}%
\end{equation}

\item \emph{checking gauge invariance of the term} $\left( ii\right) $%
\newline
Starting from the gauge transformation of the lattice variables $%
\psi_{n}^{I\left( +q\right) }$ and $\psi_{n}^{IJ\left( -q\right) }$ namely%
\begin{equation}
\begin{tabular}{lll}
&  &  \\ 
$\psi_{n}^{I\left( +q\right) }$ & $\rightarrow$ & $\  \ G(R_{n}+V_{I}^{\left(
-q\right) })\psi^{I\left( +q\right) }\left( R_{n}\right) G^{\dagger}(R_{n})$
\\ 
&  &  \\ 
$\psi_{n}^{JK\left( -q\right) }$ & $\rightarrow$ & $\  \
G(R_{n}+V_{K}^{\left( -q\right) })\psi^{JK\left( -q\right) }\left(
R_{n}\right) G^{\dagger}(R_{n}+V_{J}^{\left( -2q\right) })$ \\ 
&  & 
\end{tabular}%
\end{equation}
the putting back into 
\begin{equation}
\varepsilon_{IJK}Tr\left[ \psi^{I\left( +q\right) }(R_{n}+V_{K}^{\left(
-q\right) })\psi^{JK\left( -q\right) }\left( R_{n}\right) \right]
\end{equation}
we get%
\begin{equation}
\begin{tabular}{ll}
&  \\ 
$G{\small (R}_{{\small n}}{\small +V}_{{\small K}}^{{\small (-q)}}{\small +V}%
_{{\small I}}^{{\small (-q)}}{\small )}\psi^{I\left( +q\right) }{\small (R}_{%
{\small n}}{\small +V}_{{\small K}}^{{\small (-q)}}{\small )}G^{\dagger}%
{\small (R}_{{\small n}}{\small +V}_{{\small K}}^{{\small (-q)}}{\small )}$ $%
\  \times$ &  \\ 
&  \\ 
$G{\small (R}_{{\small n}}{\small +V}_{{\small K}}^{{\small (-q)}}{\small )}%
\psi^{JK\left( -q\right) }{\small (R}_{{\small n}}{\small )}G^{\dagger}%
{\small (R}_{{\small n}}{\small +V}_{{\small J}}^{{\small (-2q)}}{\small )}$
&  \\ 
& 
\end{tabular}%
\end{equation}
Invariance of the trace of this quantity requires%
\begin{equation}
\begin{tabular}{lll}
&  &  \\ 
$G^{\dagger \dagger}{\small (R}_{{\small n}}{\small +V}_{{\small J}}^{%
{\small (-2q)}}{\small )}G{\small (R}_{{\small n}}{\small +V}_{{\small K}}^{%
{\small (-q)}}{\small +V}_{{\small I}}^{{\small (-q)}}{\small )}$ & $=$ & $%
I_{id}$ \\ 
&  & 
\end{tabular}%
\end{equation}
requiring in turn the constraint relation%
\begin{equation}
\begin{tabular}{lllll}
&  &  &  &  \\ 
$V_{K}^{\left( -q\right) }+V_{I}^{\left( -q\right) }$ & $=$ & $V_{J}^{\left(
-2q\right) }$ & , & $I\neq J\neq K$ \\ 
&  &  &  & 
\end{tabular}%
\end{equation}
which is identically satisfied due to eqs(\ref{agr}).
\end{itemize}

\subsection{The action on $\mathcal{L}_{3D}^{su_{3}\times u_{1}}$}

Starting from the twisted Lagrangian density $\boldsymbol{L}_{twist}$ (\ref%
{caf}) namely%
\begin{equation}
\begin{tabular}{lll}
&  &  \\ 
$\boldsymbol{L}_{twist}$ & $=$ & $Tr\left[ \mathcal{F}_{ab}^{\left(
+4q\right) }\mathcal{F}^{ab{\small (-4q)}}\right] $ \\ 
&  &  \\ 
&  & $-\frac{\left( \alpha_{2}\right) ^{2}}{4\alpha_{1}\alpha_{4}}Tr\left[ 
\mathcal{F}^{\left( 0\right) }\mathcal{F}^{\left( 0\right) }\right] $ \\ 
&  &  \\ 
&  & $+\left( \frac{\alpha_{2}}{\alpha_{4}}+2\right) Tr\left[ \psi^{a\left(
+q\right) }\nabla_{a}^{\left( +2q\right) }\psi ^{{\small (-3q)}}\right] $ \\ 
&  &  \\ 
&  & $+2\frac{\alpha_{3}}{\alpha_{4}}Tr\left[ \varepsilon_{abc}\psi^{a%
{\small (+q)}}\nabla^{b\left( -2q\right) }\psi^{c\left( +q\right) }\right] $
\\ 
&  & 
\end{tabular}%
\end{equation}
and using the dictionary of subsection \textrm{6.1} between fields in
continuum and lattice variables, we can write down the gauge invariant
action on $\mathcal{L}_{3D}^{su_{3}\times u_{1}}$ following from the above
one. We find:%
\begin{equation}
\begin{tabular}{lll}
&  &  \\ 
$\mathcal{S}_{latt}$ & $=$ & $\dsum \limits_{\mathcal{L}_{3D}^{su_{3}\times
u_{1}}}Tr\left( W_{n}^{IJ\left( -4q\right) }W_{n,IJ}^{\left( +4q\right)
}\right) -\frac{\left( \alpha_{2}\right) ^{2}}{4\alpha_{1}\alpha_{4}}\dsum
\limits_{\mathcal{L}_{3D}^{su_{3}\times u_{1}}}Tr\left( W_{n}^{\left(
0\right) }W_{n}^{\dagger \left( 0\right) }\right) $ \\ 
&  &  \\ 
&  & $\left( 2+\frac{\alpha_{2}}{\alpha_{4}}\right) \dsum \limits_{\mathcal{L%
}_{3D}^{su_{3}\times u_{1}}}Tr\left[ \psi^{\left( -3q\right) }\left(
R_{n}\right) \psi^{I\left( +q\right) }(R_{n}+V_{I}^{\left( {\small -2q}%
\right) })U_{I}^{\left( +2q\right) }\left( R_{n}\right) \right] +$ \\ 
&  &  \\ 
&  & $\left( 2+\frac{\alpha_{2}}{\alpha_{4}}\right) \dsum \limits_{\mathcal{L%
}_{3D}^{su_{3}\times u_{1}}}Tr\left[ \psi^{\left( -3q\right) }\left(
R_{n}\right) U_{I}^{\left( +2q\right) }(R_{n}-V_{I}^{\left( -q\right)
})\psi^{I\left( +q\right) }(R_{n})\right] $ \\ 
&  &  \\ 
&  & $+2\frac{\alpha_{3}}{\alpha_{4}}\dsum \limits_{\mathcal{L}%
_{3D}^{su_{3}\times u_{1}}}\varepsilon_{IJK}Tr\left[ \psi^{I\left( +q\right)
}(R_{n}+V_{K}^{\left( -q\right) })\psi^{JK\left( -q\right) }\left(
R_{n}\right) \right] $ \\ 
&  & 
\end{tabular}%
\end{equation}
with $W_{n}^{IJ\left( -4q\right) },$ $W_{n,IJ}^{\left( +4q\right) },$ $%
W_{n}^{\left( 0\right) }$ and $\psi^{JK\left( -q\right) }\left( R_{n}\right) 
$ as in eqs(\ref{JK}) and where $\alpha_{i}$'s are normalization numbers.

\section{Conclusion and comments}

In this paper, we studied twisted \emph{3D} $\mathcal{N}=4$ supersymmetric
YM on a particular 3- dimensional lattice $\mathcal{L}_{3D}^{su_{3}\times
u_{1}}$ having an $SU\left( 3\right) \times U\left( 1\right) $ symmetry and
realized by the fibration 
\begin{equation*}
\mathcal{L}_{2D}^{su_{3}}\times \mathcal{L}_{1D}^{u_{1}}
\end{equation*}
with the 2- dimensional base $\mathcal{L}_{2D}^{su_{3}}=\mathbb{A}_{2}^{\ast
},$ the weight lattice of SU$\left( 3\right) $, and fiber $\mathcal{L}%
_{1D}^{u_{1}}\simeq q\mathbb{Z}$. This fibration is encoded by the
intersection matrix%
\begin{equation*}
\begin{tabular}{lllll}
&  &  &  &  \\ 
$\mathcal{J}_{ij}^{su_{3}\times u_{1}}$ & $=$ & $\left( 
\begin{array}{ccc}
\frac{2}{3}+q^{2} & \frac{1}{3}+2q^{2} & 3q^{2} \\ 
\frac{1}{3}+2q^{2} & \frac{2}{3}+4q^{2} & 6q^{2} \\ 
3q^{2} & 6q^{2} & 9q^{2}%
\end{array}
\right) ,$ & $\det \mathcal{J}_{ij}^{su_{3}\times u_{1}}$ & $=3q^{2}$ \\ 
&  &  &  & 
\end{tabular}%
\end{equation*}
with $q$ a unit charge of $U\left( 1\right) $. The $SU\left( 3\right) \times
U\left( 1\right) $ complex symmetry appearing here is one of the breaking
modes of the $SO_{E}\left( 6\right) $ symmetry of the chiral 6D $\mathcal{N}%
=1$ supersymmetric YM on $\mathbb{R}^{6}$; the usual breaking mode used in
the twisting is given by the real 
\begin{equation*}
SO_{E}\left( 3\right) \times SO_{R}\left( 3\right)
\end{equation*}
symmetry with $SO_{E}\left( 3\right) $ the isotropy group of $\mathbb{R}^{3}$
and $SO_{R}\left( 3\right) $ the R-symmetry. The group $SU\left( 3\right)
\times U\left( 1\right) $ may be therefore viewed as a complexification of
the diagonal symmetry of $SO_{E}\left( 3\right) \times SO_{R}\left( 3\right) 
$. \ 

\  \  \  \  \newline
To that purpose, we first reviewed general aspects of $SO\left( t,s\right) $
spinors in diverse dimensions; then we built the twisted \emph{3D} $\mathcal{%
N}=4$ supersymmetric \textrm{algebra (\ref{CA})} generated, in addition to
the bosonic, by 4 complex fermionic generators%
\begin{equation*}
\begin{tabular}{lll}
$Q^{\left( +3q\right) }$ & , & $Q_{a}^{\left( -q\right) }$%
\end{tabular}%
\end{equation*}
transforming respectively as a complex $SU\left( 3\right) $ singlet and a
complex $SU\left( 3\right) $ triplet carrying moreover non trivial charges
under $U\left( 1\right) $, the number $q$ is a non zero unit charge of $%
U\left( 1\right) $; but its singular limit 
\begin{equation*}
q=0
\end{equation*}
has an interpretation on lattice; it corresponds to the projection of\ $%
\mathcal{L}_{3D}^{su_{3}\times u_{1}}$ down to the base sublattice $\mathcal{%
L}_{2D}^{su_{3}}=\mathbb{A}_{2}^{\ast}$.

\  \  \newline
Then extending ideas from covariant gauge formalism of supersymmetric YM
theories and using the gauge covariant \textrm{superfields (\ref{gcs}), we
studied the superspace formulation of the twisted gauge theory }exhibiting
manifestly invariance under $Q^{\left( +3q\right) }$. This supercharge may
be also interpreted as a particular BRST operator and the corresponding
supersymmetric transformation as BRST transformations. The derivation of the
set of gauge covariant superfields\textrm{\ (\ref{gcs})} is a key step in
our construction since only 1 of the 4 complex (8 real) supersymmetric
charges are off shell; this set is explicitly derived in the appendix, eqs(%
\ref{E1}-\ref{R3}).

\  \  \  \newline
After that, we studied the lattice version of twisted $3D$ $\mathcal{N}=4$
supersymmetric YM living on $\mathcal{L}_{3D}$ given by the fibration 
\begin{equation}
\begin{tabular}{lll}
$\mathcal{L}_{1D}^{u\left( 1\right) }$ & $\rightarrow$ & $\  \  \mathcal{L}%
_{3D}^{su_{3}\times u_{1}}$ \\ 
&  & $\  \  \  \  \downarrow$ \\ 
&  & $\mathcal{L}_{2D}^{su_{3}}=\mathbb{A}_{2}^{\ast}$%
\end{tabular}
\label{D3}
\end{equation}
To achieve the lattice construction, we performed the 3 following steps:

\begin{description}
\item[$\left( a\right) $] developed a method of engineering the crystal $%
\mathcal{L}_{3D}^{su_{3}\times u_{1}}$ with a manifestly $SU\left( 3\right)
\times U\left( 1\right) $ symmetry. This lattice is given by the fibration (%
\ref{D3}); the shape of the base sublattice $\mathbb{A}_{2}^{\ast}$,
corresponds to the projection $q=0$, and is completely given by the inverse
of the Cartan matrix of $SU\left( 3\right) $. The 3D lattice $\mathcal{L}%
_{3D}^{su_{3}\times u_{1}}$ is a twist of the weight lattice $\mathbb{A}%
_{3}^{\ast}$ of $SU\left( 4\right) \simeq SO\left( 6\right) $ 
\begin{equation*}
\begin{tabular}{lll}
$\mathcal{L}_{3D}^{su_{3}\times u_{1}}$ & $\sim$ & $\ $\emph{twist\ of} $%
\mathbb{A}_{3}^{\ast}$ \\ 
&  & $\  \  \  \ $%
\end{tabular}%
\end{equation*}

\item[$\left( b\right) $] worked out the dictionary eqs(\ref{dic}-\ref{cid})
between objects $\mathcal{O}_{cont}$ living in continuum and their analogue $%
\mathcal{O}_{lattice}$ on the lattice $\mathcal{L}_{3D}^{su_{3}\times u_{1}}$%
. The objects include the twisted fields, the coordinates and the
supersymmetric generators.

\item[$\left( c\right) $] built the lattice action $\mathcal{S}_{lattice}$
that is invariant under:\newline
$\left( i\right) $ the $U\left( N\right) $ gauge symmetry,\newline
$\left( ii\right) $ the complex scalar supersymmetric charge $Q^{\left(
+3q\right) }$, \newline
$\left( iii\right) $ the $SU\left( 3\right) \times U\left( 1\right) $
symmetry of $\mathcal{L}_{3D}^{su_{3}\times u_{1}}.$
\end{description}

\  \  \  \  \newline
We conclude this study by making 2 comments; one concerning the reduction to 
\emph{2D} $\mathcal{N}=4$ dimensions; and the other regarding the extension
of the construction to twisted maximal supersymmetric YM in \emph{5D }$%
\mathcal{N}=4$\emph{\ }and\emph{\ 4D} $\mathcal{N}=4$\emph{\ }dimensions.

\begin{description}
\item[$1)$] \emph{Reduction down to} \emph{2D}\newline
The twisted \emph{2D} $\mathcal{N}=4$ SYM, that uses the following $SU\left(
3\right) $ packaging of the fields%
\begin{equation*}
\begin{tabular}{lllll}
&  &  &  &  \\ 
{\small bosons} & : & $\  \  \mathcal{G}^{a}$ & , & $\mathcal{\bar{G}}_{a},\  \ 
$ \\ 
{\small fermions} & : & $\  \  \xi$ & , & $\  \  \psi^{a}$ \\ 
&  &  &  & 
\end{tabular}%
\end{equation*}
lives on the lattice $\mathbb{A}_{2}^{\ast}$; it follows from the \emph{3D} $%
\mathcal{N}=4$ analysis by taking the limit $q=0$. However, to exhibit the
decomposition%
\begin{equation*}
\begin{tabular}{lll}
$\mathbb{Q}_{2\times2}$ & $=$ & $QI+Q_{\mu}\gamma^{\mu}+Q_{\mu \nu}\gamma^{%
\left[ \mu \nu \right] }$ \\ 
$4$ & $=$ & $1+2+1$ \\ 
&  & 
\end{tabular}%
\end{equation*}
using 2 dimensional gamma matrices $\gamma^{\mu}$ to decompose $\mathbb{Q}%
_{2\times2}$ in a similar way to the splitting eq(\ref{44}), we have to
break the $SU\left( 3\right) \times U\left( 1\right) $ symmetry down to 
\begin{equation*}
SU\left( 2\right) \times U\left( 1\right) \times U\left( 1\right)
\end{equation*}
As a consequence of this breaking, twisted \emph{3D} $\mathcal{N}=4$ algebra
leads to a particular class of twisted \emph{2D} $\mathcal{N}=4$
supersymmetry with generators as follows%
\begin{equation*}
\begin{tabular}{lllllll}
&  &  &  &  &  &  \\ 
$SU\left( 3\right) \times U\left( 1\right) $ &  & $\rightarrow$ &  & 
\multicolumn{3}{l}{$SU\left( 2\right) \times U\left( 1\right) _{diag}$} \\ 
$\  \  \  \ Q^{{\small (+3q)}}$ &  &  &  & $Q^{{\small (+3q)}}$ &  &  \\ 
$\  \  \  \ Q_{a}^{{\small (-q)}}$ &  &  &  & $Q_{\alpha}^{{\small (-p-q)}}$ & $%
,$ & $Q^{{\small (+2p-q)}}$ \\ 
$\  \  \  \ P_{a}^{{\small (+2q)}}$ &  &  &  & $P_{\alpha}^{{\small (-p+2q)}}$
& , & $Z^{{\small (+2p+2q)}}$ \\ 
&  &  &  &  &  & 
\end{tabular}%
\end{equation*}
where $U\left( 1\right) _{diag}$ is the diagonal subgroup of $U\left(
1\right) \times U\left( 1\right) $. This superalgebra has two complex $%
SU\left( 2\right) $ scalar supercharges\ $Q^{{\small (+3q)}}$, $Q^{{\small %
(+2p-q)}}$ and an isodoublet $Q_{\alpha}^{{\small (-p-q)}}$ obeying amongst
others the anticommutation relations%
\begin{equation*}
\begin{tabular}{lll}
&  &  \\ 
$\left \{ Q^{{\small (+3q)}},Q_{\alpha}^{{\small (-p-q)}}\right \} $ & $=$ & 
$2P_{\alpha}^{{\small (-p+2q)}}$ \\ 
$\left \{ Q^{{\small (+3q)}},Q^{{\small (+2p-q)}}\right \} $ & $=$ & $2Z^{%
{\small (+2p+2q)}}$ \\ 
&  & 
\end{tabular}%
\end{equation*}
where $P_{\alpha}^{{\small (-p+2q)}}$ refers to bosonic translations and
where the charge $Z^{{\small (+2p+2q)}}$ can be taken equal to zero ($Z^{%
{\small (+2p+2q)}}=0$) if we want to realize both scalar supersymmetries $Q^{%
{\small (+3q)}}$, $Q^{{\small (+2p-q)}}$ on lattice. \newline
The field spectrum describing the on shell degrees of freedom of the twisted
2D $\mathcal{N}=4$ supersymmetry, that follows from the reduction of the
twisted 3D $\mathcal{N}=4$ SYM is, up to some details, given by%
\begin{equation*}
\begin{tabular}{lllllllll}
&  &  &  &  &  &  &  &  \\ 
$SO_{E}\left( 6\right) $ & $:$ & \multicolumn{7}{l}{$\  \  \  \  \  \  \  \  \  \  \  \
\  \  \  \  \ SU\left( 2\right) \times U\left( 1\right) _{diag}$} \\ 
&  &  &  &  &  &  &  &  \\ 
$\  \mathcal{A}_{M}$ & :$\  \  \  \  \  \  \  \  \ $ & $\mathcal{G}^{\alpha \left(
-2q+p\right) }$ &  & $\mathcal{\bar{G}}_{\alpha}^{\left( +2q-p\right) }$ & 
& $\phi^{\left( -2q-2p\right) }$ &  & $\bar{\phi}^{\left( +2p+2q\right) }$
\\ 
$\  \Psi^{A}$ & : & $\psi^{\alpha{\small (+q+p)}}$ &  & $\psi^{\left(
-3q\right) }$ &  & $\psi^{{\small (+q-2p)}}$ &  &  \\ 
&  &  &  &  &  &  &  & 
\end{tabular}%
\end{equation*}
The complex bosonic fields%
\begin{equation*}
\begin{tabular}{lll}
$\mathcal{G}^{\alpha \left( -2q+p\right) }$ & , & $\phi^{\left(
-2q-2p\right) }$%
\end{tabular}%
\end{equation*}
transform respectively in the representation $2_{-2q+p}$ and $1_{-2q-2p}$ of 
$SU\left( 2\right) \times U\left( 1\right) _{diag}$. Similarly, the complex
fermionic fields 
\begin{equation*}
\begin{tabular}{lllll}
$\psi^{\alpha{\small (+q+p)}}$ & , & $\psi^{\left( -3q\right) }$ & , & $%
\psi^{{\small (+q-2p)}}$%
\end{tabular}%
\end{equation*}
transform respectively in $2_{+q+p}$, $1_{-3q}$ and $1_{+q-2p}$. \newline
The lattice $\mathcal{L}_{2D}^{su_{2}\times u_{1}}$, on which live the
twisted lattice 2D $\mathcal{N}=4$ theory, follows by the reduction of (\ref%
{D3}) and is given by the fibration%
\begin{equation*}
\begin{tabular}{lll}
$\mathcal{L}_{1D}^{u\left( 1\right) _{diag}}$ & $\rightarrow$ & $\  \ 
\mathcal{L}_{2D}^{su_{2}\times u_{1}}$ \\ 
&  & $\  \  \  \  \downarrow$ \\ 
&  & $\mathcal{L}_{1D}^{su_{2}}=\mathbb{A}_{1}^{\ast}$ \\ 
&  & 
\end{tabular}%
\end{equation*}
where the base sublattice $\mathbb{A}_{1}^{\ast}$ is the weight lattice of $%
SU\left( 2\right) $.

\item[$2)$] \emph{Extension to 5D} $\mathcal{N}=4$ \emph{supersymmetry}%
\textrm{\footnote{%
see footnote 2}}\newline
The analysis we have given in this paper extends to the case of twisted 
\emph{5D} $\mathcal{N}=4$ supersymmetric YM having \emph{16} supercharges.
The generators of the underlying twisted \emph{5D} $\mathcal{N}=4$
superalgebra carry charges under $SU\left( 5\right) \times U\left( 1\right) $
as follows 
\begin{equation*}
\begin{tabular}{lllllll}
&  &  &  &  &  &  \\ 
{\small fermionic generators} & : & $Q^{\left( +5q\right) }$ &  & $%
Q_{a}^{\left( -3q\right) }$ &  & $Q^{\left[ ab\right] \left( +q\right) }$ \\ 
$SU\left( 5\right) \times U\left( 1\right) $ & : & $1_{+5q}$ &  & $\bar {5}%
_{-3q}$ &  & $10_{+q}$ \\ 
&  &  &  &  &  & 
\end{tabular}%
\end{equation*}
The field spectrum describing the on shell degrees of freedom of this gauge
theory is obtained in 2 steps as follows: first reducing the $\mathcal{N}=1$
gauge multiplet $\left( \mathcal{A}_{M},\Psi^{A}\right) $ in \emph{10D} down
to 5D to get 
\begin{equation*}
\left( A_{\mu},B_{m},\Psi^{\alpha I}\right)
\end{equation*}
transforming into representations of 
\begin{equation*}
SO_{E}\left( 5\right) \times SO_{R}\left( 5\right)
\end{equation*}
then twisting the two $SO\left( 5\right) $ factors. In doing these steps,
one ends, up on complexification, with the complex field spectrum 
\begin{equation*}
\begin{tabular}{lllllll}
&  &  &  &  &  &  \\ 
$SO_{E}\left( 10\right) $ & $:$ & \multicolumn{5}{l}{$\  \  \  \  \  \  \  \  \
SU\left( 5\right) \times U\left( 1\right) $} \\ 
$\  \mathcal{A}_{M}$ & : & $\  \  \mathcal{G}^{a\left( -2q\right) }$ &  & 
\multicolumn{3}{l}{$\mathcal{\bar{G}}_{a}^{\left( +2q\right) }\  \ $} \\ 
$\  \Psi^{A}$ & : & $\  \  \psi^{\left( -5q\right) }$ &  & $\psi^{a\left(
+q\right) }$ &  & $\psi_{\left[ ab\right] }^{\left( -q\right) }$ \\ 
&  &  &  &  &  & 
\end{tabular}%
\end{equation*}
Applying similar techniques used in this paper, one concludes that the \emph{%
5D} lattice on which live the twisted \emph{5D} $\mathcal{N}=4$
supersymmetric YM should be given by the fibration 
\begin{equation*}
\begin{tabular}{lll}
$\mathcal{L}_{1D}^{u\left( 1\right) _{diag}}$ & $\rightarrow$ & $\  \ 
\mathcal{L}_{5D}^{su_{5}\times u_{1}}$ \\ 
&  & $\  \  \  \  \downarrow$ \\ 
&  & $\mathcal{L}_{4D}^{su_{5}}=\mathbb{A}_{4}^{\ast}$%
\end{tabular}%
\end{equation*}
with base sublattice $\mathbb{A}_{4}^{\ast}$ precisely as the one found in 
\cite{1A,2A}. More details and special features of this lattice will be
reported in a future occasion.
\end{description}

\section{Appendix: Building the set covariant superfields}

The aim of this appendix is to derive the set \textrm{(\ref{gcs})} of the
gauge covariant superfields $\Phi_{i}^{\left( q_{i}\right) }$ for describing
twisted chiral \emph{3D} $\mathcal{N}=4$ supersymmetric YM theory. A summary
of this analysis has been given in subsection 3.2.

\subsection{General on scalar supersymmetry in superspace}

First recall that the on shell degrees of freedom of the twisted chiral 
\emph{3D} $\mathcal{N}=4$ supersymmetric YM are as follows%
\begin{equation}
\begin{tabular}{llll}
Fermions & : & $\psi^{\left( -3q\right) }$ \  \  \  \  \  \  & $\psi^{a\left(
+q\right) }$ \  \  \  \  \  \  \\ 
$SU\left( 3\right) \times U\left( 1\right) $ &  & $\bar{1}_{-3q}$ & $3_{+q}$
\\ 
scale mass dim &  & 1 & 1 \\ 
&  &  &  \\ 
Bosons & : & $\mathcal{G}^{a\left( -2q\right) }$ & $\mathcal{\bar{G}}%
_{a}^{\left( +2q\right) }$ \\ 
$SU\left( 3\right) \times U\left( 1\right) $ &  & $3_{-2q}$ & $\bar {3}%
_{+2q} $ \\ 
scale mass dim &  & $\frac{1}{2}$ & $\frac{1}{2}$ \\ 
&  &  & 
\end{tabular}
\label{fs}
\end{equation}
Using the scalar Grassman variable $\theta^{\left( -3q\right) }$, associated
with the scalar supersymmetric charge $Q^{\left( +3q\right) }$, and
auxiliary fields, one may a priori combine these degrees of freedom into
particular superfields as follows%
\begin{equation}
\begin{tabular}{lll}
$\Psi^{{\small (-3q)}}$ & $=$ & $\boldsymbol{\psi}^{\left( -3q\right)
}+\theta^{\left( -3q\right) }\boldsymbol{F}^{\left( 0\right) }$ \\ 
$\mathcal{V}^{a\left( -2q\right) }$ & $=$ & $\boldsymbol{G}^{a\left(
-2q\right) }+\theta^{\left( -3q\right) }\boldsymbol{\psi}^{a{\small (+q)}}$
\\ 
$\Upsilon_{a}^{\left( -q\right) }$ & $=$ & $\boldsymbol{\gamma}_{a}^{\left(
-q\right) }+\theta^{\left( -3q\right) }\boldsymbol{G}_{a}^{\left( +2q\right)
}$ \\ 
&  & 
\end{tabular}
\label{sf}
\end{equation}
Notice that the component fields $\boldsymbol{\digamma}$ are not ordinary
fields since they depend, in addition to the bosonic coordinates z, \={z},
on extra Grassman coordinates $\vartheta^{a\left( +q\right) }$ associated
with the supersymmetric charges $Q_{a}^{\left( -q\right) }$; that is%
\begin{equation}
\boldsymbol{\digamma}=\boldsymbol{\digamma}\left( z,\vartheta \right)
\end{equation}
Explicitly, we have%
\begin{equation}
\begin{tabular}{llll}
$\boldsymbol{\psi}^{\left( -3q\right) }\left( z,\vartheta \right) $ & $=$ & $%
\psi^{\left( -3q\right) }\left( z\right) $ & $+\vartheta^{a\left( -q\right)
}\xi_{a}^{\left( -2q\right) }\left( z\right) +...$ \\ 
&  &  &  \\ 
$\boldsymbol{F}^{\left( 0\right) }\left( z,\vartheta \right) $ & $=$ & $%
F^{\left( 0\right) }\left( z\right) $ & $+\vartheta^{a\left( -q\right)
}\xi_{a}^{\left( +q\right) }\left( z\right) +...$ \\ 
&  &  &  \\ 
$\boldsymbol{G}^{a\left( -2q\right) }\left( z,\vartheta \right) $ & $=$ & $%
\mathcal{G}^{a\left( -2q\right) }\left( z\right) $ & $+\vartheta ^{b\left(
-q\right) }\xi_{b}^{a\left( -q\right) }\left( z\right) +...$ \\ 
&  &  &  \\ 
$\boldsymbol{\psi}^{a{\small (+q)}}\left( z,\vartheta \right) $ & $=$ & $%
\psi^{a{\small (+q)}}\left( z\right) $ & $+\vartheta^{b\left( -q\right)
}\Delta_{b}^{a\left( +2q\right) }\left( z\right) +...$ \\ 
&  &  &  \\ 
$\boldsymbol{\gamma}_{a}^{\left( -q\right) }\left( z,\vartheta \right) $ & $%
= $ & $\gamma_{a}^{\left( -q\right) }\left( z\right) $ & $+\vartheta
^{b\left( -q\right) }\xi_{ba}^{\left( 0\right) }\left( z\right) +...$ \\ 
&  &  &  \\ 
$\mathbf{G}_{a}^{\left( +2q\right) }\left( z,\vartheta \right) $ & $=$ & $%
\mathcal{G}_{a}^{\left( +2q\right) }\left( z\right) $ & $+\vartheta
^{b\left( -q\right) }\xi_{ba}^{\left( +3q\right) }\left( z\right) +...$ \\ 
&  &  & 
\end{tabular}
\label{cm}
\end{equation}
The dependence of these component modes into $\vartheta^{a\left( -q\right) } 
$ is eliminated at the end after integration with respect to $\theta^{\left(
-3q\right) }$ by setting $\vartheta^{a\left( -q\right) }=0$.\newline
Notice moreover that the superfields (\ref{sf}) are not good candidates for
superspace formulation of scalar supersymmetric invariance. The point is
that under gauge symmetry transformations with generic group elements $G$,
the bosonic gauge superfields $U^{a\left( -2q\right) }$ and $V_{a}^{\left(
+2q\right) }$ do not transform covariantly since%
\begin{equation}
\begin{tabular}{lll}
&  &  \\ 
$\mathcal{G}^{a\left( -2q\right) }$ & $\rightarrow$ & $G\mathcal{G}^{a\left(
-2q\right) }G^{-1}+G\partial^{a\left( -2q\right) }G^{-1}$ \\ 
&  &  \\ 
$\mathcal{G}_{a}^{\left( +2q\right) }$ & $\rightarrow$ & $G\mathcal{G}%
_{a}^{\left( +2q\right) }G^{-1}+G\partial_{a}^{\left( +2q\right) }G^{-1}$ \\ 
&  & 
\end{tabular}%
\end{equation}
To overcome this difficulty, one needs to work with the gauge covariant
superfield operators%
\begin{equation}
\begin{tabular}{lll}
$\mathcal{D}^{\left( +3q\right) }$ & $=$ & $D^{\left( +3q\right)
}+ig_{YM}\Upsilon^{\left( +3q\right) }$ \\ 
&  &  \\ 
$\mathcal{D}_{a}^{\left( -q\right) }$ & $=$ & $D_{a}^{\left( -q\right)
}+ig_{YM}\Upsilon_{a}^{\left( -q\right) }$ \\ 
&  &  \\ 
$\tciLaplace_{a}^{\left( +2q\right) }$ & $=$ & $\partial_{a}^{\left(
+2q\right) }+ig_{YM}V_{a}^{\left( +2q\right) }$ \\ 
&  &  \\ 
$\tciLaplace^{a\left( -2q\right) }$ & $=$ & $\partial^{a\left( -2q\right)
}+ig_{YM}U^{a\left( -2q\right) }$ \\ 
&  & 
\end{tabular}
\label{gym}
\end{equation}
and their graded commutators from which we learn the set of gauge covariant
superfields \textrm{(\ref{gcs})}; this set is constructed below.\ 

\subsection{Gauge covariant superfields}

We first give our result regarding the set of gauge covariant superfields;
then we turn to derive it explicitly.

\subsubsection{\emph{the set of superfields}}

Twisted chiral \emph{3D} $\mathcal{N}=4$ supersymmetric YM exhibiting
manifestly the supercharge $Q^{\left( +3q\right) }$ is described in
superspace by the following superfields%
\begin{equation}
\begin{tabular}{lllllll}
{\small Fermionic sector} & : & $\Psi^{\left( -3q\right) }$ &  & $\Phi
_{ab}^{\left( +q\right) }$ &  & $\Psi^{a\left( +q\right) }$ \\ 
$SU\left( 3\right) \times U\left( 1\right) $ & : & $\bar{1}_{-3q}$ &  & $%
3_{+q}$ &  & $3_{+q}$ \\ 
scale mass dim &  & 1 &  & 1 &  & 1 \\ 
&  &  &  &  &  &  \\ 
{\small Bosonic sector} & : & $\mathbb{J}^{\left( 0\right) }$ &  & $\mathbb{E%
}^{ab\left( -4q\right) }$ &  & $\mathbb{F}_{ab}^{\left( +4q\right) }$ \\ 
$SU\left( 3\right) \times U\left( 1\right) $ & : & $1_{0}$ &  & $\bar {3}%
_{-4q}$ &  & $3_{+4q}$ \\ 
scale mass dim &  & $\frac{3}{2}$ &  & $\frac{3}{2}$ &  & $\frac{3}{2}$ \\ 
&  &  &  &  &  & 
\end{tabular}
\label{E1}
\end{equation}
obeying constraint relations to be deriver later on. Their $\theta$-
expansion are given by%
\begin{equation}
\begin{tabular}{lll}
&  &  \\ 
$\Psi^{\left( -3q\right) }$ & $=$ & $\psi^{\left( -3q\right)
}+\theta^{\left( -3q\right) }F^{\left( 0\right) }$ \\ 
$\Phi_{ab}^{\left( +q\right) }$ & $=$ & $\phi_{ab}^{\left( +q\right)
}+\theta^{\left( -3q\right) }\mathcal{F}_{ab}^{\left( +4q\right) }$ \\ 
$\Psi^{a\left( +q\right) }$ & $=$ & $\psi^{a\left( +q\right)
}+\theta^{\left( -3q\right) }f^{a\left( +2q\right) }$ \\ 
&  &  \\ 
$\mathbb{J}^{\left( 0\right) }$ & $=$ & $\mathcal{J}^{\left( 0\right)
}+\theta^{\left( -3q\right) }\nabla_{a}^{\left( +2q\right) }\psi^{a\left(
+q\right) }$ \\ 
$\mathbb{E}^{ab\left( -4q\right) }$ & $=$ & $\mathcal{F}^{ab\left(
-4q\right) }+\theta^{\left( -3q\right) }\left[ \nabla^{a\left( -2q\right)
}\psi^{b\left( +q\right) }-\nabla^{b\left( -2q\right) }\psi^{a\left(
+q\right) }\right] $ \\ 
$\mathbb{F}_{ab}^{\left( +4q\right) }$ & $=$ & $\mathcal{F}_{ab}^{\left(
+4q\right) }+\theta^{\left( -3q\right) }\varkappa_{ab}^{\left( +7q\right) } $
\\ 
&  & 
\end{tabular}
\label{R3}
\end{equation}
In these relations $\psi^{\left( -3q\right) },$ $\psi^{a\left( +q\right) }$
are the twisted fermionic fields of the on shell \textrm{spectrum (\ref{fs})}%
; and $\mathcal{J}^{\left( 0\right) },$ $\mathcal{F}^{ab\left( -4q\right) },$
$\mathcal{F}_{ab}^{\left( +4q\right) }$ as follows%
\begin{equation}
\begin{tabular}{lll}
&  &  \\ 
$\mathcal{F}_{ab}^{\left( +4q\right) }$ & $=$ & $\frac{1}{ig_{YM}}\left[
\nabla_{a}^{\left( +2q\right) },\nabla_{b}^{\left( +2q\right) }\right] $ \\ 
&  &  \\ 
$\mathcal{E}^{ab\left( -4q\right) }$ & $=$ & $\frac{1}{ig_{YM}}\left[
\nabla^{a\left( -2q\right) },\nabla^{b\left( -2q\right) }\right] $ \\ 
&  &  \\ 
$\mathcal{J}^{\left( 0\right) }$ & $=$ & $\frac{1}{ig_{YM}}\left[
\nabla_{a}^{\left( +2q\right) },\nabla^{a\left( -2q\right) }\right] $ \\ 
&  & 
\end{tabular}%
\end{equation}
with gauge covariant derivatives as in eq(\ref{ggg}) and gauge coupling
constant $g_{YM}$ scaling like $\left( mass\right) ^{\frac{1}{2}}$.

\subsubsection{\emph{Deriving eqs(\protect \ref{E1})}}

We begin by the superspace realization of the twisted chiral \emph{3D} $%
\mathcal{N}=4$ algebra generated by 
\begin{equation}
\begin{tabular}{lllll}
$D^{{\small (+3q)}},$ & $D_{a}^{{\small (-q)}},$ & $\partial_{a}^{{\small %
(+2q)}},$ & $\partial^{a{\small (-2q)}}$ & 
\end{tabular}
\label{d}
\end{equation}
obeying the anticommutation relations 
\begin{equation}
\begin{tabular}{lll}
$\left \{ D^{{\small (+3q)}},D_{a}^{{\small (-q)}}\right \} $ & $=$ & $%
2\partial_{a}^{{\small (+2q)}}$ \\ 
$\left \{ D_{a}^{{\small (-q)}},D_{b}^{{\small (-q)}}\right \} $ & $=$ & $0$
\\ 
$\left \{ D^{{\small (+3q)}},D^{{\small (+3q)}}\right \} $ & $=$ & $0$%
\end{tabular}%
\end{equation}
To implement gauge symmetry, we covariantize the supersymmetric derivatives (%
\ref{d}) which become%
\begin{equation}
\begin{tabular}{lll}
$\mathcal{D}^{{\small (+3q)}}$ & $=$ & $D^{{\small (+3q)}}+i\Upsilon ^{%
{\small (+3q)}}$ \\ 
$\mathcal{D}_{a}^{{\small (-q)}}$ & $=$ & $D_{a}^{{\small (-q)}}+i\Upsilon
_{a}^{{\small (-q)}}$ \\ 
$\tciLaplace_{a}^{{\small (+2q)}}$ & $=$ & $\partial_{a}^{{\small (+2q)}%
}+iV_{a}^{{\small (+2q)}}$ \\ 
$\tciLaplace^{a{\small (-2q)}}$ & $=$ & $\partial^{a{\small (-2q)}}+iU^{a%
{\small (-2q)}}$ \\ 
&  & 
\end{tabular}
\label{dl}
\end{equation}
where $\Upsilon_{i}^{{\small (q}_{i}{\small )}},$ $\Upsilon^{{\small (+3q)}%
}, $ $V_{a}^{{\small (+2q)}}$, $U^{a{\small (-2q)}}$ are gauge connexions.
These superfield operators transform covariantly under arbitrary gauge
transformation superfield matrices $\boldsymbol{G}$ like%
\begin{equation}
\begin{tabular}{lll}
$\mathcal{D}^{{\small (+3q)}}$ & $\rightarrow$ & $\boldsymbol{G}\mathcal{D}^{%
{\small (+3q)}}\boldsymbol{G}^{-1}$ \\ 
$\mathcal{D}_{a}^{{\small (-q)}}$ & $\rightarrow$ & $\boldsymbol{G}\mathcal{D%
}_{a}^{{\small (-q)}}\boldsymbol{G}^{-1}$ \\ 
$\tciLaplace_{a}^{{\small (+2q)}}$ & $\rightarrow$ & $\boldsymbol{G}%
\tciLaplace_{a}^{{\small (+2q)}}\boldsymbol{G}^{-1}$ \\ 
$\tciLaplace^{a{\small (-2q)}}$ & $\rightarrow$ & $\boldsymbol{G}%
\tciLaplace^{a{\small (-2q)}}\boldsymbol{G}^{-1}$ \\ 
&  & 
\end{tabular}%
\end{equation}
with%
\begin{equation}
\begin{tabular}{lll}
$\boldsymbol{G}$ & $=$ & $\boldsymbol{G}(z,\bar{z},\vartheta^{a\left(
+q\right) };\theta^{\left( -3q\right) })$%
\end{tabular}%
\end{equation}
which, upon expanding in $\theta^{\left( -3q\right) }$- series, reads also%
\begin{equation*}
\begin{tabular}{lll}
$\boldsymbol{G}$ & $=$ & $\boldsymbol{g}+\theta^{\left( -3q\right)
}\varsigma^{\left( +3q\right) }$%
\end{tabular}%
\end{equation*}
with%
\begin{equation}
\begin{tabular}{lll}
$\  \  \  \  \boldsymbol{g}$ & $=$ & $\boldsymbol{g}(z,\bar{z}%
,\vartheta^{a\left( +q\right) })$ \\ 
$\varsigma^{\left( +3q\right) }$ & $=$ & $\varsigma^{\left( +3q\right) }(z,%
\bar{z},\vartheta^{a\left( +q\right) })$%
\end{tabular}
\label{Gg}
\end{equation}
These gauge covariant derivatives (\ref{dl}) are not independent; they obey
some constraint relations required by supersymmetry; in particular the
conventional ones%
\begin{equation}
\begin{tabular}{lll}
$\left \{ \mathcal{D}^{{\small (+3q)}},\mathcal{D}_{a}^{{\small (-q)}}\right
\} $ & $=$ & $2\tciLaplace_{a}^{{\small (+2q)}}$ \\ 
$\left \{ \mathcal{D}_{a}^{{\small (-q)}},\mathcal{D}_{b}^{{\small (-q)}%
}\right \} $ & $=$ & $0$ \\ 
$\left \{ \mathcal{D}^{{\small (+3q)}},\mathcal{D}^{{\small (+3q)}}\right \} 
$ & $=$ & $0$ \\ 
&  & 
\end{tabular}%
\end{equation}
and%
\begin{equation}
\begin{tabular}{lll}
$\left[ \mathcal{D}^{{\small (+3q)}},\tciLaplace_{a}^{{\small (+2q)}}\right] 
$ & $=$ & $0$ \\ 
&  & 
\end{tabular}
\label{fo}
\end{equation}
Being the basic gauge covariant objects of the twisted SYM theory, eqs(\ref%
{dl}) allow to build gauge covariant superfields by taking graded
commutators. The gauge covariant superfields with small scaling mass
dimension are of particular interest; we have:

\begin{description}
\item[1)] \emph{Bosonic superfields }\newline
The bosonic gauge covariant superfields that scale as $\left( mass\right) ^{%
\frac{4-1}{2}}$ are given by the commutators of $\tciLaplace_{a}^{{\small %
(+2q)}}$ and $\tciLaplace^{a\left( -2q\right) }$ as follows%
\begin{equation}
\begin{tabular}{lll}
&  &  \\ 
$\mathbb{J}^{{\small (0)}}$ & $=$ & $\frac{1}{ig_{YM}}\left[
\tciLaplace_{a}^{{\small (+2q)}},\tciLaplace^{a\left( -2q\right) }\right] $
\\ 
$\mathbb{E}^{ab{\small (-4q)}}$ & $=$ & $\frac{1}{ig_{YM}}\left[
\tciLaplace^{a\left( -2q\right) },\tciLaplace^{b\left( -2q\right) }\right] $
\\ 
$\mathbb{F}_{ab}^{{\small (+4q)}}$ & $=$ & $\frac{1}{ig_{YM}}\left[
\tciLaplace_{a}^{{\small (+2q)}},\tciLaplace_{b}^{{\small (+2q)}}\right] $
\\ 
&  & 
\end{tabular}%
\end{equation}
with $g_{YM}$ the gauge coupling constant scaling as $\left( mass\right) ^{%
\frac{1}{2}}$. \newline
Because of (\ref{fo}), the superfield $\mathbb{F}_{ab}^{{\small (+4q)}}$
obeys the remarkable property%
\begin{equation}
\mathcal{D}^{{\small (+3q)}}\mathbb{F}_{ab}^{{\small (+4q)}}=0  \label{fab}
\end{equation}
but the two others do not%
\begin{equation}
\begin{tabular}{lll}
$\mathcal{D}^{{\small (+3q)}}\mathbb{J}^{{\small (0)}}$ & $\neq$ & $0$ \\ 
&  &  \\ 
$\mathcal{D}^{{\small (+3q)}}\mathbb{E}^{ab{\small (-4q)}}$ & $\neq$ & $0$%
\end{tabular}
\label{R1}
\end{equation}
From these constraint eqs, we learn that $\mathbb{F}_{ab}^{{\small (+4q)}}$
should be a highest component of a superfield while $\mathbb{F}_{ab}^{%
{\small (+4q)}}$ and $\mathbb{E}^{ab{\small (-4q)}}$ are good candidates for
superspace formulation of twisted chiral supersymmetric YM.

\item[2)] \emph{Fermionic superfields}\newline
These fermionic gauge covariant superfields we need scale as $\left(
mass\right) ^{\frac{3-1}{2}}$; and are given by 
\begin{equation}
\begin{tabular}{lll}
&  &  \\ 
$\Psi^{{\small (-3q)}}$ & $=$ & $\frac{1}{ig_{YM}}\left[ \mathcal{D}_{a}^{%
{\small (-q)}},\tciLaplace^{a\left( -2q\right) }\right] $ \\ 
$\Psi^{a{\small (+q)}}$ & $=$ & $\frac{1}{ig_{YM}}\left[ \mathcal{D}^{%
{\small (+3q)}},\tciLaplace^{a\left( -2q\right) }\right] $ \\ 
$\Phi_{ab}^{{\small (+q)}}$ & $=$ & $\frac{1}{ig_{YM}}\left[ \mathcal{D}%
_{a}^{{\small (-q)}},\tciLaplace_{b}^{{\small (+2q)}}\right] $ \\ 
&  & 
\end{tabular}%
\end{equation}
with $\Psi^{a{\small (+q)}}$ obeying the property%
\begin{equation}
\begin{tabular}{lll}
$\mathcal{D}^{{\small (+3q)}}\Psi^{a{\small (+q)}}$ & $=$ & $0$ \\ 
&  & 
\end{tabular}
\label{do}
\end{equation}
but%
\begin{equation}
\begin{tabular}{lll}
$\mathcal{D}^{{\small (+3q)}}\boldsymbol{\Omega}^{{\small (-3q)}}$ & $\neq$
& $0$ \\ 
&  &  \\ 
$\mathcal{D}^{{\small (+3q)}}\Phi_{ab}^{{\small (+q)}}$ & $\neq$ & $0$ \\ 
&  & 
\end{tabular}
\label{R2}
\end{equation}
Here also, we learn that $\Psi^{a{\small (+q)}}$ is a highest component of a
superfield while $\Psi^{{\small (-3q)}}$ and $\Phi_{ab}^{{\small (+q)}}$ are
good candidates for the superspace formulation of twisted chiral
supersymmetric YM.

\item[3)] \emph{relations between fermionic and bosonic superfields}\newline
Using the anticommutation relations of the twisted chiral superalgebra, one
finds that the fermionic and bosonic gauge covariant superfields constructed
above are not completely independent; they are related through constraint
relations; in particular%
\begin{equation}
\begin{tabular}{lll}
&  &  \\ 
$\mathcal{D}^{{\small (+3q)}}\Psi^{{\small (-3q)}}$ & $=$ & $2\mathbb{J}%
^{\left( 0\right) }-\mathcal{D}_{a}^{{\small (-q)}}\Psi^{a{\small (+q)}}$ \\ 
&  &  \\ 
$\mathcal{D}^{{\small (+3q)}}\Phi_{ab}^{{\small (+q)}}$ & $=$ & $2\mathbb{F}%
_{ab}^{{\small (+4q)}}$ \\ 
&  &  \\ 
$\mathcal{D}^{{\small (+3q)}}\mathbb{E}^{ab{\small (-4q)}}$ & $=$ & $%
\tciLaplace^{a\left( -2q\right) }\Psi^{b{\small (+q)}}-\tciLaplace^{b\left(
-2q\right) }\Psi^{a{\small (+q)}}$ \\ 
&  &  \\ 
$\tciLaplace_{b}^{{\small (+2q)}}\Psi^{{\small (-3q)}}$ & $=$ & $%
-\tciLaplace^{a\left( -2q\right) }\Phi_{ba}^{{\small (+q)}}$ \\ 
&  & 
\end{tabular}
\label{bc}
\end{equation}
Acting on the first relation by $\mathcal{D}^{{\small (+3q)}}$ and using the
identity $\mathcal{D}^{{\small (+3q)}}\mathcal{D}^{{\small (+3q)}}=0$, we
get another constraint relation on the $\mathbb{J}^{\left( 0\right) }$
superfield%
\begin{equation}
\begin{tabular}{lll}
$\mathcal{D}^{{\small (+3q)}}\mathbb{J}^{\left( 0\right) }$ & $=$ & $%
\tciLaplace_{a}^{{\small (+2q)}}\Psi^{a{\small (+q)}}$ \\ 
&  & 
\end{tabular}
\label{cb}
\end{equation}
Doing the same thing for the second relation, we end with the constraint
relation (\ref{fab}). \newline
In what follows, we choose a particular frame for the gauge fields to build
the $\theta$- expansions of the superfields 
\begin{equation}
\begin{tabular}{lll}
$\Psi^{{\small (-3q)}}$ \  \  \  \  \  & $\Psi^{a{\small (+q)}}$ \  \  \  \  \  & $%
\Phi_{ab}^{{\small (+q)}}$ \\ 
$\mathbb{J}^{{\small (0)}}$ & $\mathbb{E}^{ab{\small (-4q)}}$ & $\mathbb{F}%
_{ab}^{{\small (+4q)}}$ \\ 
&  & 
\end{tabular}%
\end{equation}
solving the constraint relations (\ref{bc}-\ref{cb}).
\end{description}

\  \  \  \  \  \  \  \  \ 

B) \emph{Gauge fixing choice}\newline
To make explicit computations in superspace, we start from the
supersymmetric gauge covariant derivatives of eqs(\ref{dl}); then make the
gauge fixing choice%
\begin{equation}
\Upsilon^{\left( +3q\right) }=0  \label{GF}
\end{equation}
leading to $\mathcal{D}^{{\small (+3q)}}=D^{{\small (+3q)}}$ and then 
\begin{equation}
\mathcal{D}^{{\small (+3q)}}=\frac{\partial}{\partial \theta^{\left(
-3q\right) }}
\end{equation}
This particular choice also allows to expand (\ref{R1}-\ref{R2}) as in eq(%
\ref{R3}). To establish this result, notice first that eq(\ref{GF})
corresponds to reducing the set of gauge transformations (\ref{Gg}) down to
the subset of superfield matrices $\boldsymbol{G}$ having no dependence in $%
\theta^{\left( -3q\right) },$ that is 
\begin{equation}
\begin{tabular}{lll}
$D^{{\small (+3q)}}\boldsymbol{G}$ & $=$ & $0$%
\end{tabular}%
\end{equation}
By substituting back into (\ref{Gg}), the superfield matrix $\boldsymbol{G}$
reduces to $\boldsymbol{g}$ with 
\begin{equation}
\begin{tabular}{lll}
$\boldsymbol{g}$ & $\boldsymbol{=}$ & $\boldsymbol{g}(z,\bar{z},\vartheta
^{a\left( +\right) })$ \\ 
&  & 
\end{tabular}%
\end{equation}
and the gauge covariant derivatives behaving generally like%
\begin{equation}
\begin{tabular}{lll}
$\mathcal{D}^{\left( +3q\right) }$ & $\rightarrow$ & $\boldsymbol{g\mathcal{D%
}^{\left( +3q\right) }g}^{-1}$ \\ 
&  &  \\ 
$\mathcal{D}_{a}^{{\small (-q)}}$ & $\rightarrow$ & $\boldsymbol{g\mathcal{D}%
_{a}^{{\small (-q)}}g}^{-1}$ \\ 
&  &  \\ 
$\tciLaplace_{a}^{{\small (+2q)}}$ & $\rightarrow$ & $\boldsymbol{%
g\tciLaplace_{a}^{{\small (+2q)}}g}^{-1}$ \\ 
&  &  \\ 
$\boldsymbol{\tciLaplace}^{a\left( -2q\right) }$ & $\rightarrow$ & $%
\boldsymbol{g\tciLaplace^{a\left( -2q\right) }g}^{-1}$ \\ 
&  & 
\end{tabular}%
\end{equation}
become%
\begin{equation}
\begin{tabular}{lll}
$\mathcal{D}^{{\small (+3q)}}$ & $=$ & $\frac{\partial}{\partial
\theta^{\left( -3q\right) }}$ \\ 
&  &  \\ 
$\mathcal{D}_{a}^{{\small (-q)}}$ & $=$ & $\frac{\partial}{\partial
\vartheta^{a{\small (+q)}}}+\theta^{\left( -3q\right) }\nabla_{a}^{{\small %
(+2q)}}$ \\ 
&  &  \\ 
$\tciLaplace_{a}^{{\small (+2q)}}$ & $=$ & $\nabla_{a}^{{\small (+2q)}}$ \\ 
&  &  \\ 
$\tciLaplace^{a{\small (-2q)}}$ & $=$ & $\nabla^{a{\small (-2q)}%
}+i\theta^{\left( -3q\right) }\lambda^{a{\small (+q)}}$ \\ 
&  & 
\end{tabular}%
\end{equation}
with $\nabla_{a}^{{\small (+2q)}}$, $\nabla^{a{\small (-2q)}}$ given by (\ref%
{gym}). Substituting these expressions back into eqs(\ref{bc}-\ref{bc}), one
ends with the following $\theta$- expansions%
\begin{equation}
\begin{tabular}{lll}
$\Psi^{{\small (-3q)}}$ & $=$ & $\psi^{\left( -3q\right) }+\theta^{\left(
-3q\right) }F^{\left( 0\right) }$ \\ 
&  &  \\ 
$\mathbb{J}^{{\small (0)}}$ & $=$ & $\mathcal{J}^{\left( {\small 0}\right)
}+\theta^{\left( -3q\right) }\nabla_{a}^{{\small (+2q)}}\psi^{a{\small (+q)}%
} $ \\ 
&  &  \\ 
$\mathbb{E}^{ab{\small (-4q)}}$ & $=$ & $\mathcal{E}^{ab{\small (-4q)}%
}+\theta^{\left( -3q\right) }\left( \nabla^{a\left( -2q\right) }\psi^{b%
{\small (+q)}}-\nabla^{b\left( -2q\right) }\psi^{a{\small (+q)}}\right) $ \\ 
&  &  \\ 
$\Phi_{ab}^{{\small (+q)}}$ & $=$ & $\phi_{ab}^{{\small (+q)}%
}+\theta^{\left( -3q\right) }\mathcal{F}_{ab}^{{\small (+4q)}}$ \\ 
&  & 
\end{tabular}
\label{y}
\end{equation}
and remarkably $\Psi^{a{\small (+q)}}$ has no $\theta^{\left( -3q\right) }$
dependence 
\begin{equation}
\begin{tabular}{lll}
$\Psi^{a{\small (+q)}}$ & $=$ & $\psi^{a{\small (+q)}}$ \\ 
&  & 
\end{tabular}
\label{yy}
\end{equation}
with component field modes as in eqs(\ref{cm}). We also have the constraint
relations%
\begin{equation}
\begin{tabular}{lll}
$\mathcal{D}^{{\small (+3q)}}\Psi^{{\small (-3q)}}+\mathcal{D}_{a}^{{\small %
(-q)}}\Psi^{a{\small (+q)}}$ & $=$ & $2\mathbb{J}^{\left( 0\right) }$ \\ 
$\  \  \  \  \  \  \  \  \  \  \  \  \  \  \  \  \  \  \  \tciLaplace_{b}^{{\small (+2q)}}\Psi^{%
{\small (-3q)}}$ & $=$ & $-\tciLaplace^{a\left( -2q\right) }\Phi_{ba}^{%
{\small (+q)}}$ \\ 
&  & 
\end{tabular}
\label{JO}
\end{equation}
The first constraint is solved as 
\begin{equation}
\begin{tabular}{lll}
$\mathbb{J}^{\left( 0\right) }$ & $=$ & $\mathcal{J}^{\left( 0\right)
}+\theta^{\left( -3q\right) }\nabla_{a}^{{\small (+2q)}}\psi^{a\left(
+q\right) }$%
\end{tabular}
\label{oj}
\end{equation}
and the second leads to 
\begin{equation}
\begin{tabular}{lll}
$\nabla_{b}^{{\small (+2q)}}\psi^{\left( -3q\right) }$ & $=$ & $%
\nabla^{a\left( -2q\right) }\phi_{ab}^{{\small (+q)}}$%
\end{tabular}
\label{sd}
\end{equation}

\  \  \  \  \  \ 

\begin{acknowledgement}
: I thank M. Rausch Traubenberg for helpful discussions on the properties of
Majorana spinors in diverse dimensions. This work is supported by
URAC/O9/CNR.
\end{acknowledgement}

\end{document}